\documentclass[aps,prd,nofootinbib,superscriptaddress, preprint]{revtex4-1}
\pdfoutput=1
\usepackage[utf8]{inputenc}
\usepackage{amsmath,amssymb}
\usepackage{epsfig}
\usepackage{graphicx}
\usepackage[usenames,dvipsnames]{color}
\usepackage{float}
\usepackage[colorlinks,citecolor=blue]{hyperref}
\usepackage{color}
\usepackage{subfigure}

\begin{document}
\title{Dark Sector Assisted Low Scale Leptogenesis from \\ Three Body Decay}

\author{Debasish Borah}
\email{dborah@iitg.ac.in}
\affiliation{Department of Physics, Indian Institute of Technology Guwahati, Assam 781039, India}

\author{Arnab Dasgupta}
\email{arnabdasgupta@protonmail.ch}
\affiliation{Institute of Convergence Fundamental Studies , Seoul-Tech, Seoul 139-743, Korea}

\author{Devabrat Mahanta}
\email{devab176121007@iitg.ac.in}
\affiliation{Department of Physics, Indian Institute of Technology Guwahati, Assam 781039, India}

\begin{abstract}
We study the possibility of realising leptogenesis from three body decay, dark matter (DM) and neutrino mass in a minimal framework. We propose a first of its kind model to implement the idea of leptogenesis from three body decay where CP asymmetry arises from interference of multiple $1 \rightarrow 3$ diagrams using resummed propagators along with DM. The standard model is extended by three heavy singlet fermions, one scalar singlet and one scalar doublet with appropriate discrete charges. Two of these singlet fermions not only play non-trivial roles in generating light neutrino mass at radiative level in scotogenic fashion, but also act as mediators in three body decay of the third singlet fermion leading to desired CP asymmetry through interference of such diagrams. With just one additional field compared to the minimal scotogenic model, we show that successful leptogenesis can occur at a scale as low as approximately 1 TeV which is much lower than the leptogenesis scale found for minimal scotogenic model. Also, the realisation of this three body decay leptogenesis naturally leads to a two component scalar singlet-doublet dark matter scenario offering a rich phenomenology. Apart from having interesting interplay of different couplings involved in processes related to both leptogenesis and dark matter, the model can also be tested at different experiments due to the existence of its particle spectrum at TeV scale.
\end{abstract}
\maketitle

\section{Introduction}
\label{sec1}
The observed asymmetry between matter and antimatter in the present universe has been a longstanding puzzle in particle physics and cosmology. The excess of baryon over antibaryons is so huge that almost all the visible matter in the universe is in the form of baryons only. It is often quantified in terms of baryon to photon ratio, which, according to Planck 2018 data \cite{Tanabashi:2018oca, Aghanim:2018eyx} is
\begin{equation}
\eta_B = \frac{n_{B}-n_{\bar{B}}}{n_{\gamma}} = 6.1 \times 10^{-10}.
\label{etaBobs}
\end{equation}
This excess derived from the measurements of cosmic microwave background (CMB) anisotropies matches very well with the predictions of big bang nucleosynthesis (BBN). The observed excess gives rise to a puzzle because we expect the universe to be started in a baryon symmetric manner. Even if we start with an initial asymmetry, the cosmic inflationary phase will make it negligible in latter epochs of the universe. A baryon symmetric universe can evolve into an asymmetric one dynamically if certain conditions, known as Sakharov's conditions \cite{Sakharov:1967dj} are satisfied. They are namely, (1) baryon number (B) violation, (2) C and CP violation and (3) departure from thermal equilibrium. While all these criteria can be satisfied, in principle, in the standard model (SM) of particle physics and an expanding Friedman-Lemaitre-Robertson-Walker (FLRW) universe, it falls way short of the required amount to produce the huge asymmetry. This has led to different beyond standard model (BSM) proposals out of which the most popular scenario is to consider the existence of some heavy particles whose out-of-equilibrium and B, C, CP violating decays can produce the baryon asymmetry of the universe (BAU) \cite{Weinberg:1979bt, Kolb:1979qa}. Instead of generating baryon asymmetry directly, one can generate an asymmetry in the leptonic sector first through similar lepton number (L) violating decays which can later be converted into the observed baryon asymmetry through $(B+L)$-violating EW sphaleron transitions~\cite{Kuzmin:1985mm}. First proposed by Fukugita and Yanagida more than thirty years back \cite{Fukugita:1986hr}, this alternate way has come to be known as leptogenesis, a review of which can be found in \cite{Davidson:2008bu}. An interesting feature of this scenario is that the required lepton asymmetry can be generated through CP violating out-of-equilibrium decays of the same heavy fields that take part in the seesaw mechanism~\cite{Minkowski:1977sc, Mohapatra:1979ia, Yanagida:1979as, GellMann:1980vs, Glashow:1979nm, Schechter:1980gr} that explains the origin of tiny neutrino masses~\cite{Tanabashi:2018oca}, another observed phenomenon the SM fails to address.

While the asymmetric nature of visible matter has been a longstanding puzzle, another feature of the overall matter component of the present universe adds more to this puzzle. It turns out that only approximately $20\%$ of the present universe's matter density is composed of baryons or visible matter while the rest comes from a mysterious, non-luminous, non-baryonic form of matter, popularly known as dark matter (DM). This is strongly supported by both astrophysical and cosmological observations \cite{Zwicky:1933gu, Rubin:1970zza, Clowe:2006eq, Aghanim:2018eyx, Tanabashi:2018oca}. Similar to the baryon asymmetry, DM abundance is also quantified in terms of a dimensionless quantity as \cite{Aghanim:2018eyx}:
$\Omega_{\text{DM}} h^2 = 0.120\pm 0.001$
at 68\% confidence level (CL). Here $\Omega_{\rm DM}=\rho_{\rm DM}/\rho_{\rm critical}$ is the density parameter of DM and $h = \text{Hubble Parameter}/(100 \;\text{km} ~\text{s}^{-1} 
\text{Mpc}^{-1})$ is a dimensionless parameter of order unity. $\rho_{\rm critical}=3H^2/(8\pi G)$ is the critical density while $H$ is the Hubble parameter. Since none of the SM particles can satisfy the criteria for being a DM candidate, several BSM proposals have been put forward out of which the weakly interacting massive particle (WIMP) is the most widely studied one \cite{Kolb:1990vq, Arcadi:2017kky}. In the WIMP framework, a DM particle having mass around the electroweak scale and interactions similar to the weak interactions gets thermally produced in the early universe, followed by its decoupling from the thermal bath leading to a freeze-out abundance remarkably close to the observed DM abundance. This coincidence is often referred to as the \textit{WIMP Miracle}.

Motivated by the above two phenomena and neutrino mass which SM fails to explain, we consider a scenario where all three phenomena can be explained in a unified manner. One popular scenario, which can accommodate all these three phenomena is the scotogenic framework proposed by Ma in 2006 \cite{Ma:2006km}. In the minimal version of this framework, he SM is extended by two or three copies of $Z_2$ odd fermions singlet under SM gauge symmetries, and an additional scalar field similar to the Higgs doublet of the SM, but odd under the unbroken $Z_2$ symmetry. The salient feature of this framework is the way it connects the origin of light neutrino masses and DM. The unbroken $Z_2$ symmetry leads to a stable DM candidate while the $Z_2$ odd particles generate light neutrino masses at one loop level. Apart from this, the out-of-equilibrium decay of the heavy singlet fermions can also lead to successful leptogenesis at a scale as low as 10 TeV. Such low scale leptogenesis with hierarchical right handed neutrinos has been discussed by several authors \cite{Hambye:2009pw, Racker:2013lua, Clarke:2015hta, Hugle:2018qbw, Borah:2018rca, Mahanta:2019gfe, Mahanta:2019sfo, Sarma:2020msa} while quasi-degenerate right handed neutrino scenario was discussed in earlier works \cite{Kashiwase:2012xd, Kashiwase:2013uy}. For hierarchical right handed neutrinos, such a low scale leptogenesis is a significant improvement over the usual Davidson-Ibarra bound $M_1 > 10^9$ GeV for vanilla leptogenesis in type I seesaw framework \cite{Davidson:2002qv}. 

In this work, we consider the possibility of lowering the scale of leptogenesis further (compared to the ones obtained in previous works) via decay heavy singlet fermions into three different particles including one SM lepton. While leptogenesis from three body decay was covered in earlier works \cite{Masiero:1992bv, Adhikari:1996mc, Sarkar:1996sn, Hambye:2001eu}, a different way of achieving the same was also discussed in different contexts by the authors of \cite{Dasgupta:2019lha, Abdallah:2019tij, Grossman:2003jv, DAmbrosio:2003nfv, Fong:2011yx}. These works either considered soft leptogenesis in a supersymmetric framework or leptogenesis due to CP asymmetry arising from interference of from multiple $2 \rightarrow 2$ or $1 \rightarrow n (n \geq 3)$ diagrams with resummed propagators. While a concrete model to realise leptogenesis from such $ 2 \rightarrow 2$ processes along with dark matter was proposed in \cite{Dasgupta:2019lha}, there has been no concrete model yet, as far as we are aware of, to realise leptogenesis from three body decay where non-zero CP asymmetry arises due to interference of multiple $1 \rightarrow 3$ decay diagrams with resummed propagators together with dark matter. Here we try to implement this idea in a minimal extension of the scotogenic model to achieve successful leptogenesis at a low scale. Such an extension is required as in the minimal scotogenic model we can not have non-zero CP asymmetry from three body decay of right handed neutrinos. We show that successful leptogenesis can be achieved at a scale as low as 2 TeV in this scenario, which gets lowered to even below 1 TeV after including the lepton flavour effects. While, the usual two body decay of right handed neutrinos in scotogenic model can also contribute to lepton asymmetry, we show that the contribution from three body decay dominates in the low mass regime. While building such a setup, we also find that the model naturally predicts a two component dark matter scenario. We discuss interplay of different couplings involved in leptogenesis as well dark matter and show the consistency between the possibility of low scale leptogenesis and correct DM relic density in agreement with all experimental constraints including light neutrino masses and mixing.

This paper is organised as follows. In section \ref{sec2}, we discuss our model in details followed by discussion of leptogenesis and dark matter in section \ref{sec3} and \ref{sec4} respectively. We discuss our results in section \ref{sec5} and conclude in section \ref{sec6}.

\section{The Model}
\label{sec2}
We briefly discuss our model in this section. We stick to a minimal setup required to obtain the desired phenomenology. The SM particle content is extended by three singlet chiral fermions $N_{1,2}, \psi$ and two scalar fields $\eta, S$ which transform non-trivially under the additional $Z_2 \times Z'_2$ symmetry of the model. This additional discrete symmetry is chosen to remove the unwanted terms so that the desired leptogenesis and dark matter phenomenology can be ensured. UV completion of our model can explain the origin of such discrete gauge symmetries from spontaneous breaking of gauge symmetries at high energy scale, for example see \cite{Borah:2012qr, Adhikari:2015woo, Nanda:2017bmi, Barman:2019aku, Biswas:2019ygr, Nanda:2019nqy} and references therein. As we discuss in details below, with the addition of one extra field compared to the minimal scotogenic model \cite{Ma:2006km}, we can achieve much richer phenomenology with a new way of generating lepton asymmetry at a scale, lower compared to the one obtained in minimal scotogenic model \cite{Hugle:2018qbw, Borah:2018rca, Mahanta:2019gfe, Mahanta:2019sfo, Sarma:2020msa}.

\begin{table}
\begin{center}
\begin{tabular}{|c| c | c | c | }
 
  \hline
  
  Particles & $SU(3)_c \times SU(2)_{L} \times U(1)_{Y}$  & $Z_{2}$ & $Z'_{2}$ \\
  \hline 
    $ Q_L$  & $(3, 2, \frac{1}{6})$  & 1  & 1  \\
  $u_R$  & $(3, 1, \frac{2}{3})$   & 1  & 1  \\
    $d_R$  & $(3, 1, -\frac{1}{3})$   & 1  & 1  \\
  \hline
  $ \ell_L$  & $(1, 2, -\frac{1}{2})$  & 1  & 1  \\
  $\ell_R$  & (1, 1, -1)   & 1  & 1  \\
  \hline
    $N_{1,2}$ & (1, 1, 0)  & -1    &  1 \\
      $\psi$  & (1, 1, 0)     &   -1  & -1 \\
      \hline
  $H$   & $(1, 2, \frac{1}{2})$  &  1  & 1  \\
  $\eta$   & $(1, 2, \frac{1}{2})$   &  -1  & 1  \\
 $S$    & (1,1,0)     & 1     & -1 \\

  \hline
\end{tabular}
\end{center}
 \caption{Particle content of the model.}
 \label{tab1}
\end{table}

The particle content of the model is shown in Table \ref{tab1} along with their transformations under the symmetries of the model. The Yukawa Lagrangian can be written as

\begin{equation}
    \mathcal{L}=  -Y_u \overline{Q_L} \Tilde{H} u_R -Y_d \overline{Q_L} H d_R -Y_e \overline{\ell_L} H \ell_R-h_{i\alpha} (\overline{\ell_L})_i \Tilde{\eta} N_{\alpha}- \dfrac{1}{2}  M_{\alpha} \overline{N^{c}_{\alpha}} N_{\alpha}-y_{\alpha} \psi N_{\alpha} S - \dfrac{1}{2} m_{\psi} \overline{\psi^c} \psi
    \label{yukawaL}
\end{equation}
The scalar potential is given by
\begin{align}
V&=\mu^{2}_H H^{\dagger}H + \mu^2_{\eta} \eta^{\dagger} \eta + \frac{1}{2}m^{2} S^2 + \dfrac{\lambda_{1}}{2} (H^{\dagger}H)^{2} + \dfrac{\lambda_{2}}{2} (\eta^{\dagger} \eta)^{2}+ \dfrac{1}{2} \lambda_{S} S^{4}+ \lambda_{3} (H^{\dagger}H)(\eta^{\dagger}\eta) \nonumber \\
& + 
    \lambda_{4} (H^{\dagger} \eta)(\eta^{\dagger}H) + \dfrac{\lambda_{5}}{2} [ (H^{\dagger}\eta)(H^{\dagger}\eta)+ (\eta^{\dagger}H)(\eta^{\dagger}H)] +
   \frac{\lambda_{6}}{2}(H^{\dagger}H)S^{2}+
    \lambda_{7}(\eta^{\dagger}\eta)S^{2}.
    \label{scalarpot}
\end{align}
After the electroweak symmetry breaking (EWSB), the two scalar doublets of the model can be written in the following form in the unitary gauge:
\begin{equation}
H \ = \ \begin{pmatrix} 0 \\  \frac{ v +h }{\sqrt 2} \end{pmatrix} , \qquad \eta \ = \ \begin{pmatrix} \eta^\pm\\  \frac{\eta_{R}+i\eta_{I}}{\sqrt 2} \end{pmatrix} \, ,
\label{eq:idm}
\end{equation}
where $h$ is the SM-like Higgs boson, $\eta_{R}$, $\eta_{I}$ are the CP-even and CP-odd neutral scalars respectively, while $\eta^{\pm}$ are the charged scalars from the additional scalar doublet $\eta$. The vacuum expectation value (VEV) of the SM Higgs is denoted by $v$ while the other two scalars do not acquire any VEVs so that the $Z_2 \times Z'_2$ symmetry of the model remains unbroken.

The masses of the physical scalars at tree level can be written as
\begin{eqnarray}
m_h^2 & \ = \ & \lambda_1 v^2 , \; m_{S}^2 =  m^{2}+\lambda_{6}\dfrac{v^{2}}{2}, \; m_{\eta^\pm}^2 =  \mu^2_{\eta} + \frac{1}{2}\lambda_3 v^2 \nonumber\\
m_{\eta_{R}}^2 & \ = \ & \mu^2_{\eta} + \frac{1}{2}(\lambda_3+\lambda_4+\lambda_5)v^2 \ = \ m^2_{\eta^\pm}+
\frac{1}{2}\left(\lambda_4+\lambda_5\right)v^2  , \nonumber\\
m_{\eta_{I}}^2 & \ = \ & \mu^2_{\eta} + \frac{1}{2}(\lambda_3+\lambda_4-\lambda_5)v^2 \ = \ m^2_{\eta^\pm}+
\frac{1}{2}\left(\lambda_4-\lambda_5\right)v^2 \, .
\label{mass_relation}
\end{eqnarray}
Without any loss of generality, we consider $ \lambda_5 <0$ so that the CP-even scalar is lighter than the CP-odd one. Thus $\eta_R$ is the lightest component of the scalar doublet $\eta$ and also lighter than the singlet fermions $N_{1,2}$. Similarly the singlet scalar $S$ is chosen to be lighter than $\psi$. This ensures $\eta_R, S$ to be the lightest $Z_2$-odd and $Z'_2$-odd particles respectively and hence viable dark matter candidates of the model.

As can be seen from the Yukawa Lagrangian in equation \eqref{yukawaL}, there is no tree level contribution to light neutrino masses, simply because they couple to the heavy neutrinos $N_i$ only via the second scalar doublet $\eta$ which does not acquire any VEV. However, light neutrino masses can arise at radiative level as originally proposed in the context of minimal scotogenic model \cite{Ma:2006km}. In our setup, the additional scalar doublet $\eta$ and the singlet fermions $N_{1,2}$ will go inside the loop which generates the light neutrino masses, the expression for which can be evaluated as ~\cite{Ma:2006km, Merle:2015ica}
\begin{align}
(M_{\nu})_{ij} \ & = \ \sum_k \frac{h_{ik}h_{jk} M_{k}}{32 \pi^2} \left ( \frac{m^2_{\eta_R}}{m^2_{\eta_R}-M^2_\alpha} \: \text{ln} \frac{m^2_{\eta_R}}{M^2_k}-\frac{m^2_{\eta_I}}{m^2_{\eta_I}-M^2_k}\: \text{ln} \frac{m^2_{\eta_I}}{M^2_k} \right) \nonumber \\ 
& \ \equiv  \ \sum_k \frac{h_{ik}h_{jk} M_{k}}{32 \pi^2} \left[L_k(m^2_{\eta_R})-L_k(m^2_{\eta_I})\right] \, ,
\label{numass1}
\end{align}
where 
$M_k$ is the mass eigenvalue of the mass eigenstate $N_k$ in the internal line and the indices $i, j = 1,2,3$ run over the three neutrino generations and $k=1,2$ takes into account of two $N_{\alpha}$. The loop function $L_k(m^2)$ is defined as 
\begin{align}
L_k(m^2) \ = \ \frac{m^2}{m^2-M^2_k} \: \text{ln} \frac{m^2}{M^2_k} \, .
\label{eq:Lk}
\end{align}
From the expressions for physical scalar masses given in equations \eqref{mass_relation}, we can write $m^2_{\eta_R}-m^2_{\eta_I}=\lambda_5 v^2$. Therefore, in the limit $\lambda_5 \to 0$, the neutral components of inert doublet $\eta$ become mass degenerate. Also, a vanishing $\lambda_5$ implies vanishing light neutrino masses which is expected as the $\lambda_5$-term in the scalar potential~\eqref{scalarpot} breaks lepton number by two units, when considered together with the fermion Yukawa Lagrangian~\eqref{yukawaL}. As we will see later, this parameter also plays crucial role in both leptogenesis and DM phenomenology. It should also be noted that the Yukawa coupling $h_{ik}$ is a $3\times 2$ matrix in flavour basis due to the existence of only two right handed neutrinos appearing in light neutrino mass. This predicts a vanishing lightest neutrino mass.

In order to ensure that the choice of Yukawa couplings as well as other parameters involved in light neutrino mass formula discussed above are consistent with the cosmological upper bound on the sum of neutrino masses, $\sum_i m_{i}\leq 0.11$ eV~\cite{Aghanim:2018eyx}, as well as the neutrino oscillation data~\cite{deSalas:2017kay, Esteban:2018azc}, it is often useful to rewrite the neutrino mass formula given in equation \eqref{numass1} in a form similar to the well known the type-I seesaw formula: 
\begin{align}
M_\nu \ = \ h {\Lambda}^{-1} h^T \, ,
\label{eq:nu2}
\end{align}
where we have introduced the diagonal matrix $\Lambda$ with elements
\begin{align}
 \Lambda_\alpha \ & = \ \frac{2\pi^2}{\lambda_5}\zeta_\alpha\frac{2M_\alpha}{v^2} \, , \\
\textrm {and}\quad \zeta_\alpha & \ = \  \left(\frac{M_{\alpha}^2}{8(m_{\eta_R}^2-m_{\eta_I}^2)}\left[L_\alpha(m_{\eta_R}^2)-L_\alpha(m_{\eta_I}^2) \right]\right)^{-1} \, . \label{eq:zeta}
\end{align}
The light neutrino mass matrix~\eqref{eq:nu2} which is complex symmetric by virtue of Majorana nature, can be diagonalised by the usual Pontecorvo-Maki-Nakagawa-Sakata (PMNS) mixing matrix $U$ (in the diagonal charged lepton basis), written in terms of neutrino oscillation data (up to the Majorana phases) as
\begin{equation}
U=\left(\begin{array}{ccc}
c_{12}c_{13}& s_{12}c_{13}& s_{13}e^{-i\delta}\\
-s_{12}c_{23}-c_{12}s_{23}s_{13}e^{i\delta}& c_{12}c_{23}-s_{12}s_{23}s_{13}e^{i\delta} & s_{23}c_{13} \\
s_{12}s_{23}-c_{12}c_{23}s_{13}e^{i\delta} & -c_{12}s_{23}-s_{12}c_{23}s_{13}e^{i\delta}& c_{23}c_{13}
\end{array}\right) U_{\text{Maj}}
\label{PMNS}
\end{equation}
where $c_{ij} = \cos{\theta_{ij}}, \; s_{ij} = \sin{\theta_{ij}}$ and $\delta$ is the leptonic Dirac CP phase. The diagonal matrix $U_{\text{Maj}}=\text{diag}(1, e^{i\alpha}, e^{i(\zeta+\delta)})$ contains the undetermined Majorana CP phases $\alpha, \zeta$. The diagonal light neutrino mass matrix is therefore,
\begin{align}
D_\nu \ = \ U^\dag M_\nu U^* \ = \ \textrm{diag}(m_1,m_2,m_3) \, .
\end{align}   
where the light neutrino masses can follow either normal ordering (NO) or inverted ordering (IO). As mentioned earlier, the model predicts a vanishing lightest neutrino mass implying $m_1=0$ (NO) and $m_3=0$ (IO). Since the inputs from neutrino oscillation data are only in terms of the two mass squared differences and three mixing angles, it would be useful for our purpose to express the Yukawa couplings $(h)$ in terms of light neutrino parameters. This is possible through the 
Casas-Ibarra (CI) parametrisation \cite{Casas:2001sr} extended to radiative seesaw model \cite{Toma:2013zsa} which allows us to write the Yukawa coupling matrix satisfying the neutrino data as
\begin{align}
h_{i\alpha} \ = \ \left(U D_\nu^{1/2} R^{\dagger} \Lambda^{1/2} \right)_{i\alpha} \, ,
\label{eq:Yuk}
\end{align}
where $R$ is an arbitrary complex orthogonal matrix satisfying $RR^{T}=\mathbb{1}$. It is worth mentioning that, since we have only two right handed neutrinos $N_{1,2}$ taking part in generating radiative light neutrino masses, the lightest neutrino mass is vanishing. Also, in case of only two right handed neutrinos, the $R$ matrix is a function of only one complex rotation parameter $z=z_R + i z_I, z_R \in [0, 2\pi], z_I \in \mathbb{R}$ \cite{Ibarra:2003up} which can affect the results of leptogenesis as we discuss below.

\subsection{Constraints on Model Parameters}
\label{sec:constraint}
Precision measurements at LEP experiment forbids additional decay channels of the SM gauge bosons. For example, it strongly constrains the decay channel $Z \rightarrow \eta_R \eta_I$ requiring $m_{\eta_R} + m_{\eta_I} > m_Z$. Additionally, LEP precision data also rule out the region $m_{\eta_R} < 80 \; {\rm GeV}, m_{\eta_I} < 100 \; {\rm GeV}, m_{\eta_I} - m_{\eta_R} > 8 {\rm GeV}$ \cite{Lundstrom:2008ai}. We take the lower bound on charged scalar mass $m_{\eta^\pm}> 90$ GeV. If $m_{\eta_R, \eta_I} < m_h/2$, the large hadron collider (LHC) bound on invisible Higgs decay comes into play. The constraint on the Higgs invisible decay branching fraction from the ATLAS experiment at LHC is \cite{Aaboud:2019rtt}
 \begin{equation}
	\mathcal{B}(h \to \text{Invisible}) = \frac{\Gamma(h \to \text{Invisible})}{\Gamma(h \to SM) + \Gamma(h \to \text{Invisible}) } \leq 26\%
	\end{equation}
	while the recent ATLAS announcement \cite{ATLAS:2020cjb} puts a more stringent constraint at $13\%$. This can constrain the SM Higgs coupling with $\eta_R, \eta_I, S$ namely $\lambda_3+\lambda_4 \pm \lambda_5, \lambda_6$ respectively to be smaller than around $10^{-3}$ in the regime $m_{\eta_R}, m_{\eta_I}, m_S < m_h/2$ which however remains weaker than DM direct detection bounds in this mass regime (see for example, \cite{Borah:2017hgt}).

Additionally, the LHC experiment can also put bounds on the scalar masses in the model, specially the components of scalar doublet $\eta$ as they can be pair produced copiously in proton proton collisions leading to different final states which are being searched for. Depending upon the mass spectrum of its components, the heavier ones can decay into the lighter ones and a gauge boson, which finally decays into a pair of leptons or quarks. Therefore, we can have either pure leptonic final states plus missing transverse energy (MET), hadronic final states plus MET or a mixture of both. The MET corresponds to DM or light neutrinos. In several earlier works \cite{Miao:2010rg, Gustafsson:2012aj, Datta:2016nfz}, the possibility of opposite sign dileptons plus MET was discussed. In \cite{Poulose:2016lvz}, the possibility of dijet plus MET was investigated with the finding that inert scalar masses up to 400 GeV can be probed at high luminosity LHC. In another work \cite{Hashemi:2016wup}, tri-lepton plus MET final states was also discussed whereas mono-jet signatures have been studied by the authors of \cite{Belyaev:2016lok, Belyaev:2018ext}. The enhancement in dilepton plus MET signal in the presence of additional vector like singlet charged leptons was also discussed in \cite{Borah:2017dqx}. Exotic signatures like displaced vertex and disappearing or long-lived charged track for compressed mass spectrum of inert scalars and singlet fermion DM was studied recently by the authors of \cite{Borah:2018smz}. 

%
%

In addition to the collider or direct search constraints, there exists theoretical constraints also. For instance, the scalar potential of the model should be bounded from below in any field direction. This criteria leads to the following co-positivity conditions~\cite{Sher:1988mj, Kannike:2012pe, Chakrabortty:2013mha}:
\begin{align}
& \lambda_{1, 2, S} \geq 0,~\lambda_3+\sqrt{\lambda_{1} \lambda_2} \geq 0, \nonumber \\
 & \lambda_3+\lambda_4 -|\lambda_3|+\sqrt{\lambda_{1}\lambda_2} \geq 0, ~\lambda_6 + 2 \sqrt{\lambda_1 \lambda_S} \geq 0, \nonumber \\
& \sqrt{\lambda_1 \lambda_2 \lambda_S}+2(\lambda_3 \sqrt{\lambda_S}+\lambda_7 \sqrt{\lambda_1}+\frac{\lambda_6}{2} \sqrt{\lambda_2}) \nonumber \\
& +4 \sqrt{\frac{1}{2}(\lambda_3+\frac{1}{2}\sqrt{\lambda_1 \lambda_2})(\lambda_6+\sqrt{\lambda_1 \lambda_S})(\lambda_7+\frac{1}{2}\sqrt{\lambda_2 \lambda_6})} \geq 0.
\end{align}
The coupling constants appeared in above expressions are evaluated at the electroweak scale, $v$. Also, in order to avoid perturbative breakdown, all dimensionless couplings like quartic couplings $(\lambda_i, \lambda_S)$, Yukawa couplings $(Y_{ij}, h_{i\alpha}, y_i)$, gauge couplings $(g_i)$  should obey the the perturbativity conditions:
\begin{eqnarray}
|\lambda_{1, 2, 3, 4, 5, 6, 7}| < 4 \pi,~|\lambda_{S}| < 4 \pi,~ |Y_{u, d, e}, h_{ij}, y_{1,2}| < \sqrt{4 \pi},~|g_s, g, g'| < \sqrt{4\pi}
\label{eq:PerC}
\end{eqnarray}

\begin{figure}[h]
    \centering
    \includegraphics[scale=.4]{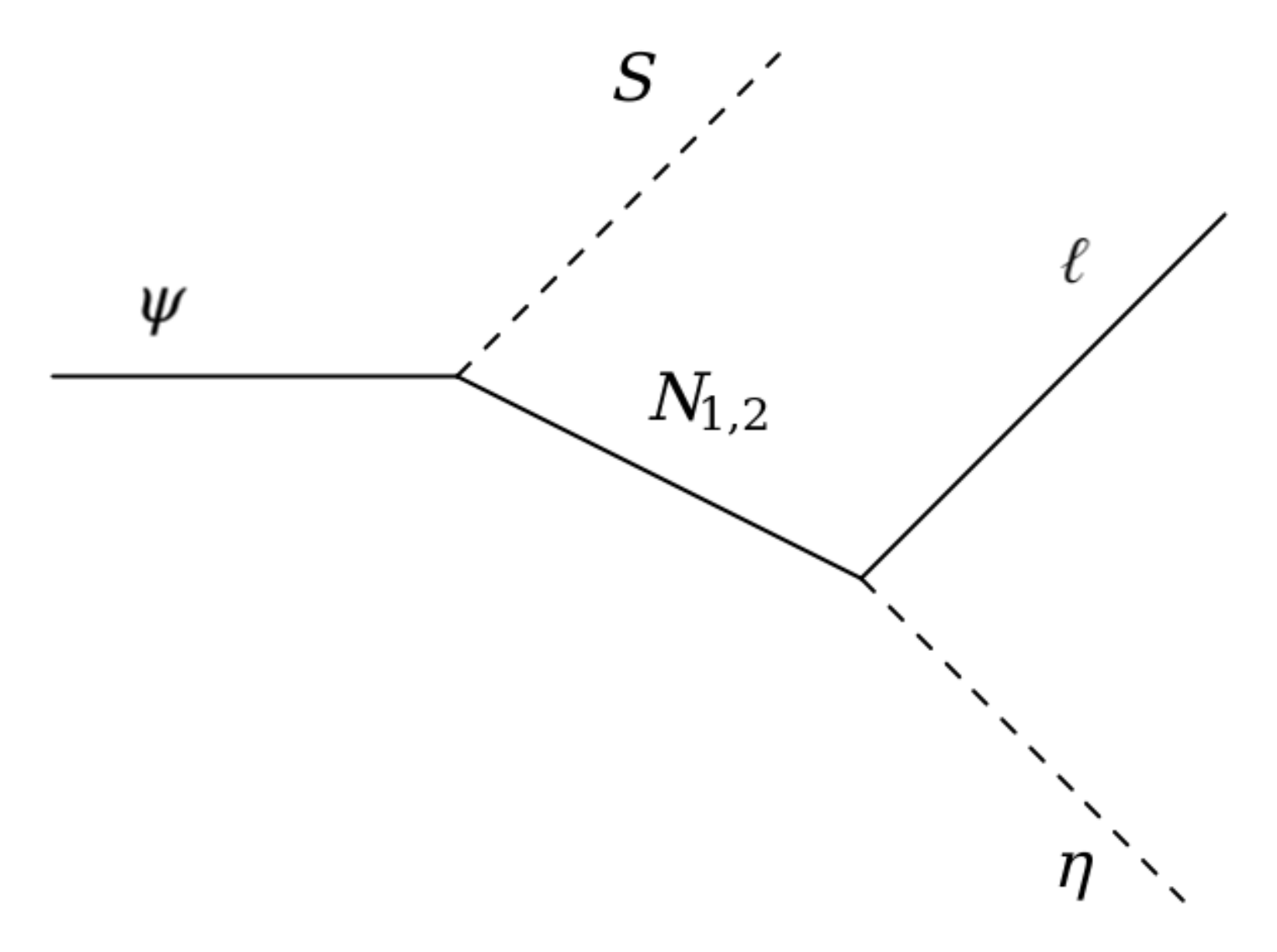}
    \caption{Three body decay of singlet fermion $\psi$.}
    \label{fig:decay}
\end{figure}

\section{Leptogenesis}
\label{sec3}
In this section, we discuss the details of a new way of generating lepton asymmetry at low scale in our model. Note that, similar to the minimal scotogenic model, here also there are two different ways of generating lepton asymmetry: out of equilibrium decay of the $N_i$~\cite{Hambye:2009pw, Kashiwase:2012xd, Kashiwase:2013uy, Racker:2013lua, Clarke:2015hta, Hugle:2018qbw, Borah:2018rca, Mahanta:2019gfe, Mahanta:2019sfo, Sarma:2020msa} or annihilation/scattering of dark sector particles \cite{Borah:2018uci, Borah:2019epq}. In \cite{Hambye:2009pw, Racker:2013lua}, the authors considered a hierarchical right handed neutrino spectrum to show that successful leptogenesis from two body decay can occur at a scale as low as a few tens of TeV. Leptogenesis with quasi-degenerate right handed neutrinos in scotogenic model was discussed in \cite{Kashiwase:2012xd, Kashiwase:2013uy}. In more recent works \cite{Hugle:2018qbw, Borah:2018rca}, successful leptogenesis was shown to be possible at a scale as low as 10 TeV even with hierarchical right handed neutrinos while requiring to be in a weak washout regime predicting a vanishingly small lightest neutrino mass. If leptogenesis occurs only from two heavy neutrinos, then the scale of leptogenesis is pushed above the TeV scale by a few order of magnitudes \cite{Mahanta:2019gfe}. This will correspond to our scenario if we do not have $\psi, S$ in our model. To summarise, it has been shown in the above mentioned works that the scale of leptogenesis in scotogenic model can be as low as a few TeV without requiring any resonance enhancement arising due to tiny mass splitting of right handed neutrinos. This significant improvement over the usual Davidson-Ibarra bound on the scale of leptogenesis $M_N > 10^9$ GeV for vanilla leptogenesis in type I seesaw framework \cite{Davidson:2002qv} makes the scotogenic model a very attractive and testable framework for leptogenesis. We also note that high scale leptogenesis in scotogenic model was also studied recently by the authors of \cite{Huang:2018vcr}.

In addition to the usual $1 \rightarrow 2$ decay or $2 \rightarrow 2$ annihilations as sources of lepton asymmetry, we can also have $1 \rightarrow 3$ decay in our model. Such three body decay as a source of lepton asymmetry was discussed earlier by several authors \cite{Masiero:1992bv, Adhikari:1996mc, Sarkar:1996sn, Hambye:2001eu} in the different contexts like radiative seesaw models, R parity violating supersymmetry and so on. Our present work is motivated two features namely, (i) dark matter particles assist in such three body decay processes and (ii) non-zero lepton asymmetry can be generated due to interference of three body decay diagrams with two different resummed propagators. We show that for our chosen regime of parameter space, such three body decay leptogenesis can be dominant over other possible sources of leptogenesis and the scale of leptogenesis can be lower than what was found by considering two body decay or annihilation processes discussed in earlier works.

In our model, we consider the three body decay of singlet fermion $\psi$ as the origin of CP asymmetry through the process shown in figure \ref{fig:decay}\footnote{In the absence of this process, the source of leptogenesis will be from decay of $N_{1,2}$ only and it was shown earlier that in such two right handed neutrino limit of scotogenic model, the scale of leptogenesis is pushed towards higher side \cite{Mahanta:2019gfe}.}. To prevent the two body decay of $\psi$ into $N_i, S$ we impose the kinematical constraint $m_{\psi} < M_{\alpha}+m_{S}$. The relevant decay processes that can generate Lepton asymmetry are $\psi \longrightarrow S l \eta$ and $N_{1} \longrightarrow \eta l$. Although $N_2$ decay can also generate lepton asymmetry in principle, we consider the asymmetry generated by $N_2$ decay or any pre-existing asymmetry to be negligible due to strong washout effects mediated either by $N_{1}$ or $N_{2}$ themselves, to be discussed below. The corresponding CP asymmetry parameters are defined as
\begin{equation}\label{epsilonpsi}
\epsilon_{\psi}  =  \dfrac{\Gamma_{\psi \longrightarrow S l_i \eta}-\Gamma_{\psi \longrightarrow S \Bar{l_i} \eta^{*}}}{\Gamma_{\psi \longrightarrow S l_i \eta}+\Gamma_{\psi \longrightarrow S \Bar{l_i} \eta^{*}}}, \;
\epsilon_{(N_{1})_i}  =  \dfrac{\Gamma_{N_{1} \longrightarrow l_i \eta}-\Gamma_{N_{1}\longrightarrow \Bar{l_i}\eta^{*}}}{\Gamma_{N_{1} \longrightarrow l_i \eta}+\Gamma_{N_{1}\longrightarrow \Bar{l_i}\eta^{*}}}
\end{equation}
The details of these CP asymmetry parameters are given in appendix \ref{appen1} and \ref{appen2} respectively. Appendix \ref{appen1} contains the details of one-loop calculation for resummed propagator in such three body decay, together with the usual tree plus one-loop level interference, giving rise to the same results. In the unflavoured regime, we sum over lepton flavours $i$ to obtain the net CP symmetries $\epsilon_{\psi}, \epsilon_{N_1}$ which will be used in the Boltzmann equations. Since we are considering sum over all lepton flavours, we are not going to show the flavour index explicitly in the following discussions.

Along with these two decay processes contributing to the creation of lepton asymmetry, there are washout processes too which tend to destroy the asymmetry created. The relevant washout processes in our model can be categorised as
\begin{itemize}
  \item Inverse decays:  $\ell \eta \longrightarrow N_{1}, \ell \eta S \longrightarrow \psi$
  \item $\Delta L=1$ scatterings: $S \ell \longrightarrow \psi \eta$,  $\ell \eta \longrightarrow \psi S$, $\psi \ell  \longrightarrow S \eta $, $\ell \eta \longrightarrow N_{1} (W^{\pm},Z)$.
     \item $\Delta L=2$ scatterings: $\ell \ell \longrightarrow N_{1}N_{1}$, $ll \longrightarrow \eta \eta$, $\eta \ell \longrightarrow \eta^{*} \Bar{\ell}$.
        \end{itemize}
       
The Boltzmann equations relevant for leptogenesis in this model can be summarised as 
\begin{align}
\dfrac{dn_{\psi}}{dz} & = -D_{\psi}(n_{\psi}-n_{\psi}^{\rm eq})+D_{N_{1}\longrightarrow \psi S}(n_{N_{1}}-n_{N_{1}}^{eq})-W_{ID_{N_{1}\longrightarrow \psi S}}n_{\psi} \nonumber \\ 
&- \dfrac{s}{H(z)z}[ (n_{\psi}n_{\eta}-n_{\psi}^{\rm eq}n_{\eta}^{\rm eq}) \langle \sigma v \rangle_{\psi \eta \longrightarrow S l}  + [n_{\psi} n_{S}-n_{\psi}^{\rm eq} n_{S}^{\rm eq}] \langle \sigma v \rangle_{\psi S \longrightarrow l \eta}   ], \\
\label{eq:4}
\end{align}
\begin{align}
\dfrac{dn_{N_{1}}}{dz} & = -D_{N_{1}}(n_{N_{1}}-n_{N_{1}}^{\rm eq})-D_{N_{1}\longrightarrow \psi S}(n_{N_{1}}-n_{N_{1}}^{\rm eq})-\dfrac{s}{H(z)z} [ (n_{N_{1}}^2-(n_{N_{1}}^{\rm eq})^{2})\langle \sigma v \rangle_{N_{1}N_{1} \longrightarrow l l} \nonumber \\
 & +  [n_{N_{1}}n_{SM}-n_{N_{1}}^{\rm eq}n_{SM}^{\rm eq}] \langle \sigma v \rangle_{\eta l \longrightarrow N_{1} (W^{\pm},Z)}   ],       \\
\label{eq:5}
\end{align}
\begin{align}\label{eq:6}
        \dfrac{dn_{B-L}}{dz} & =-\epsilon_{\psi}D_{\psi}(n_{\psi}-n_{\psi}^{\rm eq})-\epsilon_{N_{1}}D_{N_{1}}(n_{N_{1}}-n_{N_{1}}^{\rm eq})-(W_{N_{1}}+W_{\psi})n_{B-L} \nonumber \\ 
        & -\dfrac{s}{H(z)z} [ \Gamma_{S l \longrightarrow \psi \eta} + \Gamma_{l \eta \longrightarrow \psi S}+ \Gamma_{ll \longrightarrow \eta \eta}+ \Gamma_{ll \longrightarrow N_{1}N_{1}}+ \Gamma_{l \eta \longrightarrow (N_{1} W^{\pm},Z)}  + \Gamma_{\eta l \longrightarrow \eta^{*} \Bar{l}} ]n_{B-L}.
\end{align}
 
In the above equations $z=\dfrac{m_{\psi}}{T}$ and $n_{f}^{\rm eq}=\frac{z^2}{2}\kappa_2(z)$ is the equilibrium number density of $f \equiv N_1, \psi$ (with $\kappa_i(z)$ being the modified Bessel function of $i$-th kind). The quantity $D_f$ on the right hand side of above equations is
\begin{align}
D_{N_{1}} & = K_{N_{1}}z \left(\dfrac{M_{1}}{m_{\psi}} \right) \dfrac{\kappa_{1}\left[z\left(\dfrac{M_{1}}{m_{\psi}} \right)\right]}{\kappa_{2}\left[z\left(\dfrac{M_{1}}{m_{\psi}} \right)\right]} , \nonumber\\
D_{\psi} & =  K_{\psi}z \dfrac{\kappa_{1}(z)}{\kappa_{2}(z)}, \; D_{N_{1}\longrightarrow \psi S} = K_{N_{1}\longrightarrow \psi S}z \left(\dfrac{M_{1}}{m_{\psi}} \right)\dfrac{\kappa_{1}\left[z\left(\dfrac{M_{1}}{m_{\psi}} \right)\right]}{\kappa_{2}\left[z\left(\dfrac{M_{1}}{m_{\psi}} \right)\right]}.
\end{align}
Here, the decay parameters are defined as
\begin{equation}
K_{N_{1}}=  \dfrac{\Gamma_{N_{1} \longrightarrow l \eta}}{H(m_{\psi})}, \;
K_{\psi} =  \dfrac{\Gamma_{\psi \longrightarrow S l \eta}}{H(m_{\psi})}, \;
K_{N_{1}\longrightarrow \psi S} = \dfrac{\Gamma_{N_{1}\longrightarrow \psi S}}{H(m_{\psi})}
\end{equation}
with $\Gamma_{f}$ is the partial decay width of particle $f$ for the specified decay process, $H$ is the Hubble parameter. Since leptogenesis is a high scale phenomena and occurs in the radiation dominated phase of the universe, the Hubble parameter can be expressed in terms of the temperature $T$ as follows
\begin{equation}
H=\sqrt{\dfrac{8\pi^{3} g_{*}}{90}}\dfrac{T^{2}}{M_{\rm Pl}}=H(z=1)\dfrac{1}{z^{2}}
\end{equation}
where $g_{*}$ is the effective number of relativistic degrees of freedom and $M_{\rm Pl}\simeq 1.22\times10^{19}$ GeV is the Planck mass. The washout terms are given as
\begin{eqnarray}
W_{N_{1}} & \ = \ & \dfrac{1}{4}z^{3}\left( \dfrac{M_{1}}{m_{\psi}}\right)^2K_{N_{1}}\kappa_{1}\left[z \dfrac{M_{1}}{m_{\psi}}\right], \; W_{\psi}  =  \dfrac{1}{4}z^{3}K_{\psi} \kappa_{1}(z), \nonumber \\
W_{N_{1}\longrightarrow \psi S} & \ = \ & \dfrac{1}{4} z \dfrac{\kappa_{1}\left[ z\dfrac{M_{1}}{m_{\psi}} \right]}{\kappa_{2} \left[ z\dfrac{M_{1}}{m_{\psi}} \right]} K_{N_{1}\longrightarrow \psi S} \dfrac{n_{N_{1}}^{eq}}{n_{\psi}^{eq}}, \nonumber \\
\label{Definations}
\end{eqnarray}

The decay process $N_{1} \longrightarrow \psi S$ does not contribute to the CP asymmetry but can affect the abundance of $\psi, N_1$ as can be seen from the Boltzmann equations written above. The decay width for the decay $N_{1} \longrightarrow \psi S$ is given by 
\begin{equation}
  \begin{split}
  \Gamma_{N_{1}\longrightarrow \psi S} & = \dfrac{1}{16\pi M_{1}^{3}}  \sqrt{M_{1}^{4}+m_{\psi}^{4}+m_{S}^{4}-2M_{1}^{2}m_{\psi}^{2}-2m_{\psi}^{2}m_{S}^{2}} \{ (M_{1}^{2}+m_{\psi}^{2}-m_{S}^{2}) \lvert y_{1} \rvert^{2} \\ & -2{\rm Re}[y_{1}^{2}]m_{\psi}M_{1} \}.
      \end{split}
\end{equation}

As mentioned earlier, we use the Casas-Ibarra parametrisation to rewrite the Yukawa coupling $h_{ij}$ in terms of  light neutrino parameters. Also, in the case of two right handed neutrinos taking part in generating light neutrino masses in our model, the complex orthogonal matrix R is a function of only one rotation parameter $z=z_{R}+iz_{I}, z_{R}\in[0,2\pi], z_{I}\in {\rm I\!R}$ \cite{Casas:2001sr,Ibarra:2003up}. Our choice of R matrix is 
\begin{equation}
R=\begin{pmatrix}
0 & \cos{(z_{R}+iz_{I})} & \sin{(z_{R}+iz_{I})} \\
0 & -\sin{(z_{R}+iz_{I})} & \cos{(z_{R}+iz_{I})}
\end{pmatrix}
\end{equation}

Then the Yukawa matrix for normal ordering of light neutrino masses can then be explicitly written as 
\begin{equation}
h=\begin{pmatrix}
\sqrt{m_{2}}\sqrt{\Lambda_{1}} \cos(z)U_{12}+\sqrt{m_{3}}\sqrt{\Lambda_{1}} \sin(z)U_{13} & - \sqrt{m_{2}}\sqrt{\Lambda_{2}} \sin(z)U_{12}+\sqrt{m_{3}}\sqrt{\Lambda_{2}} \cos(z)U_{13}\\
\sqrt{m_{2}}\sqrt{\Lambda_{1}} \cos(z)U_{22}+\sqrt{m_{3}}\sqrt{\Lambda_{1}} \sin(z)U_{23} & - \sqrt{m_{2}}\sqrt{\Lambda_{2}} \sin(z)U_{22}+\sqrt{m_{3}}\sqrt{\Lambda_{2}} \cos(z)U_{23} \\
\sqrt{m_{2}}\sqrt{\Lambda_{1}} \cos(z)U_{32}+\sqrt{m_{3}}\sqrt{\Lambda_{1}} \sin(z)U_{33} &  -\sqrt{m_{2}}\sqrt{\Lambda_{2}} \sin(z)U_{32}+\sqrt{m_{3}}\sqrt{\Lambda_{2}} \cos(z)U_{33}.
\end{pmatrix}
\end{equation}
with $U_{ij}$ being the elements of the PMNS mixing matrix mentioned earlier. The other Yukawa coupling which affects lepton asymmetry namely, $y_i$ is not related to the origin of light neutrino mass and hence we keep it as a free parameter. The choice of this Yukawa coupling affect both leptogenesis and dark matter as we discuss in upcoming sections.

After obtaining the numerical solutions of the above Boltzmann equations \eqref{eq:4}, \eqref{eq:5} and \eqref{eq:6}, we convert the final $B-L$ asymmetry $n_{B-L}^f$ just before electroweak sphaleron freeze-out into the observed baryon to photon ratio by the standard formula 
\begin{align}
\eta_B \ = \ \frac{3}{4}\frac{g_*^{0}}{g_*}a_{\rm sph}n_{B-L}^f \ \simeq \ 9.2\times 10^{-3}\: n_{B-L}^f \, ,
\label{eq:etaB}
\end{align}
where $a_{\rm sph}=\frac{8}{23}$ is the sphaleron conversion factor (taking into account two Higgs doublets). We take the effective relativistic degrees of freedom to be $g_*=111.75$, slightly higher than that of the SM at such temperatures as we are including the contribution of the inert doublet as well as the scalar singlet too. The heavy singlet fermions $N_{1,2}, \psi$ do not contribute as they have already decoupled from the bath by this epoch. In the above expression $g_*^0=\frac{43}{11}$ is the effective relativistic degrees of freedom at the recombination epoch.

\section{Dark Matter}
\label{sec4}
As mentioned earlier, our model has two DM candidates both of which are stable due to the unbroken $Z_2 \times Z'_2$ symmetry. Although a two component DM was not part of the original motivation, it emerged naturally due to the chosen charge assignments of different particles namely, $\eta, S, \psi, N_i$ under $Z_2 \times Z'_2$ symmetry. In fact, the introduction of the second $Z_2$ symmetry, necessary to forbid direct coupling of $\psi$ with SM leptons, has given rise to the second DM component in the model. A very recent study on such two component DM with scalar doublet and scalar singlet can be found in \cite{Bhattacharya:2019fgs}. For some earlier works on multi-component dark matter, please refer to \cite{Cao:2007fy, Zurek:2008qg, Chialva:2012rq, Heeck:2012bz, Biswas:2013nn, Bhattacharya:2013hva, Bian:2013wna, Bian:2014cja, Esch:2014jpa, Karam:2015jta, Karam:2016rsz, DiFranzo:2016uzc, Bhattacharya:2016ysw, DuttaBanik:2016jzv, Klasen:2016qux, Ghosh:2017fmr, Ahmed:2017dbb, Bhattacharya:2017fid, Ahmed:2017dbb, Borah:2017hgt, Bhattacharya:2018cqu, Bhattacharya:2018cgx, Aoki:2018gjf, DuttaBanik:2018emv, Barman:2018esi, YaserAyazi:2018lrv, Poulin:2018kap, Chakraborti:2018lso, Chakraborti:2018aae, Bernal:2018aon, Elahi:2019jeo, Borah:2019epq, Borah:2019aeq, Biswas:2019ygr,Bhattacharya:2019tqq, Nanda:2019nqy, Borah:2020jzi} and references therein.

\begin{figure}[h]
\begin{center}
\includegraphics[scale=.22]{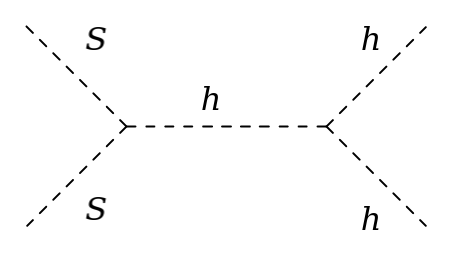}
\includegraphics[scale=.22]{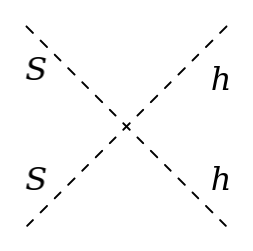}
\includegraphics[scale=.22]{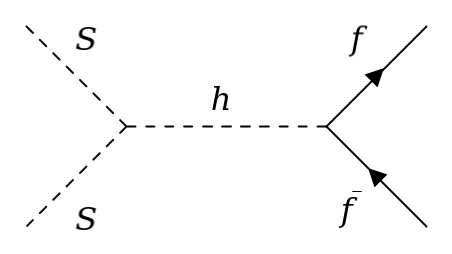}
\includegraphics[scale=.22]{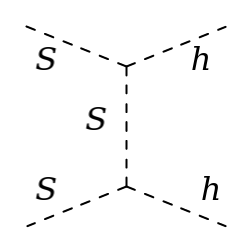}\\
\includegraphics[scale=.22]{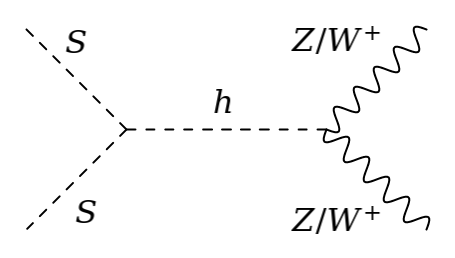}
\includegraphics[scale=.22]{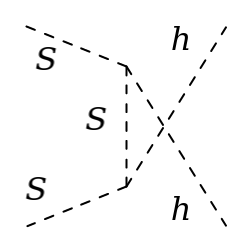}
\caption{Feynman diagrams for relevant annihilation processes for singlet scalar DM.}
\label{pure_singlet}
\end{center}
\end{figure}

\begin{figure}[h]
\begin{center}
\includegraphics[scale=.22]{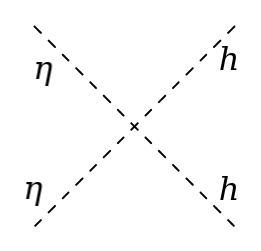}
\includegraphics[scale=.22]{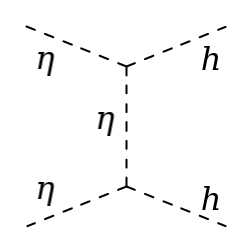}
\includegraphics[scale=.22]{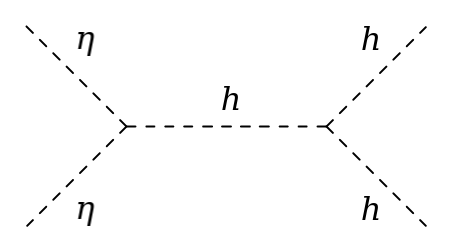}
\includegraphics[scale=.22]{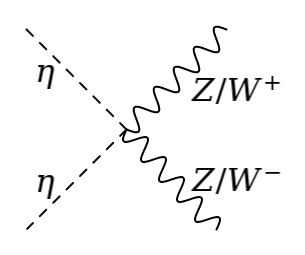}\\
\includegraphics[scale=.22]{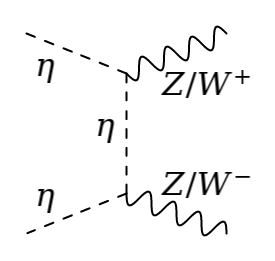}
\includegraphics[scale=.22]{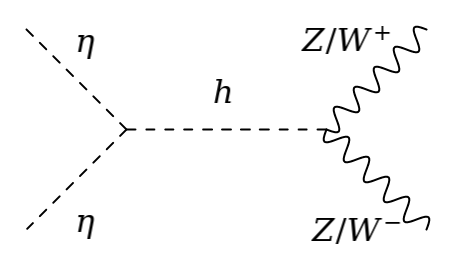}
\includegraphics[scale=.22]{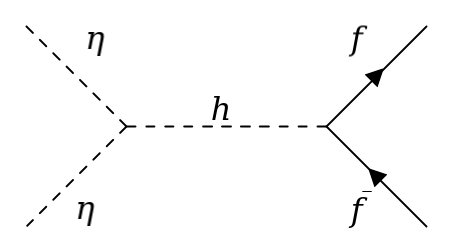}\\
\includegraphics[scale=.22]{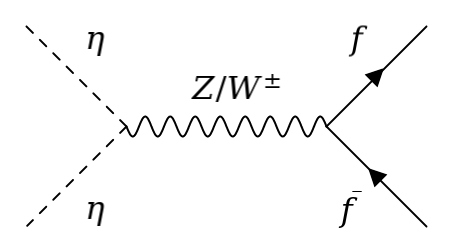}
\includegraphics[scale=.22]{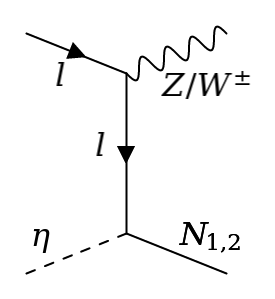}
\includegraphics[scale=.22]{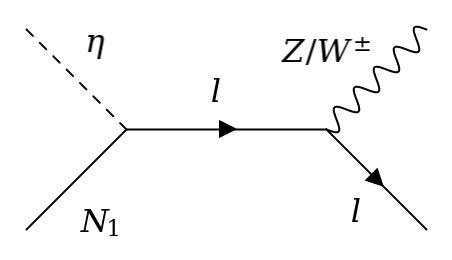}\\
\includegraphics[scale=.22]{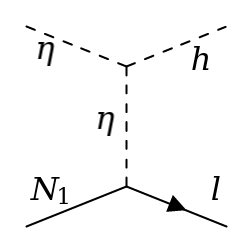}
\includegraphics[scale=.22]{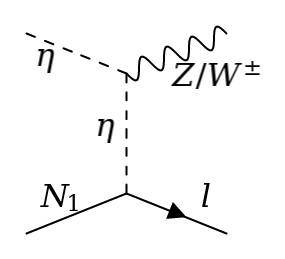}
\caption{Feynman diagrams of all the relevant processes for scalar doublet dark matter in scotogenic model. Here DM is chosen to be the real scalar component of the doublet.}
\label{pure_doublet}
\end{center}
\end{figure}

Relic abundance of two component DM in our model $\eta_R, S$ can be found by numerically solving the corresponding Boltzmann equations. Let $n_1 = n_{\eta_R}$ and
$n_2=n_{S}$ are the total
number densities of two dark matter
candidates respectively. The two coupled Boltzmann
equations in terms of $n_2$ and $n_1$ are given below
\cite{Biswas:2019ygr},   
\begin{eqnarray}
\frac{dn_{1}}{dt} + 3n_{1} H &=& 
-\langle{\sigma {\rm{v}}}_{\eta_R \eta_R \rightarrow {X \bar{X}}}\rangle 
\left(n_{1}^2 -(n_{1}^{\rm eq})^2\right)
-{\langle{\sigma {\rm{v}}}_{\eta_R \eta_R
\rightarrow S S}\rangle} \bigg(n_{1}^2 - 
\frac{(n_{1}^{\rm eq})^2}{(n_{2}^{\rm eq})^2}n_{2}^2\bigg) \,,
%
\label{boltz-eq1} \\
\frac{dn_{2}}{dt} + 3n_{2} H &=& -\langle{\sigma {\rm{v}}}
_{S S \rightarrow {X \bar{X}}}\rangle \left(n_{2}^2 -
(n_{2}^{\rm eq})^2\right) 
+ {\langle{\sigma {\rm{v}}}_{\eta_R \eta_R \rightarrow S S}\rangle} 
\bigg(n_{1}^2 - \frac{(n_{1}^{\rm eq})^2}{(n_{2}^{\rm eq})^2}
n_{2}^2\bigg)\,,
\label{boltz-eq2} 
\end{eqnarray}
where, $n^{\rm eq}_i$ is the equilibrium number density of dark matter species $i$ and $H$ denotes the Hubble parameter, defined earlier. In the annihilation processes, X denotes all particles where DM can annihilate into. In the above equations, $ \langle \sigma v \rangle $ is the thermally averaged annihilation cross section, given by~\cite{Gondolo:1990dk}
\begin{equation}
\langle \sigma v \rangle_{\rm DM DM \rightarrow X \bar{X}} \ = \ \frac{1}{8m_{\rm DM}^4T \kappa^2_2\left(\frac{m_{\rm DM}}{T}\right)} \int\limits^{\infty}_{4m_{\rm DM}^2}\sigma (s-4m_{\rm DM}^2)\sqrt{s}\: \kappa_1\left(\frac{\sqrt{s}}{T}\right) ds \, ,
\label{eq:sigmav}
\end{equation}
where $\kappa_i(x)$'s are modified Bessel functions of order $i$ mentioned before. The annihilation processes of scalar singlet scalar doublet are shown in figure \ref{pure_singlet} and \ref{pure_doublet} respectively. While for scalar singlet DM alone, there is no coannihilation processes, scalar doublet dark matter in scotogenic model can have several coannihilation processes, either with the heavier components of the doublet or fermions as shown in figure \ref{pure_doublet}. Such coannihilation effects within the framework of inert doublet model as well as scotogenic model have already been studied in details by several authors \cite{Deshpande:1977rw, Dasgupta:2014hha, Cirelli:2005uq, Barbieri:2006dq, Ma:2006fn, LopezHonorez:2006gr, Hambye:2009pw, Dolle:2009fn, Honorez:2010re, LopezHonorez:2010tb, Gustafsson:2012aj, Goudelis:2013uca, Arhrib:2013ela, Diaz:2015pyv, Ahriche:2017iar, Borah:2017dfn}. In the presence of coannihilations, one follows the recipe given by~\cite{Griest:1990kh} to calculate the relic abundance. Since scalar singlet DM has just one component, there is no such coannihilations present. Similar to the inert doublet dark matter model, scalar singlet dark matter has also been studied extensively by several authors \cite{Silveira:1985rk, McDonald:1993ex, Burgess:2000yq, Athron:2017kgt, Borah:2020nsz}.

\begin{figure}[h]
\begin{center}
\includegraphics[scale=.22]{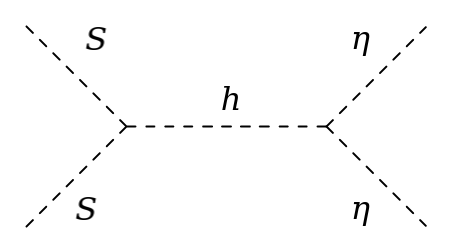}
\includegraphics[scale=.22]{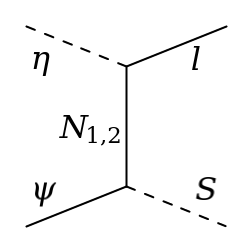}
\includegraphics[scale=.22]{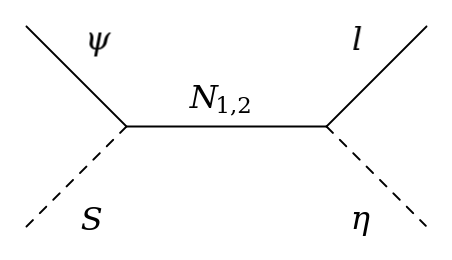}
\includegraphics[scale=.22]{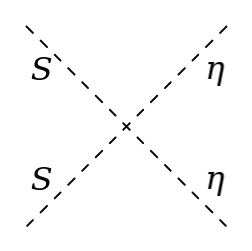}
\caption{Feynman diagrams of all the relevant processes determining the DM relic density which emerged due to the extension of the scotogenic model.}
\label{singlet_doublet}
\end{center}
\end{figure}

The second terms on the right hand side of the above Boltzmann equations specifically consider the conversions between two DM candidates $\eta_R, S$ while assuming the former to be the heavier DM component. Such a conversion can occur either directly due to the $\lambda_6$ coupling of the scalar potential given in equation \eqref{scalarpot} or via SM Higgs portal interactions. These conversion processes are shown in figure \ref{singlet_doublet}. There can be another conversion process due to the interactions shown in the Feynman diagram of figure \ref{fig:decay}. This can occur due to coannihilation processes, not shown in above Boltzmann equations. In our model, however, singlet scalar DM can, in principle, coannihilate with other particles involved in the same Feynman diagram of figure \ref{fig:decay}. Since the two DM candidates are stabilised by two separate $Z_2$ symmetries, their coannihilation can only lead to $\psi$ which is odd under both the $Z_2$ symmetries. Alternatively, one of the DM can also coannihilate with $\psi$ and convert into the other DM. These processes are shown in figure \ref{singlet_doublet}. Since we consider $\psi$ to be heavier than both the DM candidates, we do not show it in the final states.

In order to cover all the features of annihilations, coannihilations as well as conversions, we use \texttt{micrOMEGAs} \cite{Belanger:2014vza} to calculate the relic abundance of two component DM in our model. The model information has
been supplied to \texttt{micrOMEGAs} using \texttt{FeynRules}
\cite{Alloul:2013bka} while all the relevant annihilation and coannihilation cross sections
of dark matter number changing  processes required to solve
the coupled equations are calculated using \texttt{CalcHEP} \cite{Belyaev:2012qa}. While singlet scalar DM annihilates either through four point scalar interactions or SM Higgs mediated processes, the scalar doublet DM can annihilate (coannihilate) via Higgs as well as electroweak gauge boson portals apart from the four point interactions with Higgs as well as gauge bosons. Additionally, the conversion coupling $\lambda_6$ as well as Yukawa coupling $y_i$ can play significant role in individual as well as total DM relic densities.

Just like the SM Higgs boson mediates DM annihilation into SM particles, similarly, it can also mediate spin independent DM-nucleon scatterings. Different ongoing experiments like Xenon1T \cite{Aprile:2017iyp, Aprile:2018dbl}, LUX \cite{Akerib:2016vxi}, PandaX-II \cite{Tan:2016zwf, Cui:2017nnn} are trying to detect the DM in the lab-based experiments and give a strong upper bound on the spin-independent (SI) direct detection (DD) cross-section as a function of DM mass. We have extracted the SI elastic scattering cross-section for both the DM candidates from \texttt{micrOMEGAs}. DD analysis for two-component DM is slightly different from the single component scenario. To compare the result of our model with Xenon1T bound, we have multiplied the elastic scattering cross-section by the relative number density of each DM candidate and used the following conditions 

\begin{eqnarray}\nonumber
{\rm \sigma_{1}^{\rm eff}=\frac{n_{1}}{n_{1}+n_{2}} \sigma_{1}^{SI}\leq \sigma_{Xenon1T} }\\  
{\rm \sigma_{2}^{\rm eff}=\frac{n_{2}}{n_{1}+n_{2}} \sigma_{2}^{SI}\leq \sigma_{Xenon1T} }
\label{eff:DD}
\end{eqnarray} 

Further details related to the direct detection of multi component DM can be found in \cite{Herrero-Garcia:2017vrl, Herrero-Garcia:2018lga}.

\section{Results and Discussion}
\label{sec5}
In this section, we discuss our numerical results for leptogenesis as well as dark matter separately.
\subsection{Leptogenesis}
To calculate the lepton asymmetry, we first solve the coupled Boltzmann equations \eqref{eq:4}, \eqref{eq:5} and \eqref{eq:6} numerically to estimate the final B-L asymmetry. We considered two possible ranges for $N_1$ mass $M_1$. 

\begin{figure}
\includegraphics[scale=.53]{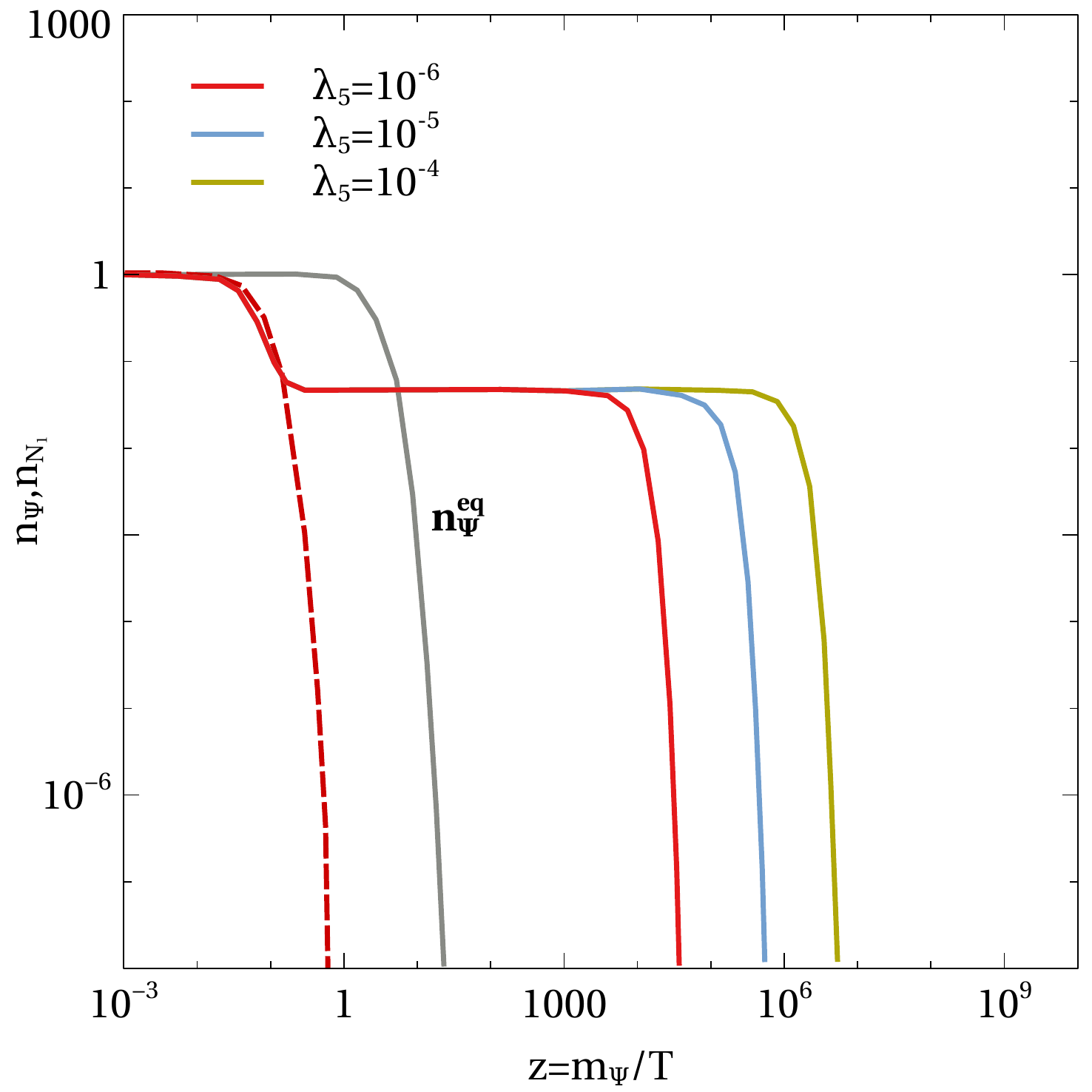}
\includegraphics[scale=.53]{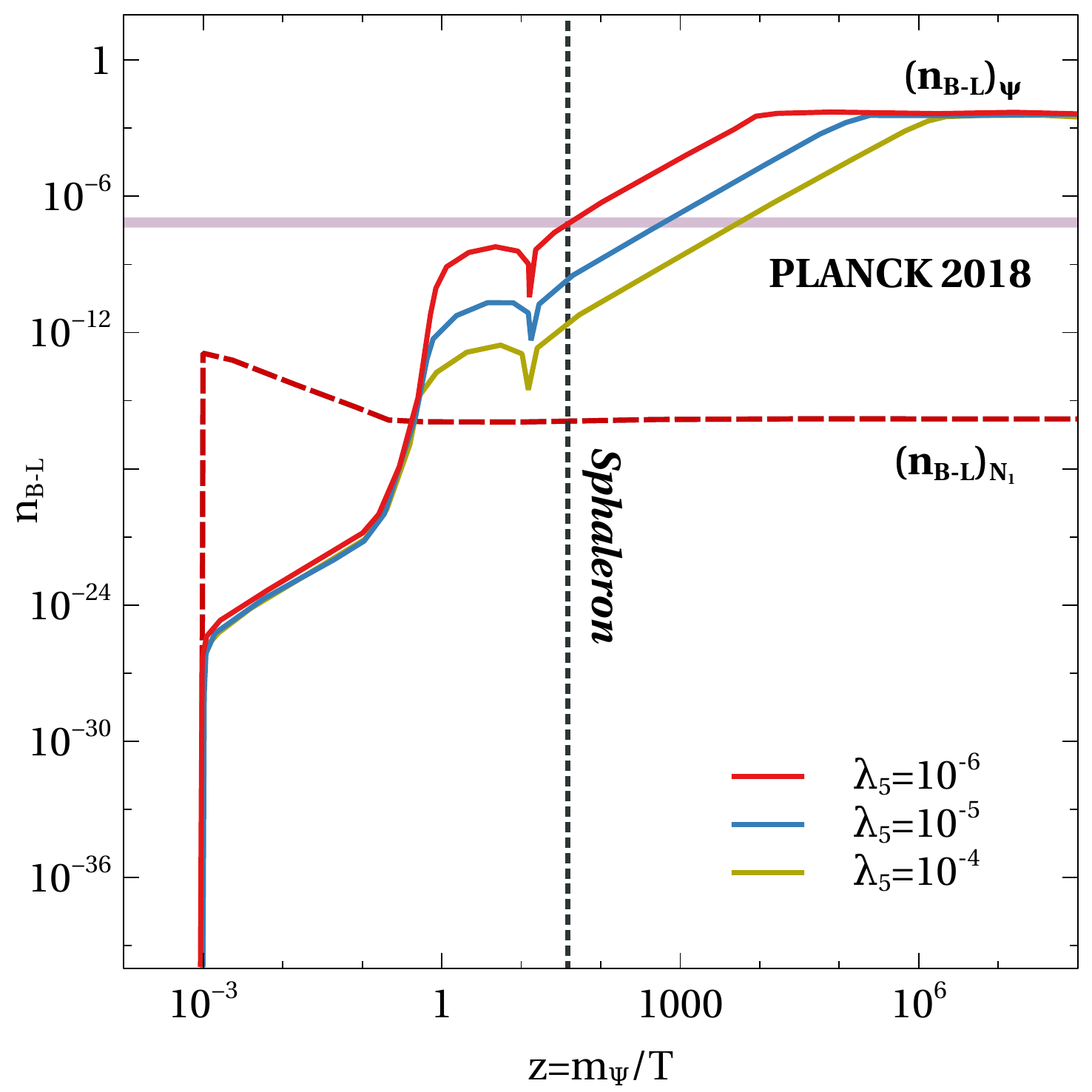} \\
\includegraphics[scale=0.53]{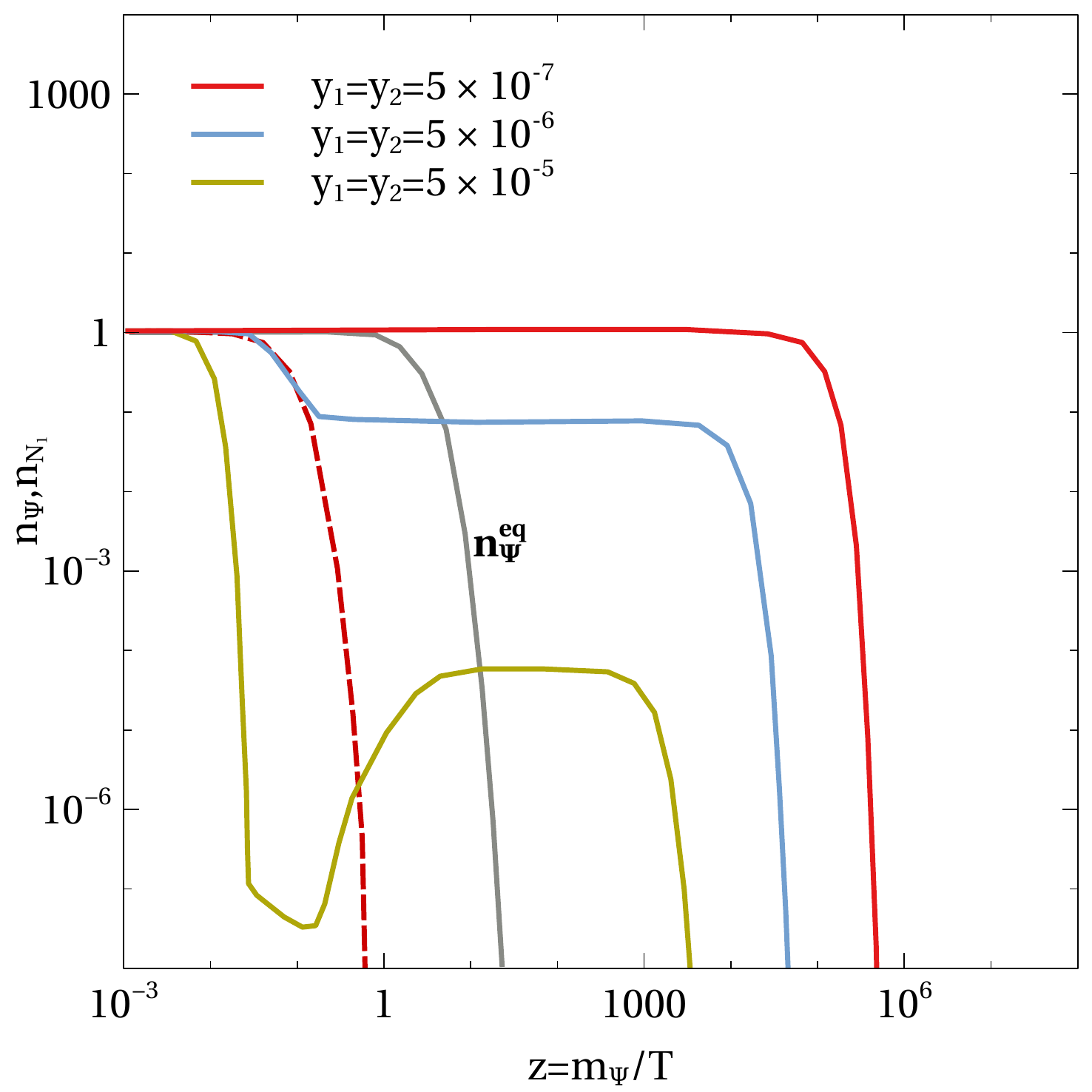}
\includegraphics[scale=0.53]{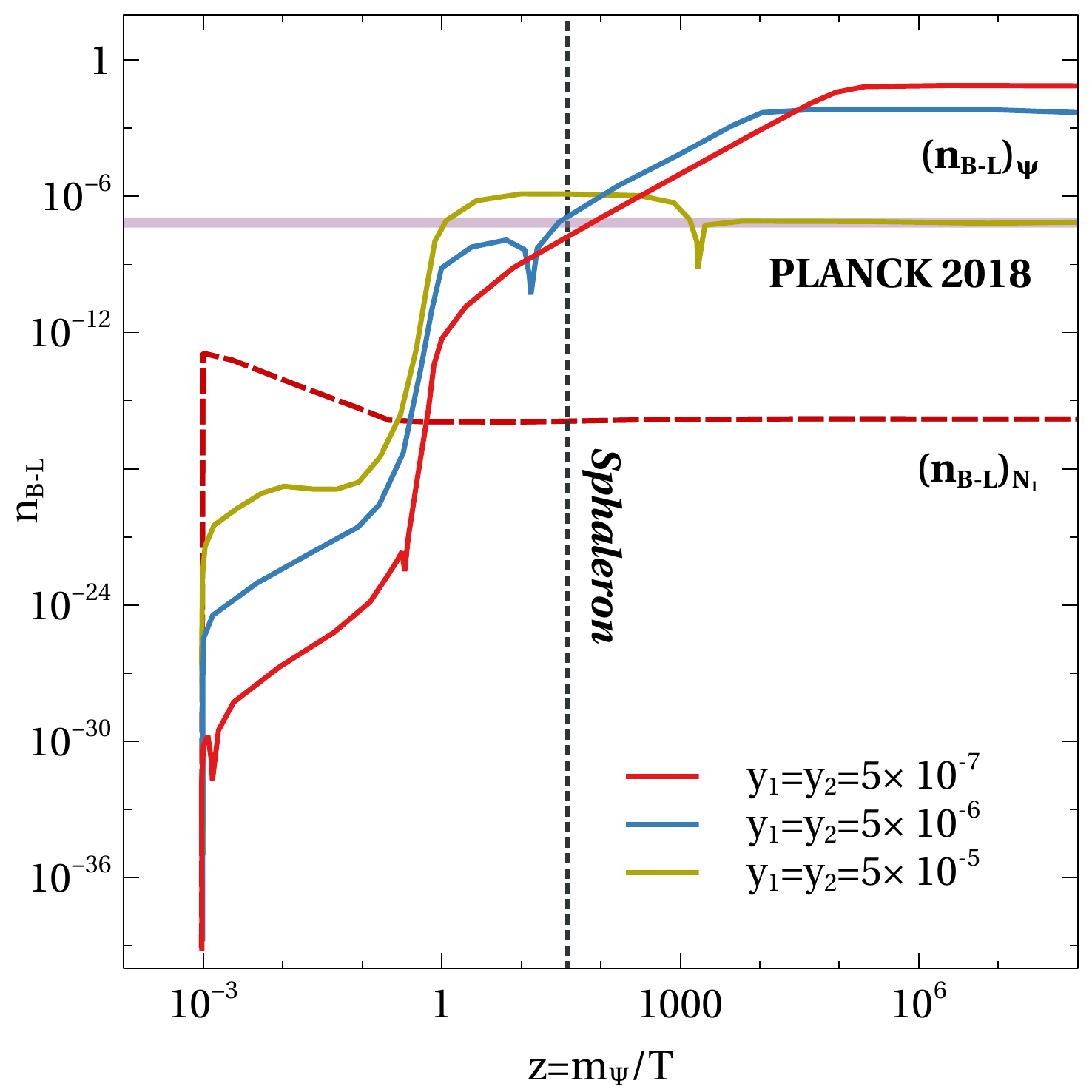}
\caption{Evolution of co-moving number densities of $\psi$ and $N_{1}$ (left panel) and $B-L$ (right panel) with $z=\dfrac{m_{\psi}}{T}$ for different values of $\lambda_5$ (upper panel), Yukawa couplings $y_{1,2}$ (lower panel). The other parameters are set at benchmark values: $M_{1}=2\times10^{5}$ GeV, $M_{2}=2\times10^{6}$ GeV, $m_{\eta}=100$ GeV, $m_{S}=500$ GeV, $m_{\psi}=5$ TeV and $y_{1}=y_{2}=5\times10^{-6}$ (upper panel) and $\lambda_{5}=10^{-6}$ (lower panel).}
\label{case1}
\end{figure}

\begin{figure}
\includegraphics[scale=.53]{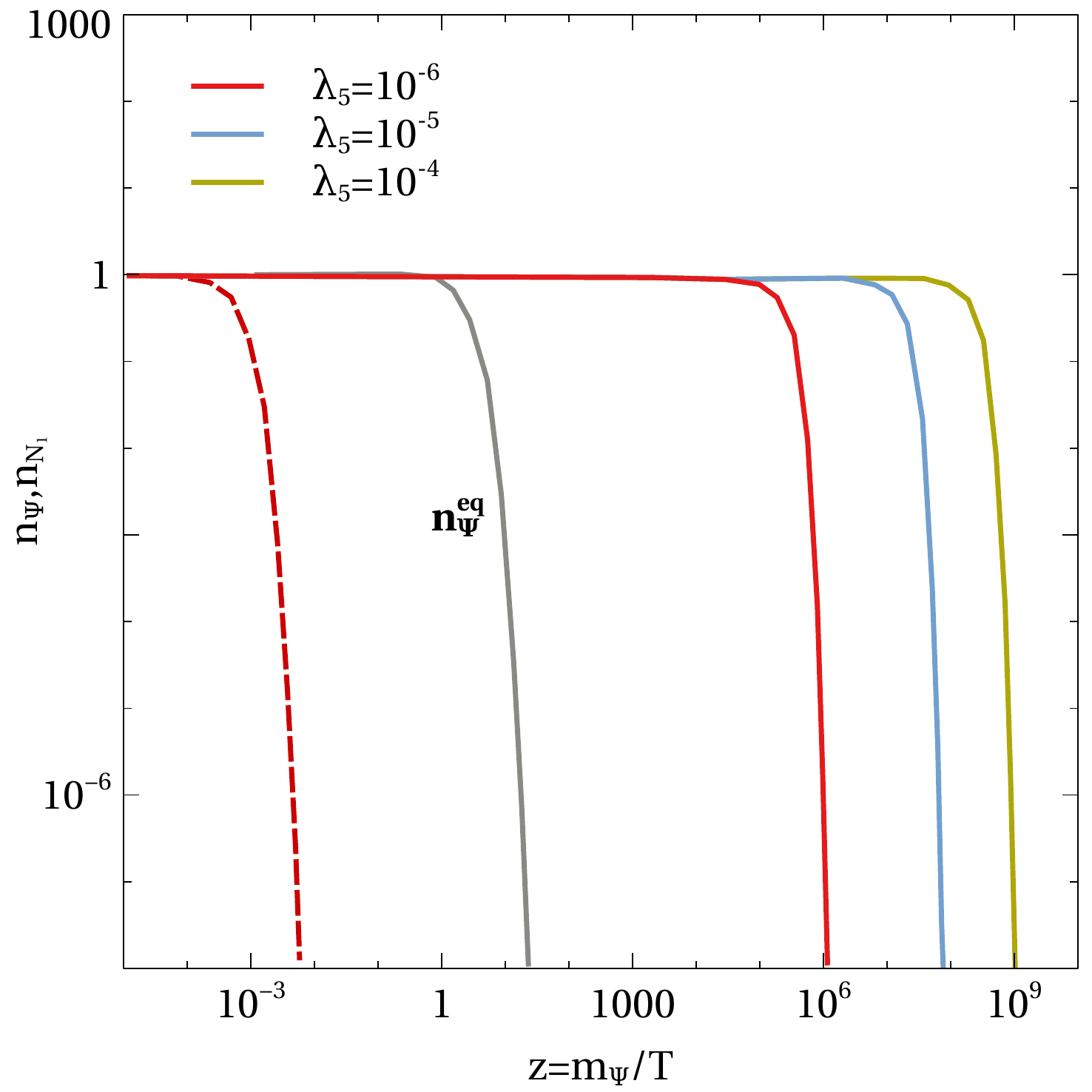}
\includegraphics[scale=.53]{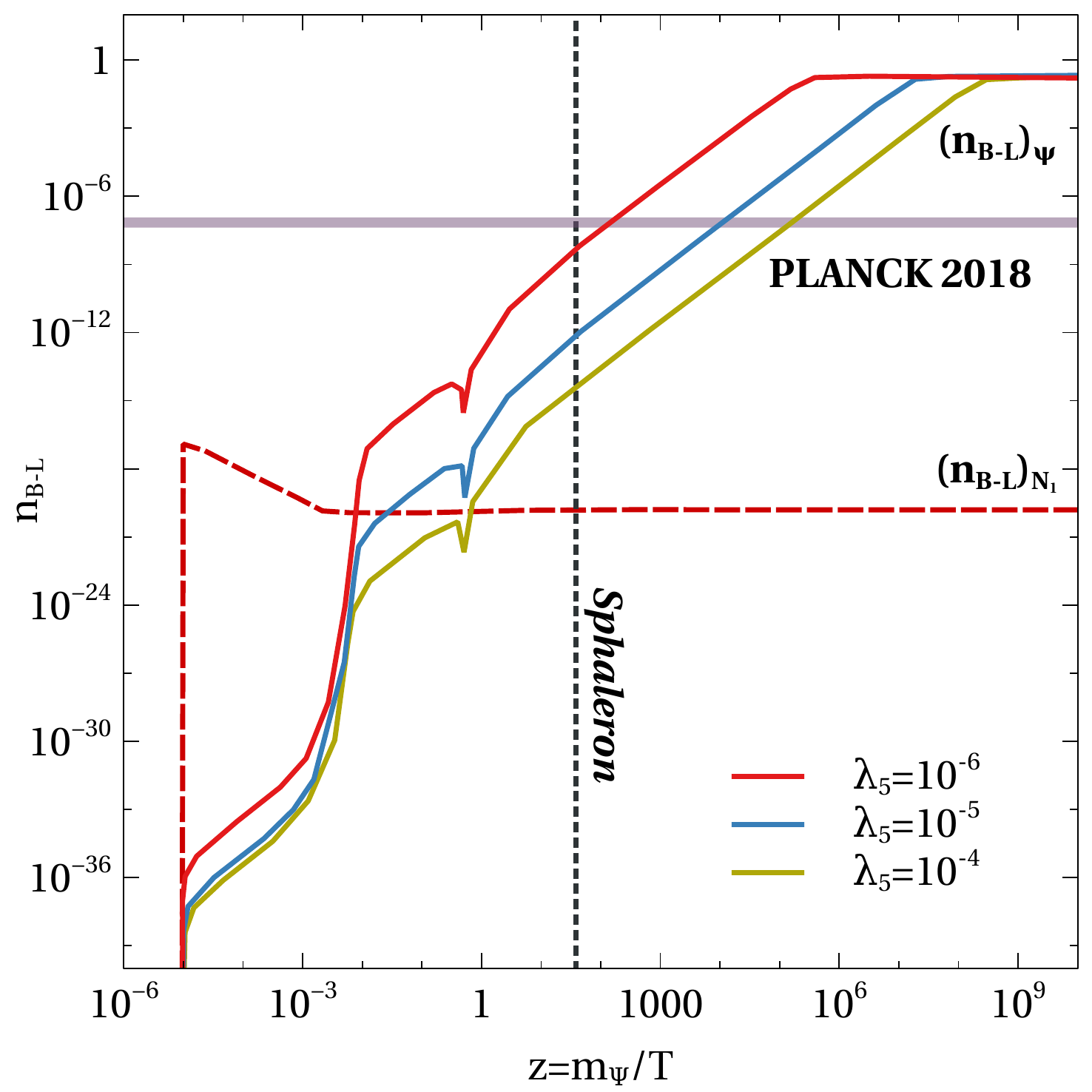} \\
\includegraphics[scale=0.53]{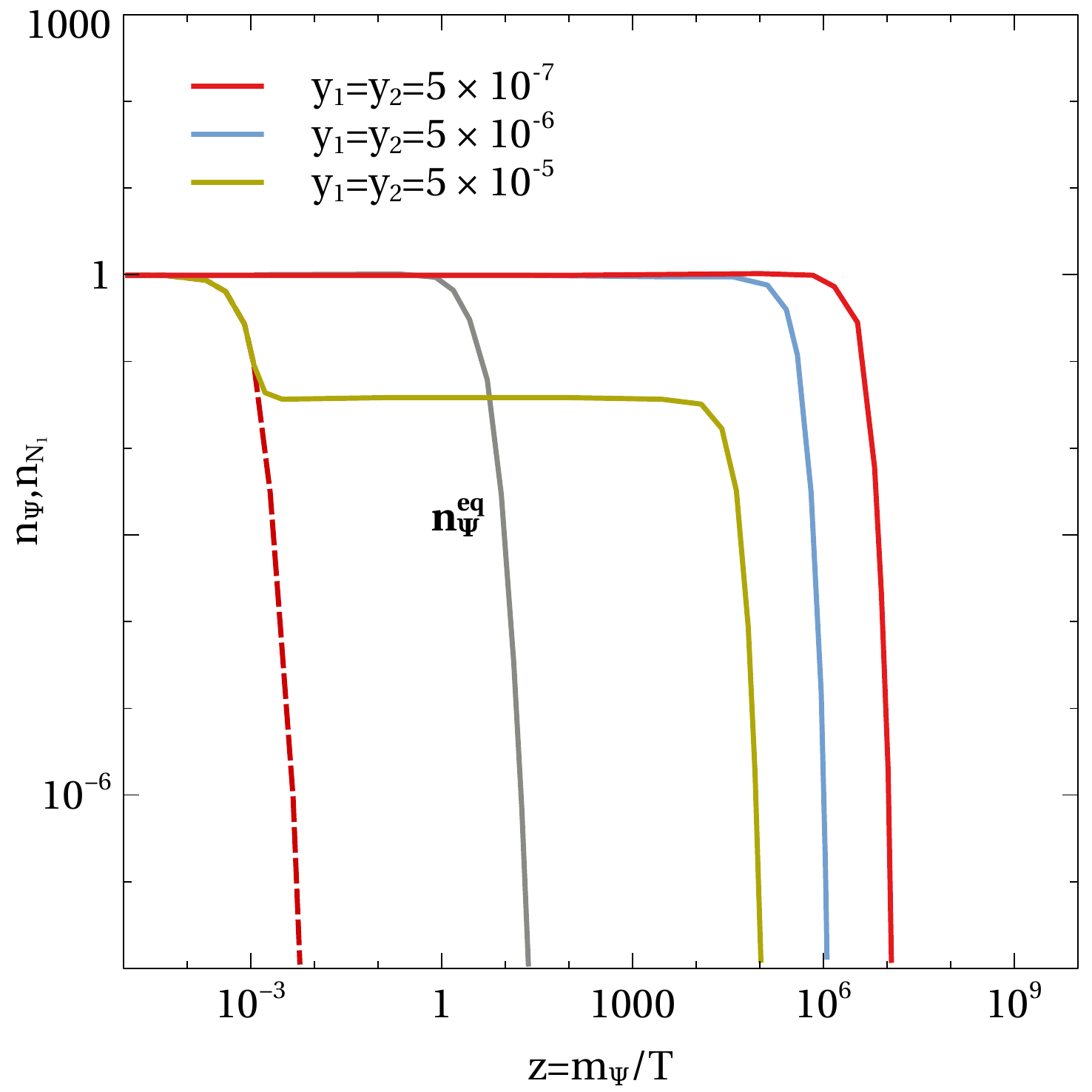}
\includegraphics[scale=0.53]{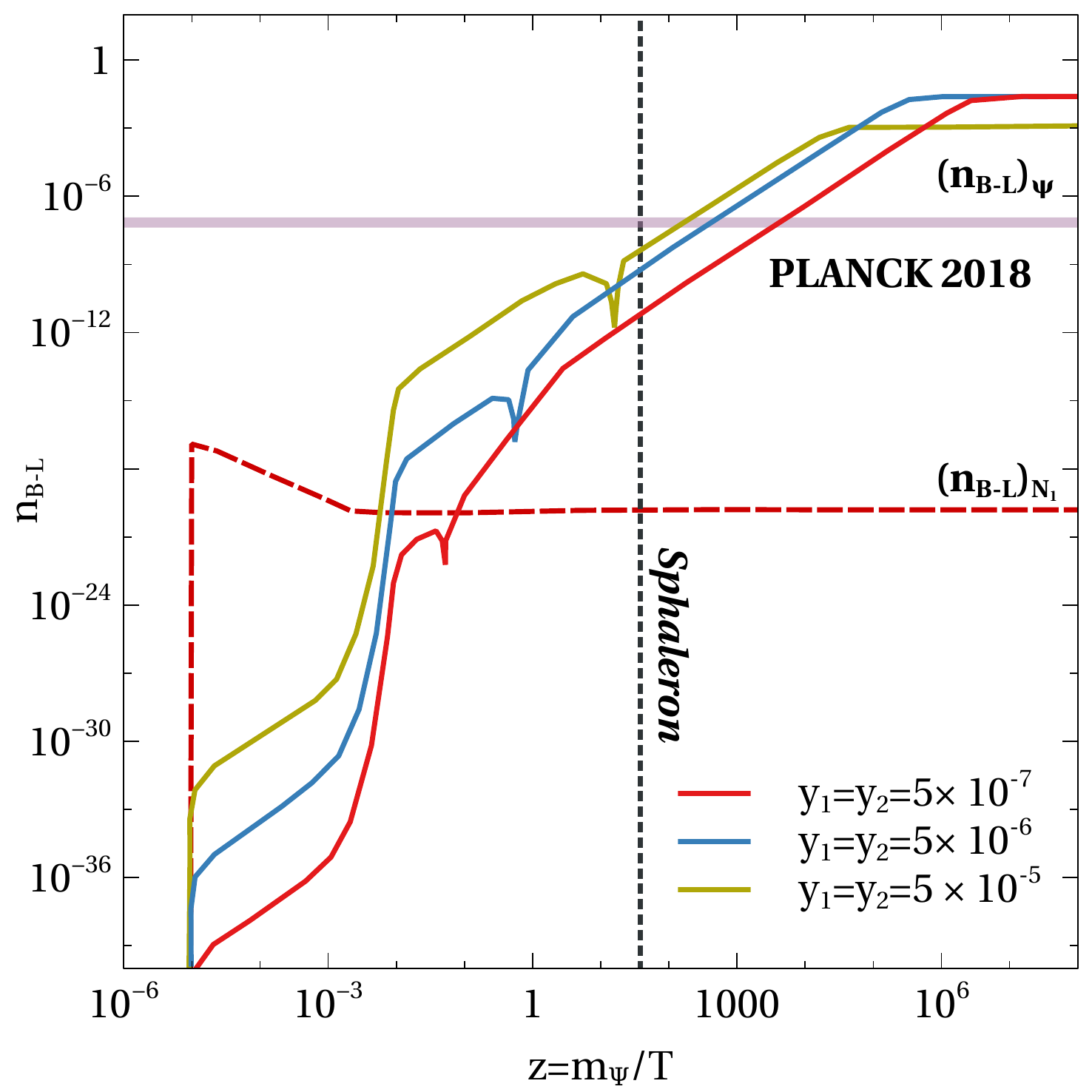}
\caption{Evolution of co-moving number densities of $\psi$ and $N_{1}$ (left panel) and $B-L$ (right panel) with $z=\dfrac{m_{\psi}}{T}$ for different values of $\lambda_5$ (upper panel), Yukawa couplings $y_{1,2}$ (lower panel). The other parameters are set at benchmark values: $M_{1}=2\times10^{7}$ GeV, $M_{2}=2\times10^{8}$ GeV, $m_{\eta}=100$ GeV, $m_{S}=500$ GeV, $m_{\psi}=5$ TeV and $y_{1}=y_{2}=5\times10^{-6}$ (upper panel) and $\lambda_{5}=10^{-6}$ (lower panel).}
\label{case2}
\end{figure}

In the first case we have chosen a benchmark as $M_{1}=2 \times 10^{5}$ GeV and in the other case we choose $M_{1}=2 \times 10^{7}$ GeV while keeping $M_{2}=10M_{1}$ and other parameters fixed for both the cases. Using the first choice of benchmark values for $M_{1}$ and $M_{2}$, in figure \ref{case1}, the evolution of the comoving number densities of $\psi$, $N_{1}$ and $B-L$ are shown with $z=\dfrac{m_{\psi}}{T}$ for different values of $\lambda_{5}$ and $y_{1,2}$. Similarly in figure \ref{case2}, the second benchmark for RHN masses is chosen and the evolution of the co-moving number densities of  $\psi$, $N_{1}$ and $B-L$ are shown with $z=\dfrac{m_{\psi}}{T}$ for different values of $\lambda_{5}$ and $y_{1,2}$. The parameter $\lambda_5$ decides the strength of Dirac Yukawa coupling of neutrinos via Casas-Ibarra parametrisation as \eqref{eq:Yuk} discussed earlier.

In the three body decay width $\Gamma_{\psi \longrightarrow l S \eta}$, apart from Dirac Yukawa couplings of active neutrinos, we also have other Yukawa couplings $y_{1,2}$ which can affect leptogenesis, without affecting neutrino mass. In this section we discuss the effect of these two types of Yukawa couplings on the asymmetry. In the upper left panel of figure \ref{case1} we show the evolution of the comoving number densities of $\psi$ (solid lines) and $N_{1}$ (dashed lines) for different benchmark values of $\lambda_5$. In the upper right panel of figure \ref{case1}, we show the evolution of $B-L$ for different $\lambda_{5}$. We have taken both $\psi$ and $N_{1}$ to be in equilibrium at very high temperatures and numerically solved the coupled Boltzmann equations upto a temperature when the asymmetry gets saturated. From the evolution of the comoving number densities of $\psi$ and $N_{1}$ we can see that initially both the particles follow their equilibrium number densities but soon after the $\psi$ abundance deviates from its equilibrium abundance while the $N_{1}$ abundance remains very close to its equilibrium number density and therefore vanishes shortly after it becomes non-relativistic. It is mainly because $N_{1}$ has very strong two-body decays ($N_{1}\longrightarrow l \eta$ and $N_{1} \longrightarrow \psi S$) and corresponding inverse decays and therefore its abundance remains very close to its equilibrium abundance. On the other hand, $\psi$ has relatively feeble three-body decay $\psi \longrightarrow S \eta l$ and a strong inverse decay $\psi S \longrightarrow N_{1}$. Because of the strong two-body inverse decay of $\psi$ the abundance of $\psi$ decreases sharply. This inverse decay stops when the temperature drops to a value such that the process $\psi S \longrightarrow N_{1}$ becomes kinematically forbidden. After this point the $\psi$ abundance gets saturated and goes out of equilibrium before finally decaying through the three-body decay at a low temperature. The effect of this strong inverse decay ($\psi S \longrightarrow N_{1}$) can also be seen in the asymmetry evolution shown on the right panel plot. The solid lines represent the asymmetry generated from the $\psi$ decay and the dashed line represents the asymmetry generated from the two-body decay of $N_{1}$. From asymmetry plots on right panel of figure \ref{case1}, it can be seen that asymmetry generated from the $N_{1}$ decay is very less compared to the one generated from the three-body decay of $\psi$, which is expected as $N_{1}$ remains very close to its equilibrium abundance. Another important point is the asymmetry generated from $N_{1}$ gets saturated at a very high temperature as $N_{1}$ abundance vanishes when it becomes non-relativistic. However, the asymmetry generated from the $\psi$ decay keep evolving upto a very low temperature because of the small decay width of $\psi$. Because of the small decay parameter for the decay $\psi \longrightarrow S \eta l$, the inverse decay rate is also very small and we are always in a weak washout regime ($K_{\psi} \ll 1$). However, the asymmetry generation from the two-body decay of $N_{1}$ is always is in strong washout regime  as we have only two RHNs (for two RHNs scenarios, $K_{N_{1}}>1$) \cite{Hugle:2018qbw, Mahanta:2019sfo}. Because of this, the resultant asymmetry is mainly determined by the three-body decay.

In the upper panel plot of figure \ref{case1} it is observed that abundance of $\psi$ become less than its equilibrium abundance because of the inverse decay $\psi \longrightarrow N_{1} S$ (while $N_1$ has strong two-body decays into both $\psi S$ and $\eta l$ final states) and therefore the asymmetry generated from the $\psi$ decay remains suppressed initially. However, when the inverse decay stops, the $\psi$ abundance saturates and finally the asymmetry becomes overabundant before its decay is complete. When the $\psi$ becomes more compared to its equilibrium abundance then the $B-L$ asymmetry starts rising steadily, as can be seen by comparing left and right panel plots of figure \ref{case1}. In upper left panel plot of figure \ref{case1} we can see that the decay of $\psi$ happens earlier for smaller value of $\lambda_{5}$, which is expected as smaller value of $\lambda_{5}$ lead to larger Dirac Yukawa couplings. For the same reason the saturation of asymmetry happens earlier for smaller value values of $\lambda_{5}$, as seen from upper right panel plot. The asymmetry generated upto the sphaleron epoch ($T_{\rm Sphaleron} \simeq 131$ GeV, shown by vertical dashed line) is important as the asymmetry generated after the sphaleron freeze-out temperature can not be converted into a baryon asymmetry. It can be seen from upper right panel plot of figure \ref{case1} that for smaller $\lambda_{5}$ the asymmetry generated upto sphaleron temperature is more compared to the ones for larger $\lambda_{5}$. In the lower left panel plot of figure \ref{case1} we observe that for larger values of the Yukawa couplings $y_{1,2}$, the effect of the inverse decay $\psi S \longrightarrow N_{1}$ is more. For larger Yukawa coupling $y_{1,2}$ the $\psi$ abundance decreases sharply because of the very strong inverse decay $\psi S \longrightarrow N_{1}$ and at later epochs also when its abundance is more compared to equilibrium abundance, smaller Yukawa leads to larger abundance as expected. For the same reason, the $B-L$ asymmetry also increases more sharply for larger Yukawa couplings leading to larger asymmetry at the epoch of sphaleron freeze-out, as seen from the lower right panel plot of figure \ref{case1}. It should be noted that, we are showing only the absolute value of $B-L$ asymmetry on the right panel plots; in reality, the points towards the left of the dip in solid lines correspond to negative asymmetry. Clearly, for large Yukawa $y_{1,2}$, the asymmetry remains negative even at the sphaleron epoch, as seen from lower right panel plot. Therefore, we can not make Yukawa coupling $y_{1,2}$ arbitrarily large to get more asymmetry at the epoch of shpaleron decoupling.

Similarly, in figure \ref{case2} we have shown the evolution $\psi$, $N_{1}$ and $B-L$ number densities with $z=m_{\psi}/T$ for different values of $\lambda_{5}$ (upper panel) and $y_{1,2}$ (lower panel) but with heavier mediator masses namely, $M_{1}=2\times 10^{7}$ GeV and $M_{2}=10M_{1}$, keeping other parameters fixed as in figure \ref{case1}. It can be seen that for the same value of $m_{\psi}, \lambda_{5}, m_{\eta}, m_{S}$ and $y_{1,2}$ the effect of the inverse decay $\psi S \longrightarrow N_{1}$ is much less in this case. This is expected as for larger value of $N_{1}$ will make the inverse decay $\psi S \longrightarrow N_{1}$ inefficient even at very high temperatures. The variation of the $B-L$ asymmetry with $\lambda_{5}$ and $y_{1,2}$ can be understood in a way similar to the figure \ref{case1} discussed earlier. Also, comparing figure \ref{case1} and figure \ref{case2} we can see that for the same set of parameters the decay of $\psi$ and generation of asymmetry occur slowly making the asymmetry less in figure \ref{case2} than in \ref{case1} at the epoch of sphaleron decoupling. This is expected as larger $N_{1,2}$ masses make the three-body decay width of $\psi$ smaller due to propagator suppression.

\begin{figure}
    \centering
    \includegraphics[scale=.53]{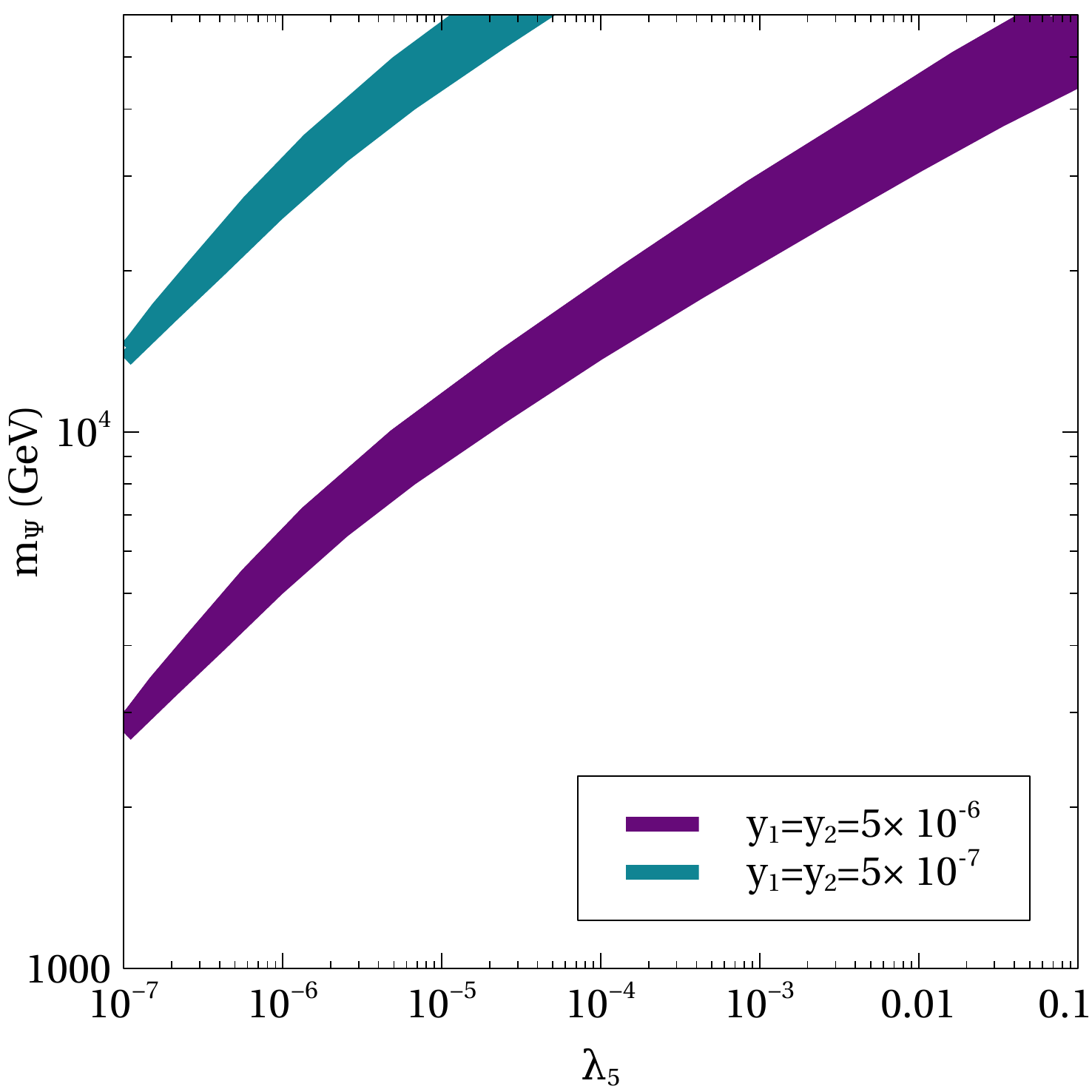}
    \includegraphics[scale=.53]{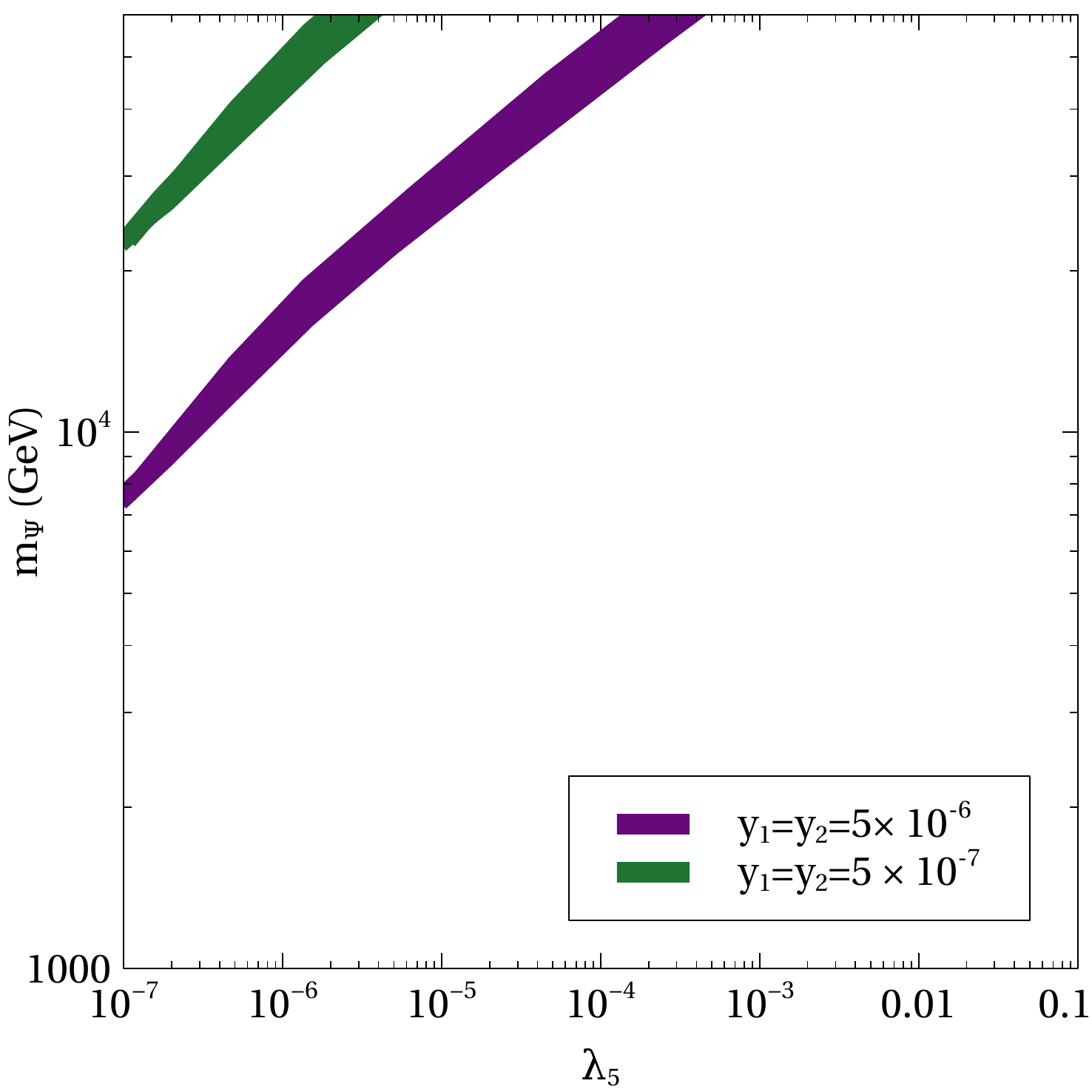}
    \caption{The variation of $m_{\psi}$ with $\lambda_{5}$, required to satisfy the observed asymmetry. For the left panel the $M_{1}=2\times 10^{5}$ GeV,  $M_{2}=2\times 10^{6}$ GeV and for the right panel $M_{1}=2\times 10^{7}$ GeV, $M_{2}=2\times 10^{8}$ GeV. The other parameters are set at $m_{\eta}=10$ GeV, $m_{S}=500$ GeV.}
    \label{scan}
\end{figure}

Finally, we perform a numerical scan to find the relevant parameter space in $m_{\psi}-\lambda_5$ plane that can give rise to the observed baryon asymmetry for both $M_{1}=2\times 10^{5}$ GeV and $M_{1}=2\times 10^{7}$ GeV. While varying these parameters, we keep the masses of other relevant particles to be fixed at $m_{S}=500$ GeV, $m_{\eta}=100$ GeV. The parameter space in $m_{\psi}-\lambda_5$ plane for benchmark choices of $y_{1,2}$ is shown in figure \ref{scan} for $M_{1}=M_2/10=2\times 10^{5}$ GeV (left panel) and for $M_{1}= M_2/10= 2 \times 10^{7}$ GeV (right panel). In figure \ref{scan} we can see that for a benchmark value of $y_{1,2}$ the mass required of $\psi$ become large for larger values of $\lambda_{5}$, which is expected as larger values of $\lambda_{5}$ make the Dirac Yukawa couplings $h_{i\alpha}$ smaller making the three body decay width of $\psi$ smaller. For similar reason, for a particular value of $\lambda_{5}$, the required mass of $\psi$ for a small Yukawa coupling $y_{1,2}$ is more compared to that for larger Yukawa coupling. However, as mentioned earlier we can not take arbitrarily large values of the new Yukawa couplings $y_{1,2}$ to lower the scale of leptogenesis. As discussed earlier, beyond a certain value of $y_{1,2}$ the inverse decay $\psi S \longrightarrow N_{1}$ will become so dominant that the asymmetry become negative at the time of sphaleron. We found that for a TeV scale leptogenesis with moderately high $M_{1,2}$ it is safe to take $y_{1,2} \leq 10^{-5}$ such that the asymmetry become positive at the sphaleron freeze-out temperature. From figure \ref{scan} we can conclude that successful TeV scale leptogenesis is possible dominantly from the three body decay for appropriate choice of the model parameters. Also we can see that the scale of leptogenesis is slightly higher in the right panel plot of \ref{scan} compared to the left plot of figure \ref{scan}. This is because with the increases in $M_{1,2}$ the the three body decay width encounters propagator suppression as discussed earlier.

\subsection{Flavour effects on leptogenesis}
\begin{figure}
    \centering
    \includegraphics[scale=.53]{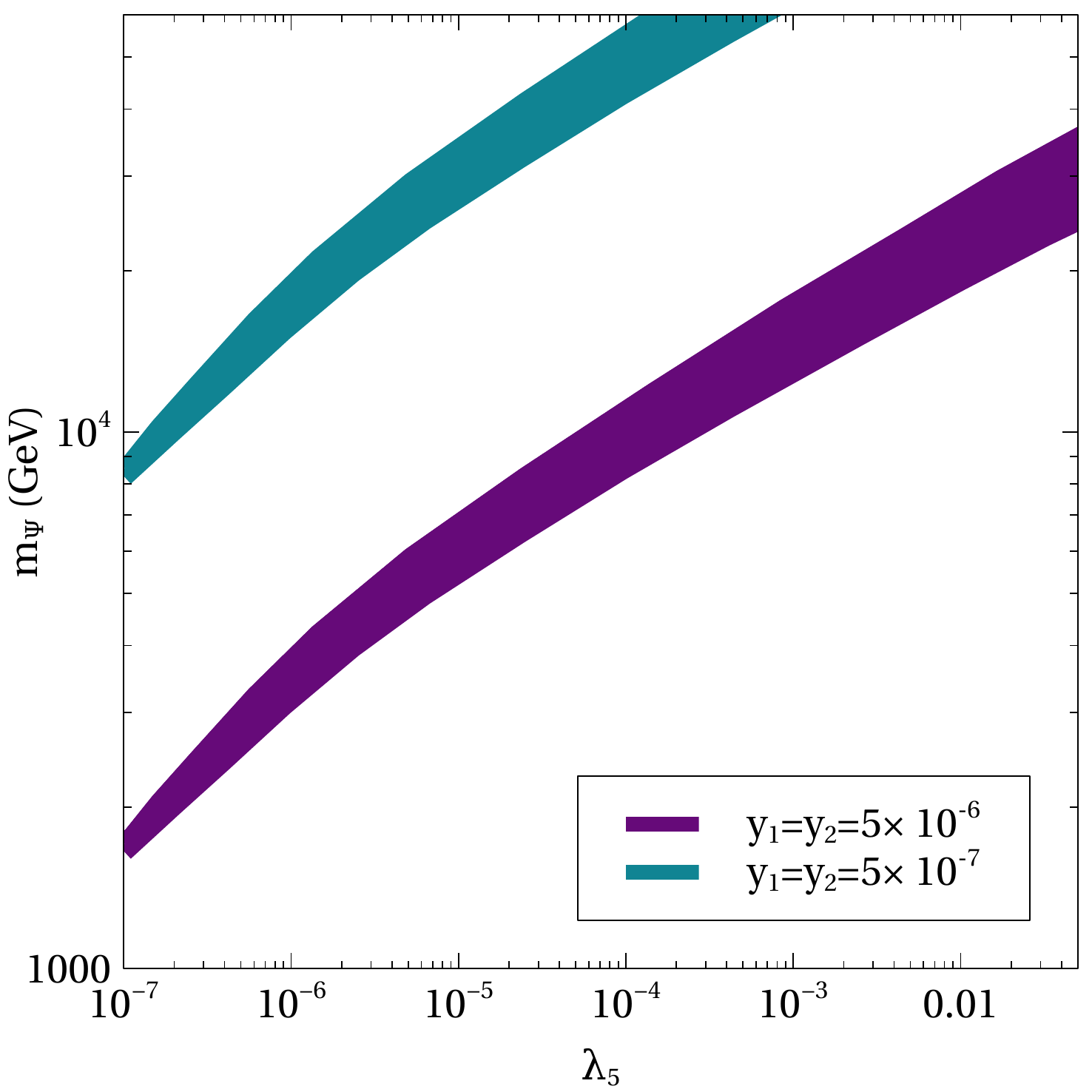}
    \includegraphics[scale=.53]{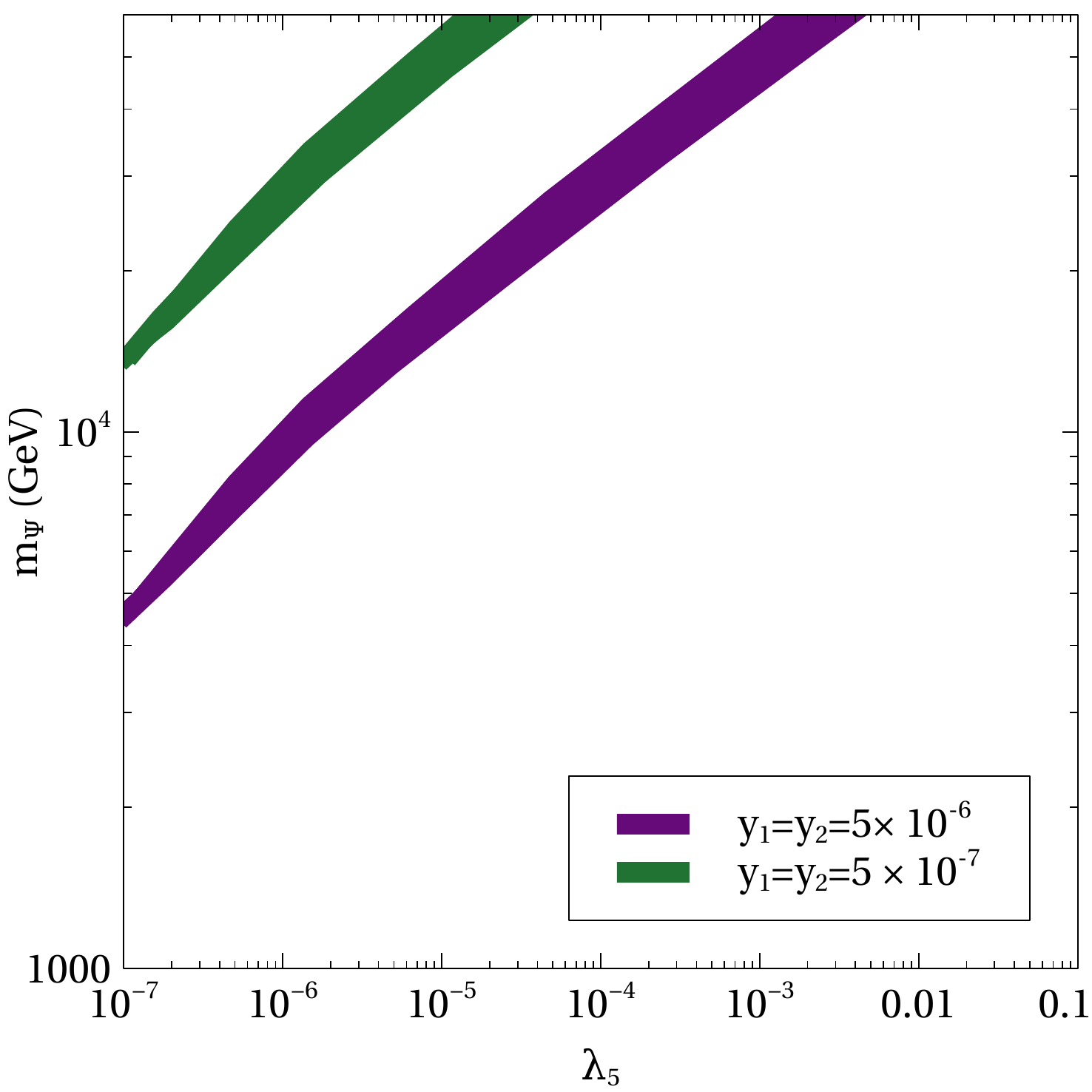}
    \caption{The variation of $m_{\psi}$ with $\lambda_{5}$, required to satisfy the observed asymmetry using lepton flavour effects. For the left panel, $M_{1}=2\times 10^{5}$ GeV and $M_{2}=2\times 10^{6}$ GeV and for the right panel, $M_{1}=2\times 10^{7}$ GeV and $M_{2}=2\times 10^{9}$ GeV. The other parameters are set at $m_{\eta}=10$ GeV, $m_{S}=500$ GeV.}
    \label{case12_scan_flav}
\end{figure}

It should be noted that while discussing leptogenesis in the above sections, we did not consider the effects of lepton flavours. Since we are considering leptogenesis at low scale, lepton flavour effects may play non-trivial roles as pointed out by several earlier works on flavoured leptogenesis \cite{Abada:2006fw, Abada:2006ea, Nardi:2006fx, Blanchet:2006be}, also summarised in a recent review article \cite{Dev:2017trv}. Adopting the notations of \cite{Blanchet:2006be}, the Boltzmann equations for flavoured leptogenesis can be written as
\begin{align}
\dfrac{dn_{\psi}}{dz} & = -D_{\psi}(n_{\psi}-n_{\psi}^{\rm eq})+D_{N_{1}\longrightarrow \psi S}(n_{N_{1}}-n_{N_{1}}^{\rm eq})-W_{ID_{N_{1}\longrightarrow \psi S}}n_{\psi} \nonumber \\ 
& \nonumber- \dfrac{s}{H(z)z}[ (n_{\psi}n_{\eta}-n_{\psi}^{\rm eq}n_{\eta}^{\rm eq})  \langle \sigma v \rangle_{\psi \eta \longrightarrow  S l   }  + (n_{\psi} n_{S}-n_{\psi}^{\rm eq} n_{S}^{\rm eq})    \langle \sigma v \rangle_{\psi S \longrightarrow l \eta} \nonumber \\ & (n_{\psi}-n_{\psi}^{\rm eq})n_{l}^{\rm eq}\langle \sigma v \rangle_{\psi l \longrightarrow \eta S}  ], 
\label{eq:4}
\end{align}

\begin{align}
\dfrac{dn_{N_{1}}}{dz} & = -D_{N_{1}}(n_{N_{1}}-n_{N_{1}}^{\rm eq})-D_{N_{1}\longrightarrow \psi S}(n_{N_{1}}-n_{N_{1}}^{\rm eq})-\dfrac{s}{H(z)z} [ (n_{N_{1}}^2-(n_{N_{1}}^{\rm eq})^{2})\langle \sigma v \rangle_{N_{1}N_{1} \longrightarrow l l} \nonumber \\
 &  +  [n_{N_{1}}n_{SM}-n_{N_{1}}^{\rm eq}n_{SM}^{\rm eq}] \langle \sigma v \rangle_{\eta l \longrightarrow N_{1} (W^{\pm},Z)}   ],       
\label{eq:5}
\end{align}

\begin{align}\label{eq:6}
        \dfrac{dn_{(B-L)_i}}{dz} & =-\epsilon_{\psi_{i}}D_{\psi}(n_{\psi}-n_{\psi}^{\rm eq})-\epsilon_{N_{1}i}D_{N_{1}}(n_{N_{1}}-n_{N_{1}}^{\rm eq})-(W_{N_{1}}P_{1i}+W_{\psi_{i}}P_{\psi i})n_{(B-L)_i} \nonumber \\ 
        & -\dfrac{s}{H(z)z} [P_{\psi i} \Gamma_{S l_{i} \longrightarrow \psi \eta} + P_{\psi i}\Gamma_{l_{i} \eta \longrightarrow \psi S}+ P_{\psi i} \sum_j P_{\psi j} \Gamma_{l_{i}l_{j} \longrightarrow \eta \eta}+ P_{\psi i} \sum_{j}P_{\psi j} \Gamma_{l_{i}l_{j} \longrightarrow N_{1}N_{1}}+ \nonumber \\ & P_{\psi i}\Gamma_{l_{i} \eta \longrightarrow (N_{1} W^{\pm},Z)}   +  P_{\psi i} \sum_{j}\Gamma_{\eta l_{i} \longrightarrow \eta^{*} \Bar{l_{j}}}+P_{\psi i}\Gamma_{\psi l_{i} \longrightarrow S \eta} ]n_{(B-L)_i},
\end{align}

where the projectors $P_{\psi i}$ and $P_{1i}$ are defined, respectively, as 
\begin{equation}
    P_{\psi i}=\dfrac{\Gamma_{\psi \longrightarrow S \eta l_{i}}}{\sum_{j}\Gamma_{\psi \longrightarrow S \eta l_{j}}}.
\end{equation}
\begin{eqnarray}
P_{1i}=\dfrac{\Gamma_{N_{1}\longrightarrow l_{i}\eta}}{\sum_{j} \Gamma_{N_{1}\longrightarrow l_j \eta}}
\end{eqnarray}
In the projectors, the denominator indicates the total decay width of $\psi, N_1$ whereas the numerators correspond to partial decay width into a particular lepton flavour. The washout terms remain same as before. We fix the benchmarks same as the ones used in scanning the parameter space of unflavoured leptogenesis. The resulting parameter space in $m_{\psi}-\lambda_5$ plane for both the cases are shown in figure \ref{case12_scan_flav}. Comparing with scan plots for unflavoured leptogenesis shown in figure \ref{scan} , it is seen that the scale of leptogenesis can be be lowered after inclusion of lepton flavour effects, as expected. Thus, successful leptogenesis can occur at a scale just above 1 TeV for $M_{1}=2\times 10^{5}$ GeV and can be as low as approximately $5$ TeV for case $M_{1}=2\times 10^{7}$ GeV.

\subsection{Dark Matter}
\begin{figure}
\centering
\includegraphics[scale=.26]{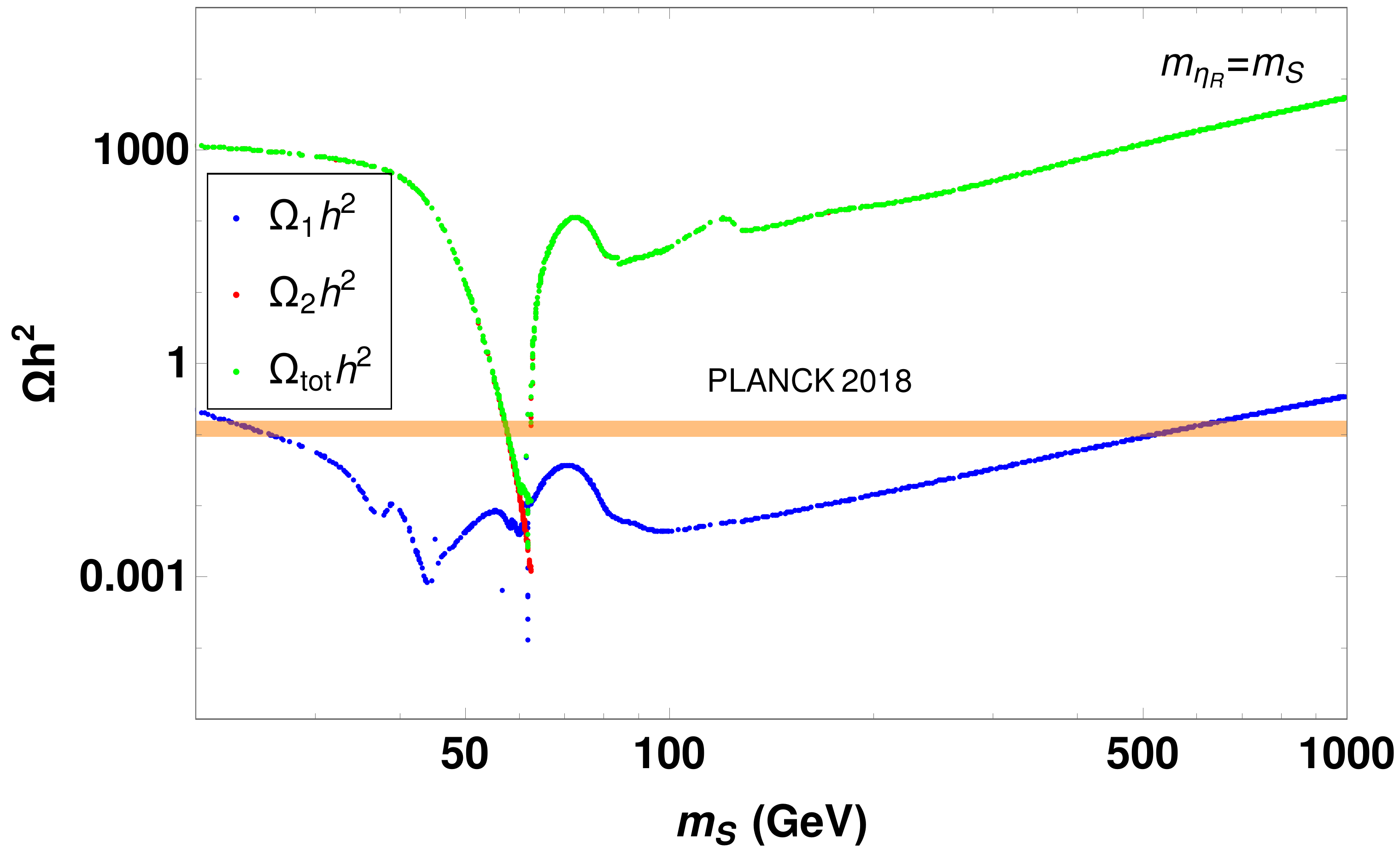}
\includegraphics[scale=.26]{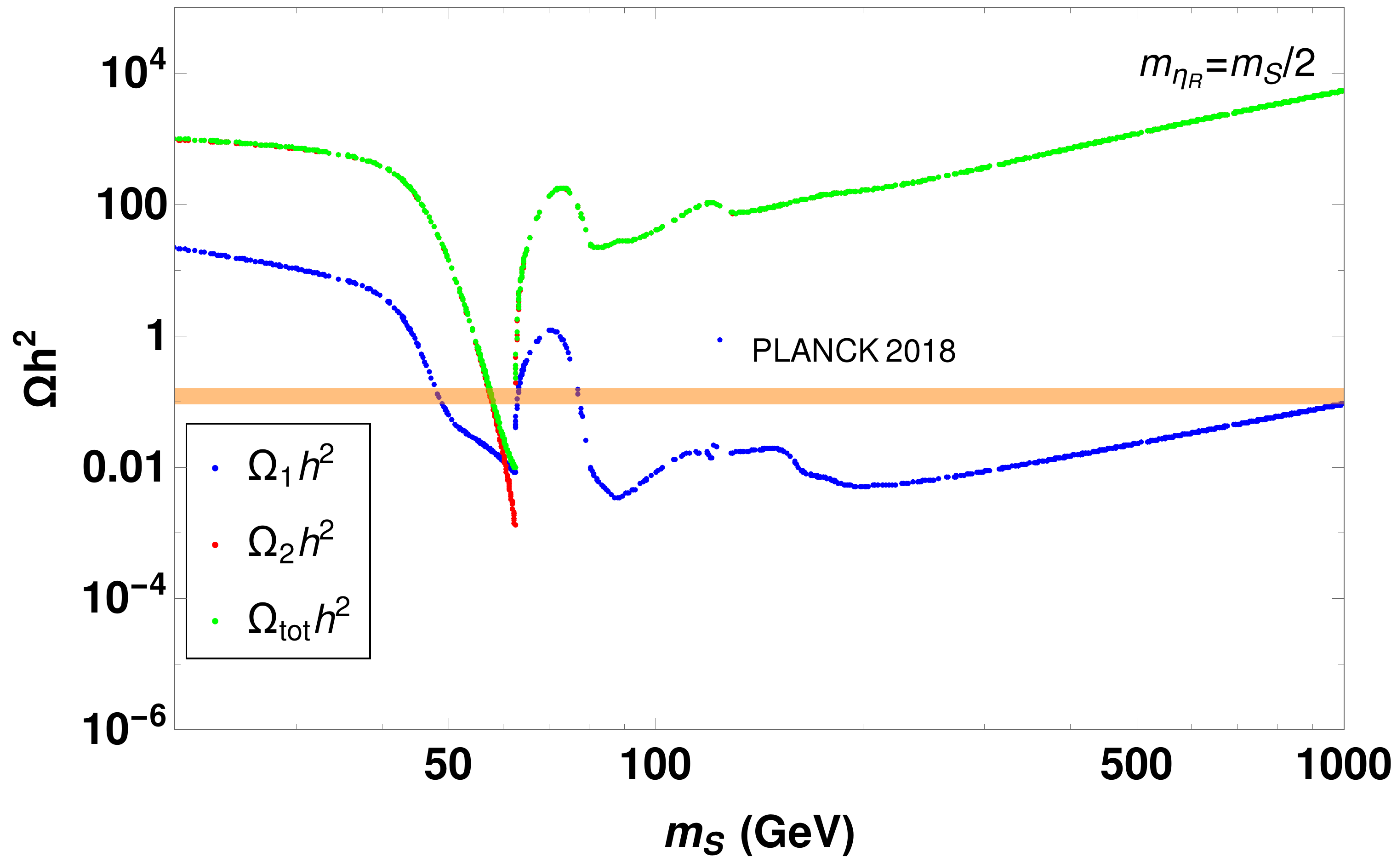} \\
\includegraphics[scale=.26]{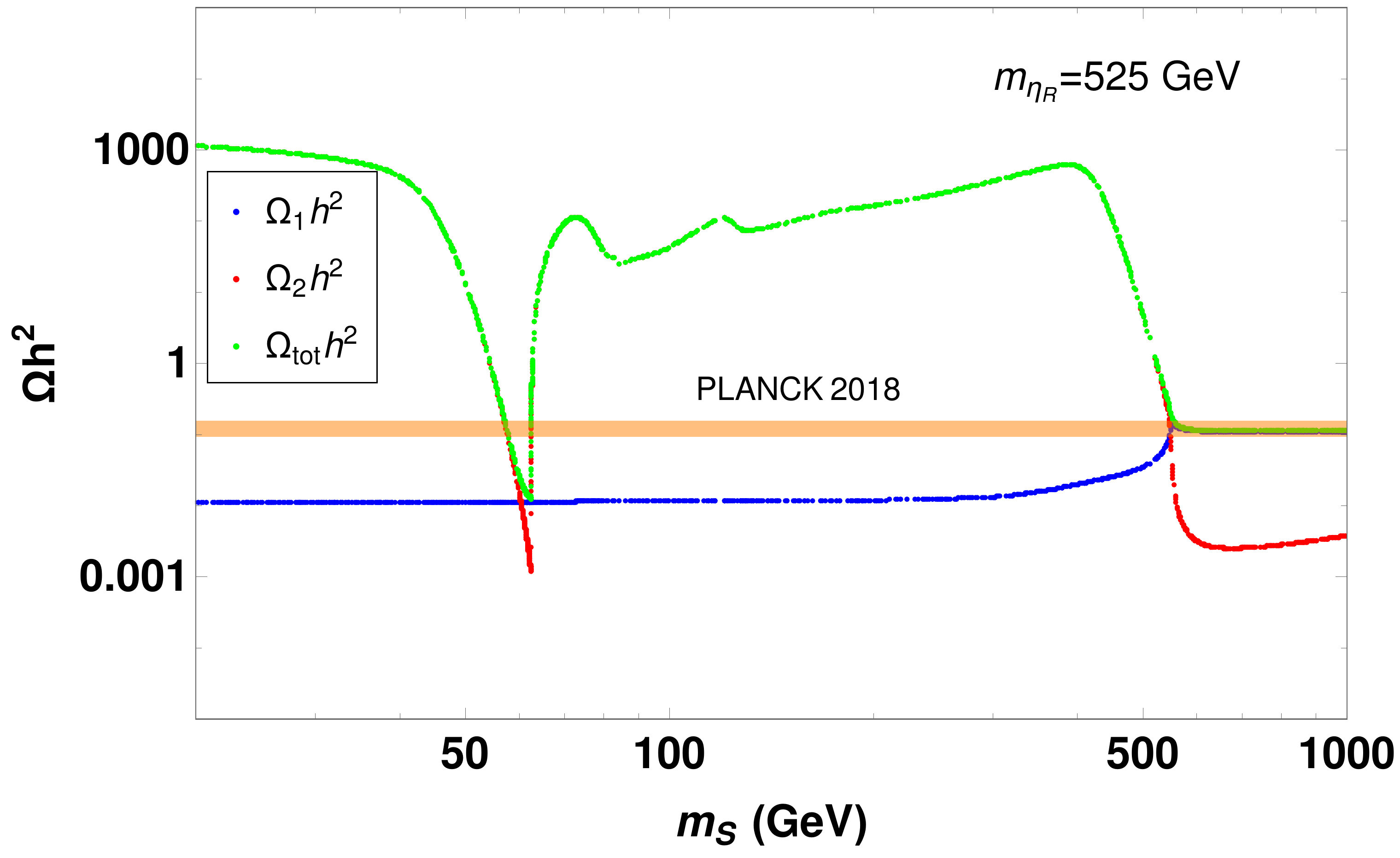}
\includegraphics[scale=.26]{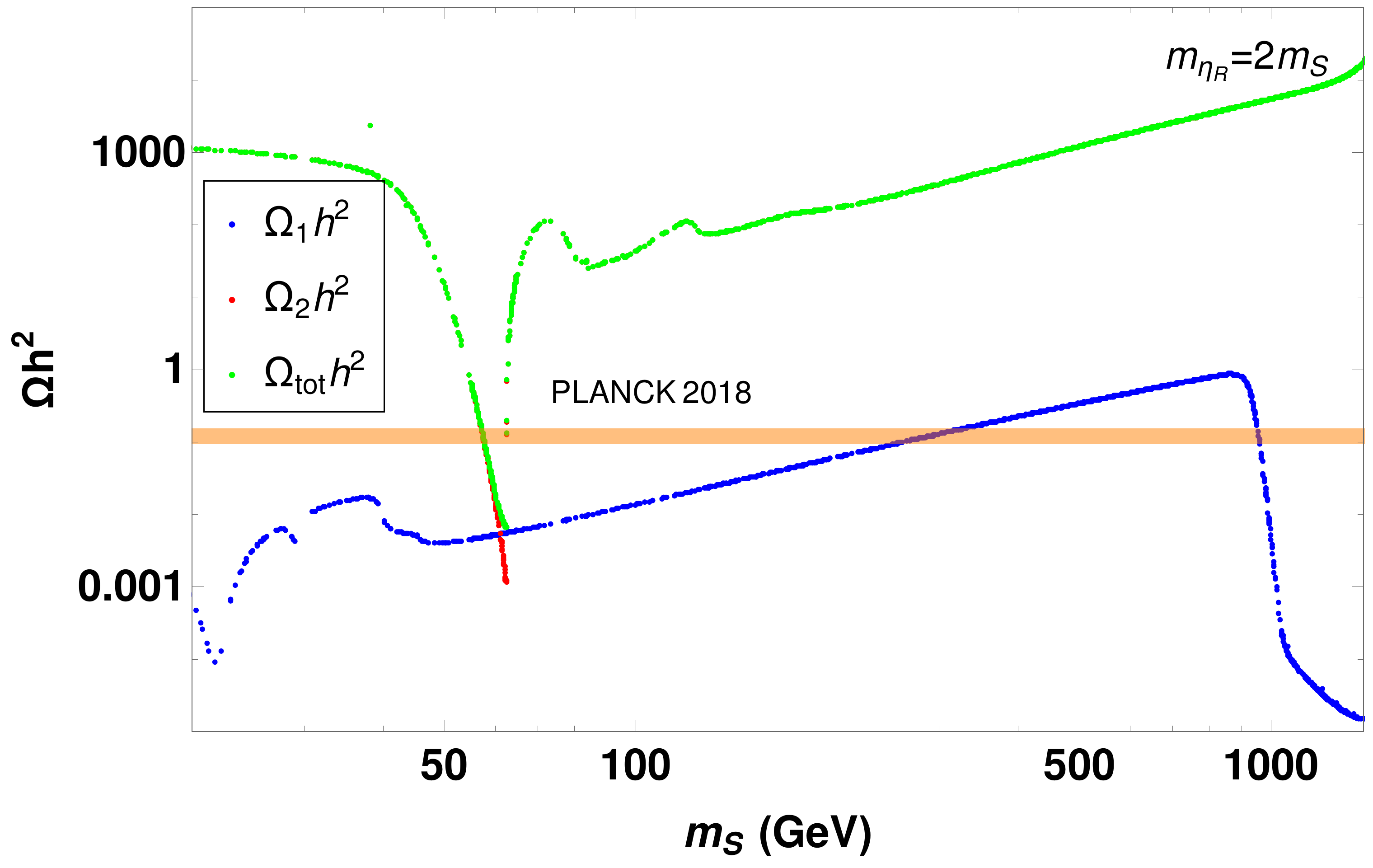}
\caption{Relic abundance versus DM mass for various mass relations between two DM candidates. The other parameters are fixed at the following benchmark values $\lambda_{L}=10^{-4}$, $\lambda_{6}=10^{-3}$, $\lambda_{7}=0$, $y_{1,2}=10^{-4}$, $\Delta m_{\eta_{I}}=2$ GeV and $\Delta m_{\eta_{\pm}}=2$ GeV. }    
\label{fig:dm1}
\end{figure}

We briefly discuss our dark matter results in this subsection. As mentioned earlier, a two component scalar singlet and scalar doublet DM has been recently discussed in details within a type I seesaw model \cite{Bhattacharya:2019fgs}. Instead of showing the details in general, here we focus on possible differences due to new couplings of these two DM candidates in relation to leptogenesis and neutrino mass as discussed above. We first discuss the behaviour of DM relic density with its mass for various possible combinations of relevant benchmark parameters. In figure \ref{fig:dm1}, we show the variation in individual and total DM relic densities for different mass relations between two DM candidates. While the overall features agree with the known results of scalar singlet and scalar doublet DM, there are some interesting differences due to inter-conversions and coannihilations here which we highlight.

In top left panel of figure \ref{fig:dm1}, the two DM candidates are assumed to have equal masses. The Higgs portal interactions of both the DM candidates are open due to the chosen non-zero couplings $\lambda_6, \lambda_L=\lambda_3+\lambda_4+\lambda_5$. Although the Higgs portal coupling of doublet DM is relatively smaller, the coannihilation channels are very efficient due to tiny mass splittings $\Delta m_{\eta_I}=m_{\eta_I}-m_{\eta_R}, \Delta m_{\eta^{\pm}}=m_{\eta^{\pm}}-m_{\eta_R}$, keeping its relic abundance suppressed compared to the singlet DM. In the top right panel plot of figure \ref{fig:dm1}, a noticeable change in doublet DM relic abundance is observed. While all relevant couplings have the same value as those on the top left panel plot, the doublet DM relic increases as singlet DM mass is twice the mass of doublet DM and hence there can be efficient conversions from singlet to doublet DM through Higgs portal interactions. Note that in both of these plots, the direct conversion coupling $\lambda_7$ is switched off and hence all possible DM conversions can occur only via Higgs portal interactions. To show the effect of DM conversion more clearly, we keep the mass of doublet DM fixed in the bottom left panel plot of figure \ref{fig:dm1}. As the singlet DM mass approaches the doublet DM mass, there is a sharp fall in its relic while at the same time the doublet relic increases due to relative conversions. In this plot, such conversions can occur via both Higgs portal and direct coupling $\lambda_7$. Finally, on the bottom right panel of figure \ref{fig:dm1}, we show one interesting feature where doublet DM relic density suddenly drops as its mass becomes close to 1.5 TeV. This particular feature is not due to DM conversions via Higgs portal or direct coupling $\lambda_7$ as that can happen at any mass, given the fact that doublet mass is twice that of singlet mass all throughout. This happens due to doublet DM coannihilation with $\psi$ whose mass is fixed at 1.5 TeV. Due to this coannihilation $\eta_R \psi \rightarrow S \ell$, the singlet relic density also increases, though it is not as prominent as the depletion of doublet relic density in the figure.

\begin{figure}
\centering
\includegraphics[scale=.26]{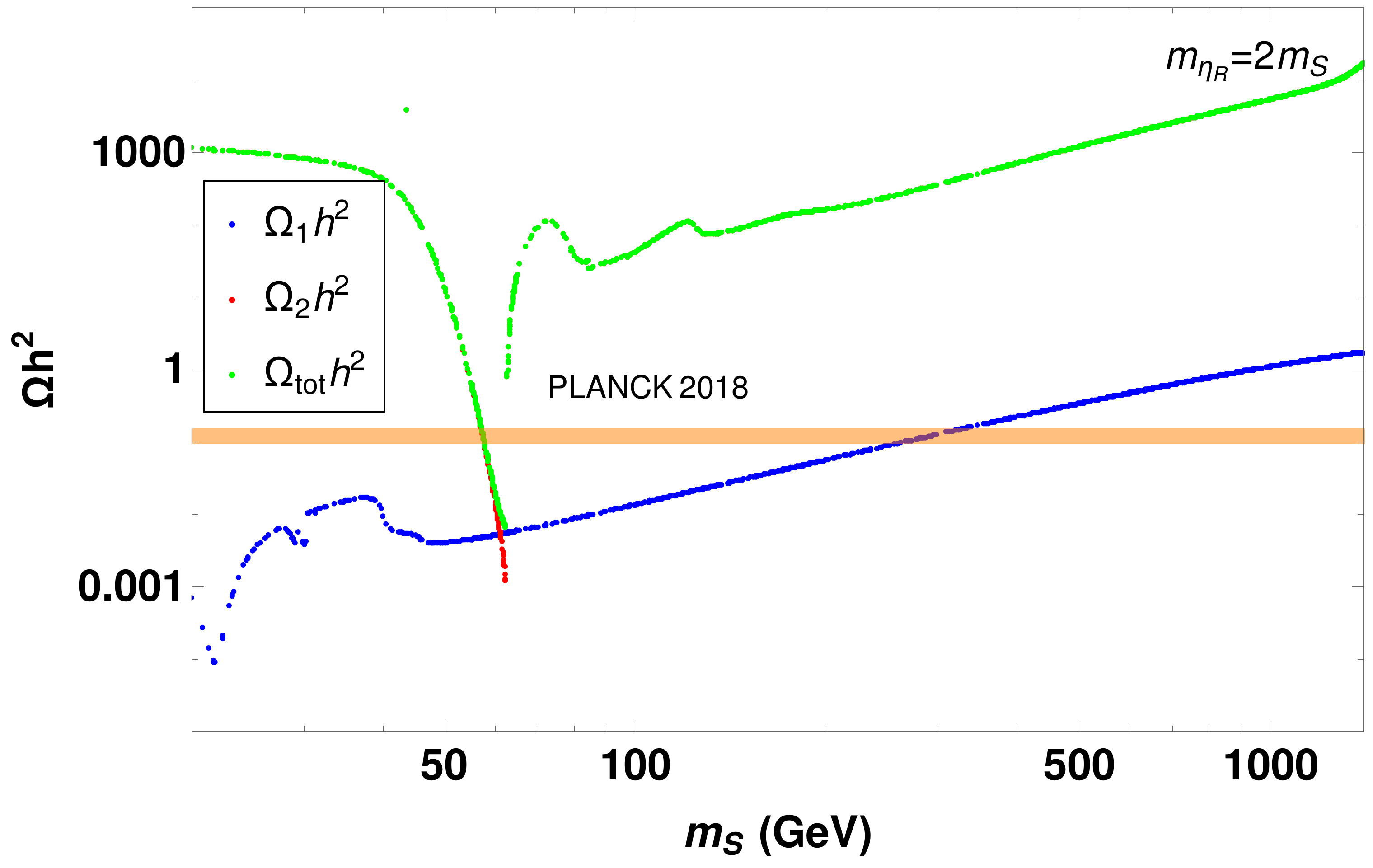}
\includegraphics[scale=.26]{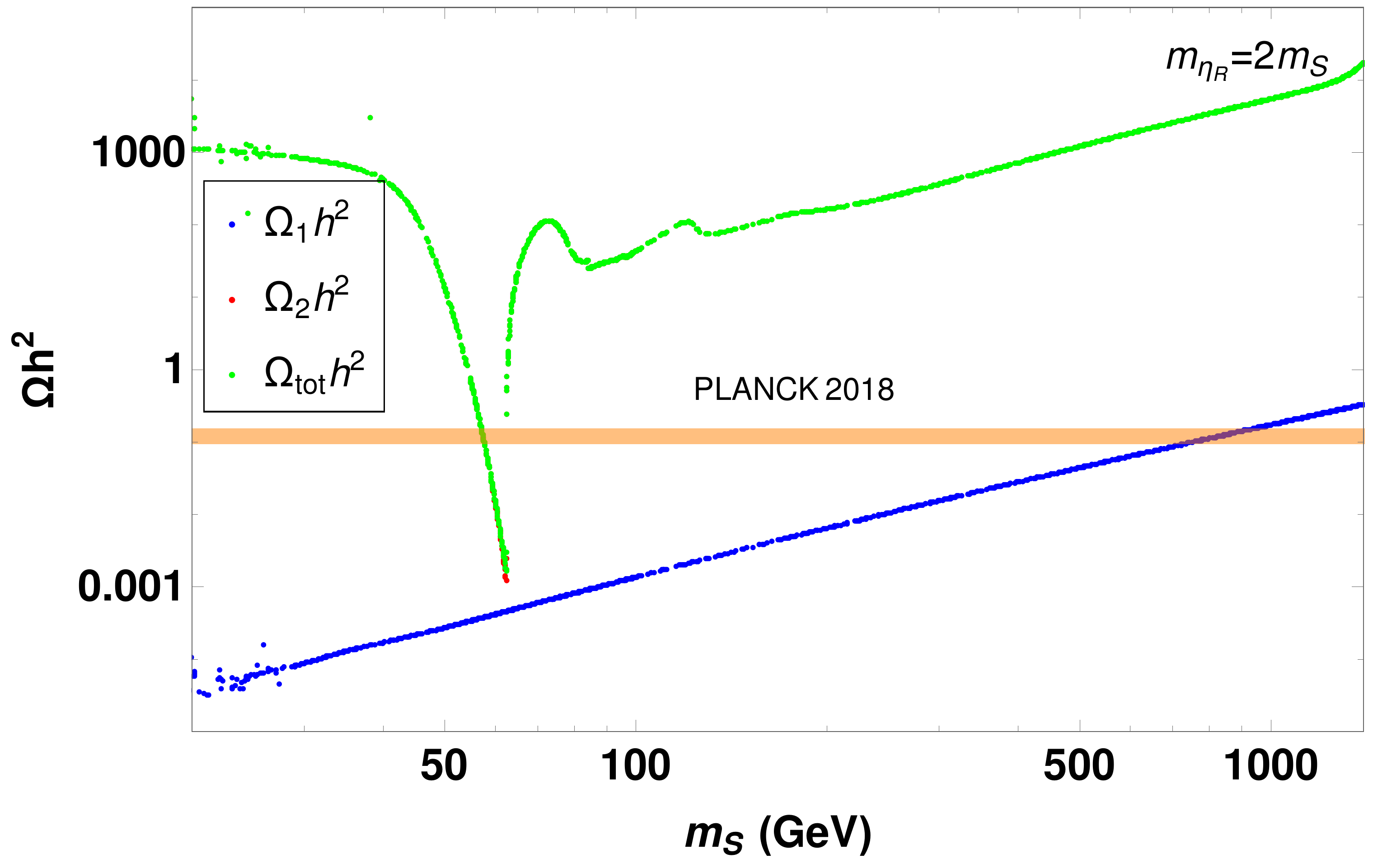} \\
\includegraphics[scale=.26]{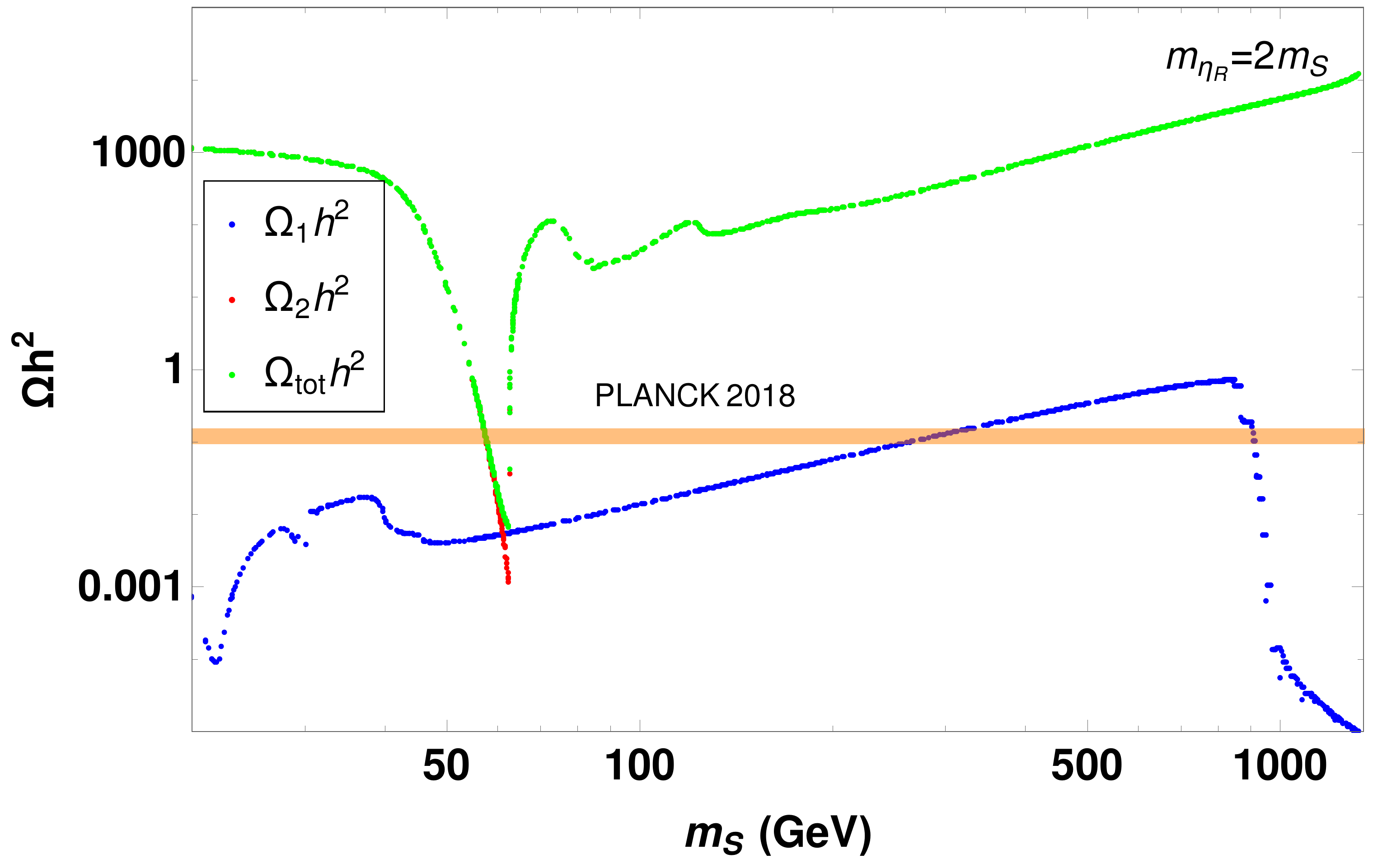}
\includegraphics[scale=.26]{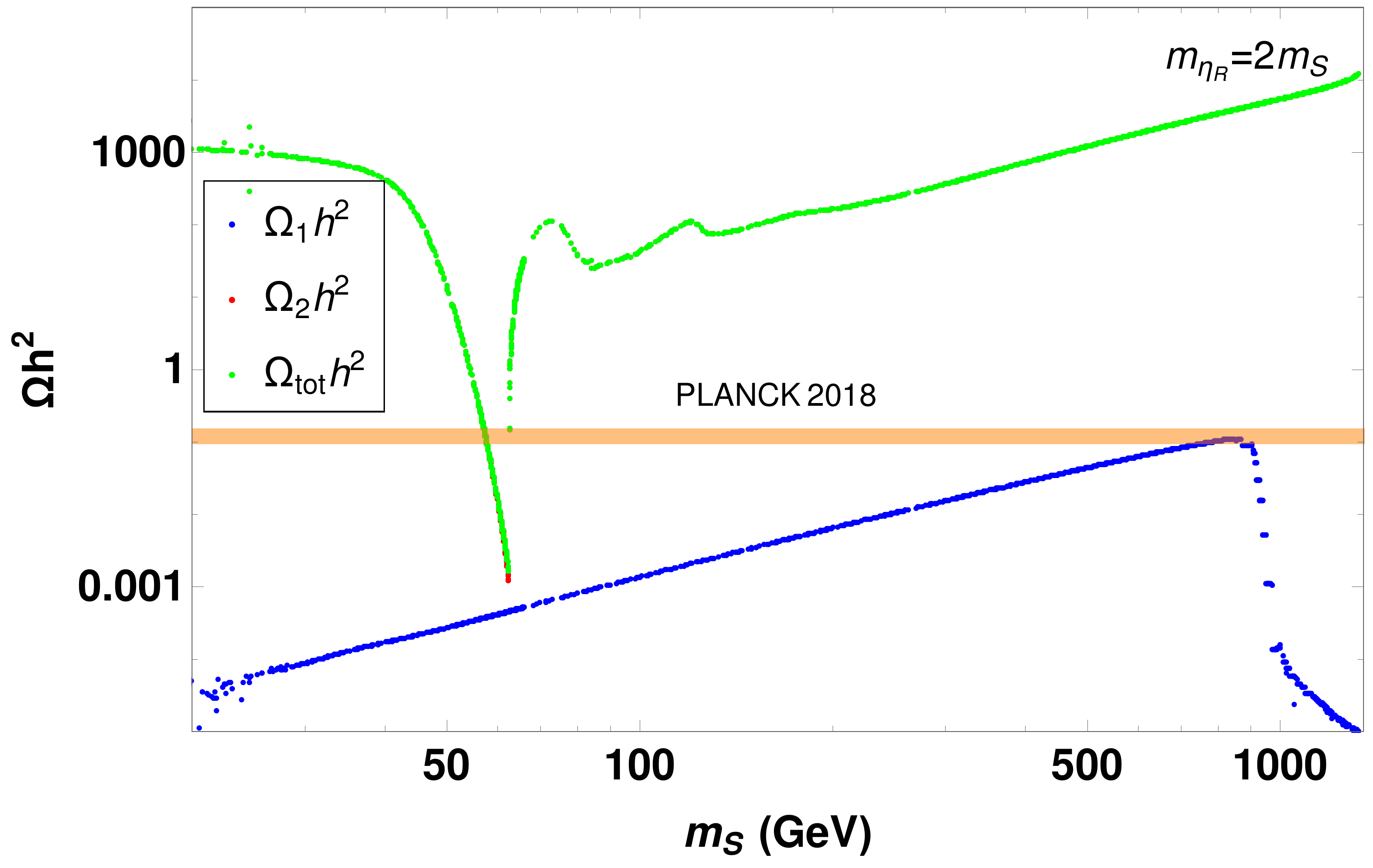}
\caption{Relic abundance versus DM mass showing the effects of direct conversion coupling $\lambda_7$ and Yukawa coupling $y_{1,2}$ of $\psi-S-N_{1,2}$ vertices. The benchmark parameters fixed for all the four plots are $\lambda_{6}=10^{-3}$ and $\lambda_{L}=10^{-4}$. The conversion coupling and the new Yukawa coupling are fixed at $ y_{1,2}=0$, $\lambda_{7}=0$ (upper left panel plot), $y_{1,2}=0$, $\lambda_{7}=1$ (upper right panel plot), $y_{1,2}=10^{-2}$, $\lambda_{7}=0$ (lower left panel plot) $y_{1,2}=10^{-2}$, $\lambda_{7}=1$ (lower right panel plot).}    
\label{fig:dm2}
\end{figure}

After discussing the general features of DM relic dependence on various relevant parameters, in figure \ref{fig:dm2}, we specifically show the effects of direct conversion coupling $\lambda_7$ and Yukawa coupling $y_{1,2}$ of $\psi-S-N_{1,2}$ vertices. Mass of doublet dark matter is assumed to be twice of singlet dark matter mass. Comparing top panel plots of figure \ref{fig:dm2} where $y_{1,2}=0$, it is seen that turning on the direct conversion coupling $\lambda_7$ leads to sharp fall in heavier DM relic density. Same effect is visible while comparing the bottom panel plots also where the effect of $y_{1,2} \neq 0$ is also shown leading to depletion of doublet DM relic as its mass approaches $m_{\psi}$.

\begin{figure}[h]
    \centering
    \includegraphics[scale=.53]{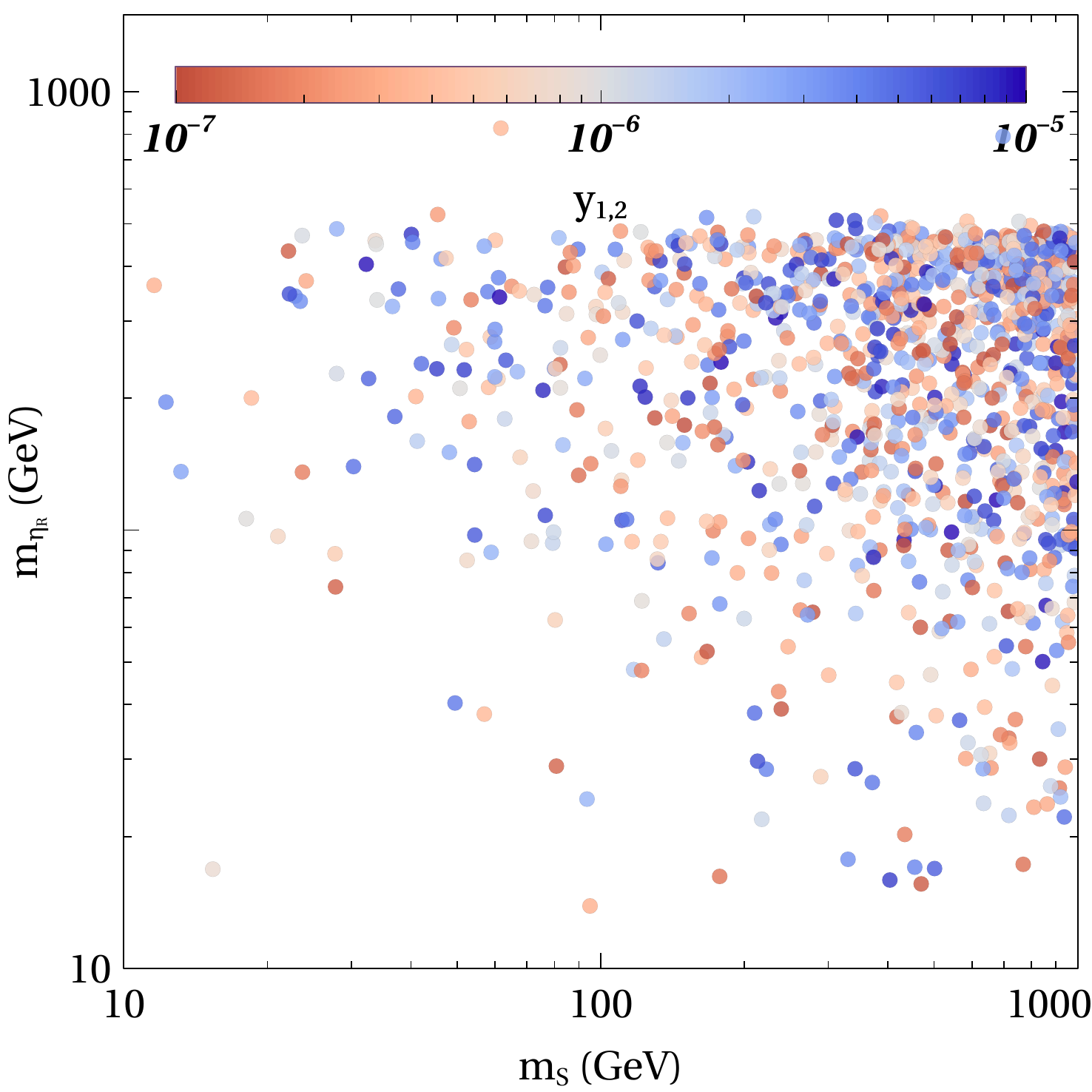}
    \includegraphics[scale=.53]{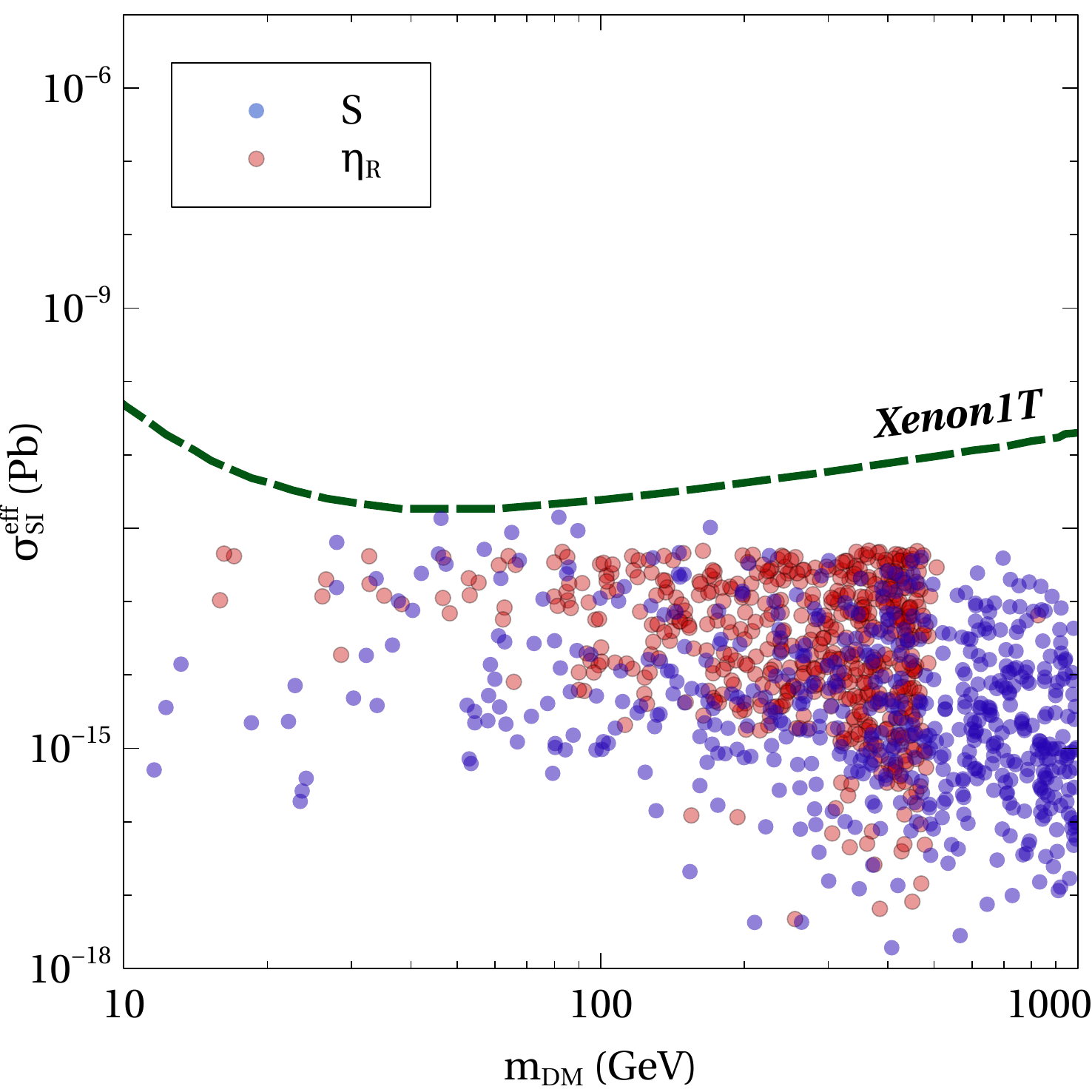}
    \caption{Scan plot showing the parameter space in $m_{S}-m_{\eta_{R}}$ plane allowed from total DM relic
abundance (left panel) and $m_{\rm DM}-\sigma_{\rm SI}$ plot for all the points satisfying the total relic (right panel). For this scan the RHN masses are set at $M_{1}=2\times 10^{5}$ GeV and $M_{2}=2\times 10^{6}$ GeV. The other important parameters are randomly varied within the ranges $10^{-7}<\lambda_{5}<10^{-1}$, $10^{-4}<\lambda_{6}<10^{-2}$, $10^{-4}<\lambda_{7}<10^{-2}$ and $10^{-7}<y_{1,2}<10^{-5}$. }
    \label{fig:DMscan1}
\end{figure}

\begin{figure}
    \centering
    \includegraphics[scale=.53]{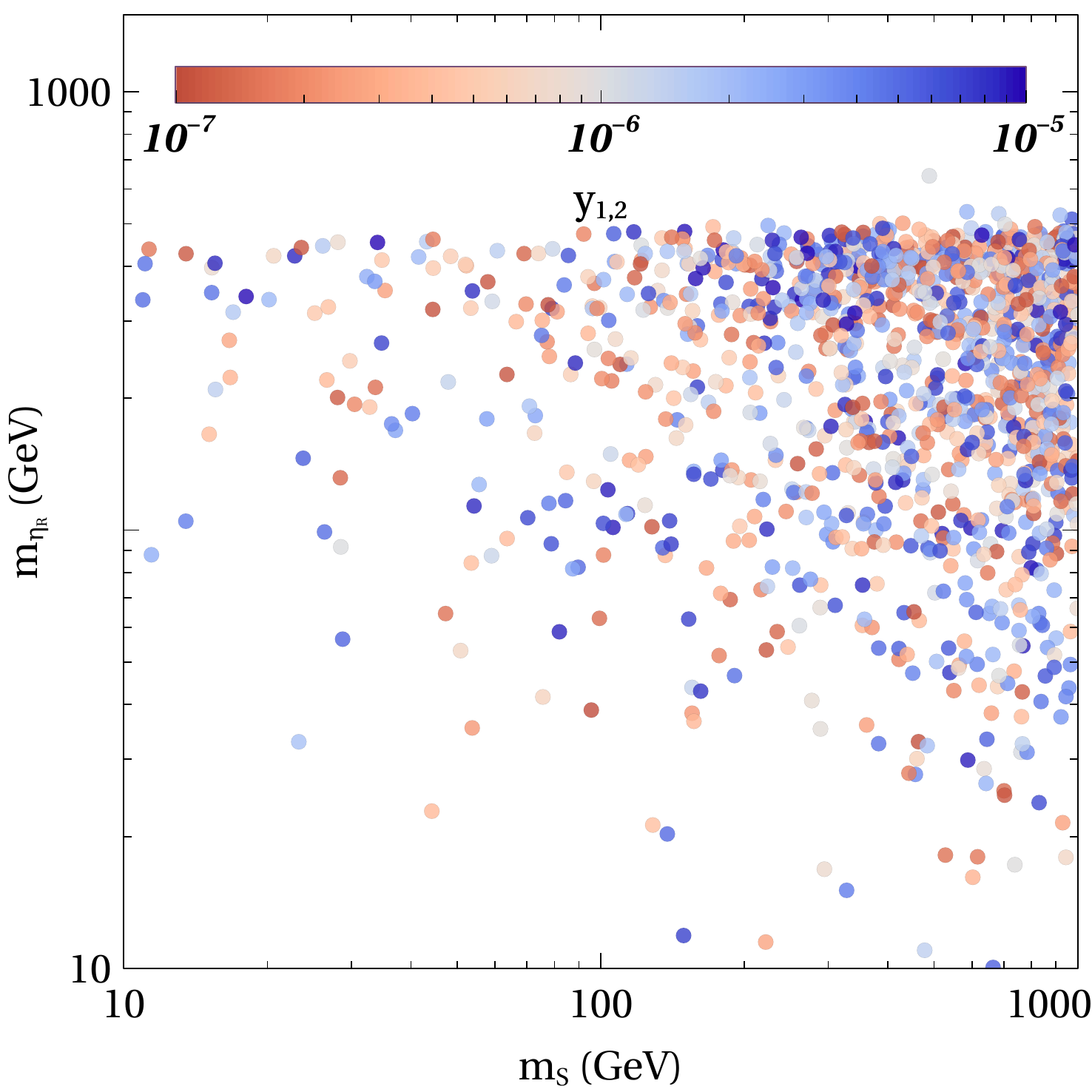}
    \includegraphics[scale=.53]{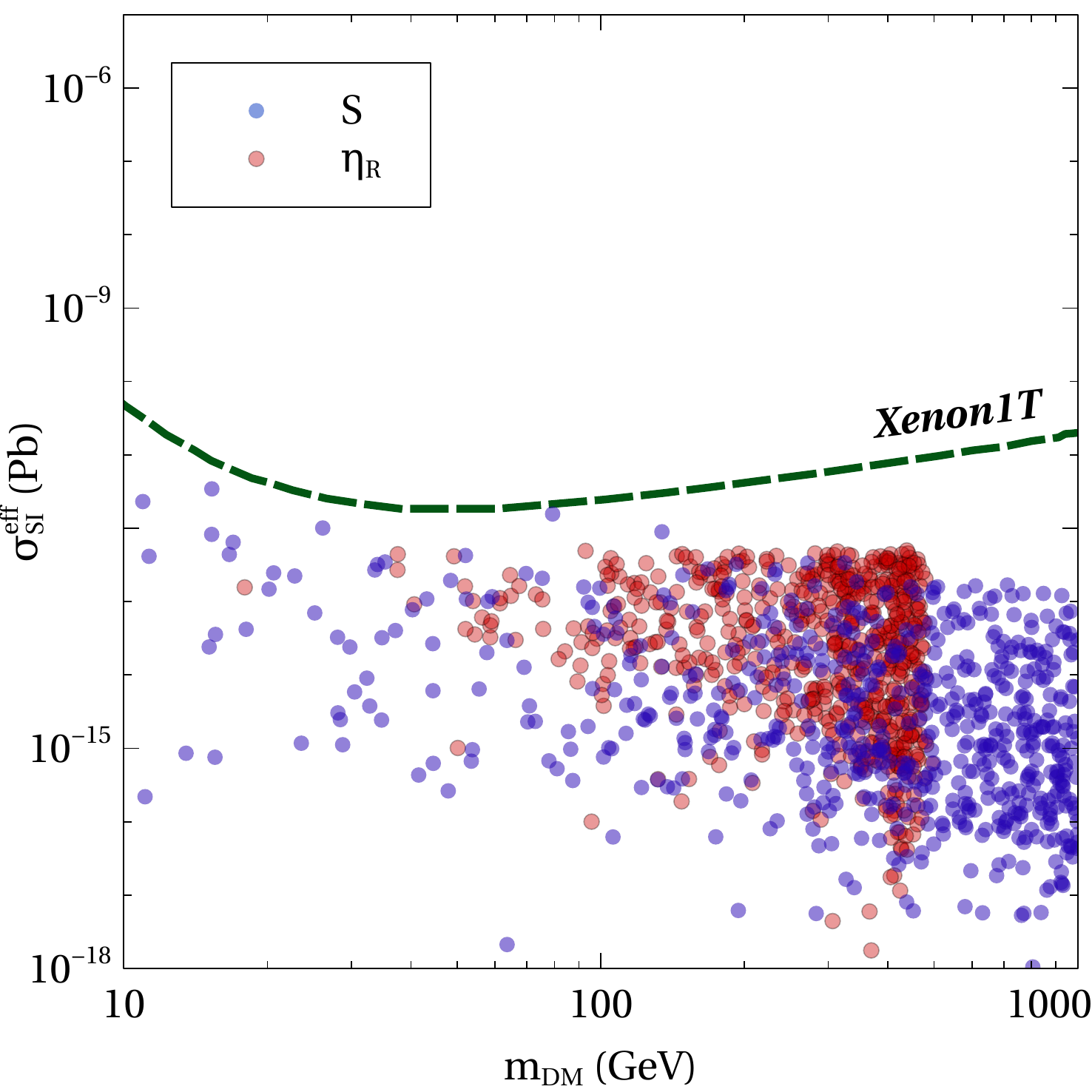}
    \caption{Scan plot showing the parameter space in $m_{S}-m_{\eta_{R}}$ plane allowed from total DM relic
abundance (left panel) and $m_{\rm DM}-\sigma_{\rm SI}$ plot for all the points satisfying the total relic (right panel). For this scan the RHN masses are set at $M_{1}=2\times 10^{7}$ GeV and $M_{2}=2\times 10^{8}$ GeV. The other important parameters are randomly varied within the ranges $10^{-7}<\lambda_{5}<10^{-1}$, $10^{-4}<\lambda_{6}<10^{-2}$, $10^{-4}<\lambda_{7}<10^{-2}$ and $10^{-7}<y_{1,2}<10^{-5}$.}
    \label{fig:DMscan2}
\end{figure}

To find the relevant parameter space of DM that gives rise to the observed relic density, we perform a numerical scan of the relevant parameter space favoured from the requirement of successful leptogenesis. For the first case ($M_{1}=2\times 10^{5}$ GeV), the parameter space in terms of two DM masses is shown on left panel plot of figure \ref{fig:DMscan1} . To be in agreement with the parameter space chosen for leptogenesis, here we fix $m_{\eta_{R}}<m_{\psi}$ and vary other parameters in the range $10\, {\rm GeV}<m_{S},m_{\eta}<1000$ GeV, $10^{-6}<\lambda_{5}<10^{-1}$, $10^{-4}<\lambda_{6},\lambda_{7}<10^{-2}$, $10^{-7}<y_{1,2}<10^{-5}$. While singlet DM masses are evenly distributed across the range, there seems to be an upper bound on doublet DM mass near 500 GeV. This is due to the chosen mass splitting within doublet components. As earlier studies of inert scalar doublet DM shows \cite{Deshpande:1977rw, Dasgupta:2014hha, Cirelli:2005uq, Barbieri:2006dq, Ma:2006fn, LopezHonorez:2006gr, Hambye:2009pw, Dolle:2009fn, Honorez:2010re, LopezHonorez:2010tb, Gustafsson:2012aj, Goudelis:2013uca, Arhrib:2013ela, Diaz:2015pyv, Ahriche:2017iar, Borah:2017dfn}, for such small mass splitting, the DM is overproduced in the high mass regime. While underproduction of one DM component in our model can be compensated by the second DM component, overabundance of one is difficult to reconcile with. Choosing a larger mass splitting within inert doublet components will allow more region of parameter space in terms of doublet DM mass. The right panel plot of figure \ref{fig:DMscan1} shows the spin independent DM-nucleon scattering rate of both the DM components, compared against the latest bound from Xenon1T experiment \cite{Aprile:2018dbl}. Clearly, all the points satisfy the direct detection bounds. This is due to the fact that, we have kept the Higgs portal coupling of both the DM candidates fixed at small value. We varied $10^{-6}<\lambda_{6}<10^{-2}$ and fixed $\lambda_{L}=10^{-3}$. Since tree level DM-nucleon scattering arises through Higgs portal couplings only, the corresponding rates remain low enough to survive Xenon1T bounds. The colour code on left panel plot of figure \ref{fig:DMscan1} shows the value of $y=\lvert y_1 \rvert = \lvert y^*_2 \rvert$. Similar scan plot for case 2 is shown on the left panel of figure \ref{fig:DMscan2}. While we notice a similar upper bound on doublet DM mass due to chosen mass splitting, the parameter space remains safe from direct detection bounds.

\section{Conclusion}
\label{sec6}
We have proposed a model to implement the idea of leptogenesis from three body decay of a heavy particle where non-zero CP asymmetry arises due to interference of multiple three body decay diagrams with resummed propagators along with dark matter. Adopting a minimal framework to implement the idea, we augment the standard model of particle physics by three singlet fermions and two scalar fields: one singlet and one doublet. While two of these singlet fermions and the additional scalar doublet help in generating light neutrino masses one loop level, the other two particles help in realising the desired three body decay leptogenesis. The two singlet fermions taking part in radiative neutrino mass generation also act like mediators in two different three body decay diagrams the interference of which results in the required non-zero CP asymmetry. It turns out that this setup automatically gives rise to a two component dark matter scenario in terms of scalar singlet and neutral component of scalar doublet. After deriving the particle spectrum of the model and applying the theoretical as well as experimental bounds, we calculate the CP asymmetry from three body decay of heavy singlet fermion by considering interference of two different diagrams. We then solve the Boltzmann equations relevant for leptogenesis incorporating the sources of lepton asymmetry as well as washouts to obtain the parameter space that can give rise to successful leptogenesis. While both two body decay of right handed neutrino $N_1$ (similar to scotogenic model) and three body decay of the new singlet fermion $\psi$ introduced in our model can contribute to lepton asymmetry, we check that in the low scale leptogenesis scenario we focus, the contribution from two body decay remains sub-dominant. After analysing the role or effects of some key parameters on generation of lepton asymmetry, we performed a numerical scan and show that successful leptogenesis can occur at a scale as low as 3 TeV. This is a factor of around 3 times lower than the scale of leptogenesis in minimal scotogenic model considering two body decay of hierarchical heavy neutrinos studied in earlier works \cite{Hambye:2009pw, Kashiwase:2012xd, Kashiwase:2013uy, Racker:2013lua, Clarke:2015hta, Hugle:2018qbw, Borah:2018rca, Mahanta:2019gfe, Mahanta:2019sfo, Sarma:2020msa}. In fact, we also checked that after incorporating lepton flavour effects, the scale can be as low as 2 TeV. Such low scale leptogenesis possibility could have tantalising prospects of being probed at ongoing or near future experiments. This difference in scale of leptogenesis from minimal scotogenic model arises due to three body decay as a dominant source and also due to the freedom in choosing $y_{\alpha} \psi  N_{\alpha} S$ coupling in \eqref{yukawaL} which does not play any role in generating light neutrino masses. Lowering of leptogenesis scale due to introduction of such new couplings (not related to origin of neutrino mass) have been explored in earlier works also. For example, in \cite{Alanne:2018brf}, the scale of leptogenesis in a scalar singlet extension of type I seesaw model was shown to be as low as 500 GeV even with hierarchical right handed neutrinos which is significantly lower than the scale of leptogenesis in usual type I seesaw model.

After finding the parameter space that gives rise to successful TeV scale leptogenesis, we calculate the relic abundance of two DM components. Since such two component scalar DM have been already studied in earlier works, we focus primarily on the role of new parameters involving the two DM candidates in our model which also play non-trivial roles in leptogenesis. We first analyse these effects with benchmark choices of parameters and finally show the parameter space of two DM masses that is consistent with correct relic abundance and direct detection rates. Such a low scale model with two component DM, successful leptogenesis and light neutrino masses should face further scrutiny with future data from collider, neutrino, cosmology as well as rare decay experiments looking for charged lepton flavour violation, neutrinoless double beta decay etc. While neutrinoless double beta decay contribution will effectively arise from light neutrino contributions only and will remain below the current experimental sensitivity of KamLAND-Zen experiment, i.e., $\lvert m_{ee} \rvert \leq (0.061-0.165)\; {\rm eV}$ \cite{KamLAND-Zen:2016pfg} for vanishing lightest neutrino mass. While charged lepton flavour violation like $\mu \rightarrow e \gamma, \mu \rightarrow 3e$ and $\mu \rightarrow e$ (Ti) conversion in scotogenic models can be sizeable and saturate experimental upper bounds on corresponding branching ratios for fermion DM scenario \cite{Vicente:2014wga, Borah:2018smz, Borah:2020wut}, in our model they are likely to be suppressed as the singlet fermions $N_{1,2}$ are heavier than the scale of leptogenesis. Another interesting prospect of probing our model can be in the form of gravitational waves from a strongly first order phase transition (SFOPT). In a recent work \cite{Borah:2020wut}, it was shown that in the minimal scotogenic model, the criteria of SFOPT constrains the scalar sector a lot, leading to a scalar DM parameter space in tension with direct detection bounds. Due to the presence of an additional singlet scalar in our model whose mass is not as constrained as the inert doublet components, the SFOPT criteria is likely to be satisfied with more freedom. We leave a detailed study of this model from SFOPT point of view to future works.

\acknowledgments
DB acknowledges the support from Early Career Research Award from DST-SERB, Government of India (reference number: ECR/2017/001873).

\appendix
\section{CP asymmetry from three body decay of $\psi$}
\label{appen1}
\begin{figure}[h]
 \includegraphics[scale=0.3]{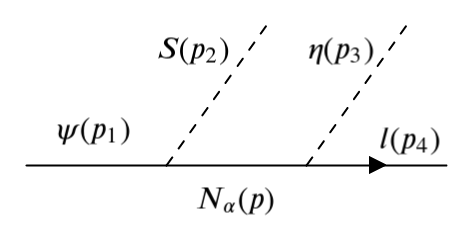}
 \includegraphics[scale=0.3]{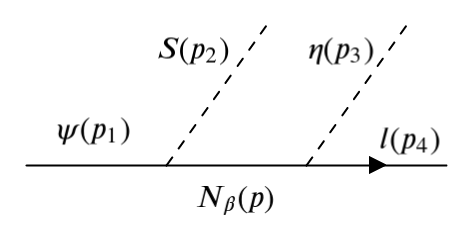}
 \caption{Two processes for the three body decay.}
 \label{fig:3decay}
\end{figure}

Let us start deriving the most general expression of asymmetry from an out of equilibrium process. The usual amplitude for a process is given as 

\begin{equation}
i \mathcal{M}_{i \rightarrow f}=\mathcal{E}\times \mathcal{A} \times \omega,
\end{equation}  

where $\mathcal{E}$ comprises of all the couplings and $\omega$ comprises of  all the wave functions of outgoing and incoming particles. Finally $\mathcal{A}$ contains all the rest of the term of the amplitude. For a non-zero CP asymmetry to be created one need at least two amplitudes for a process as shown in figure \ref{fig:3decay}.

Then the total amplitude for the process can be written as,
\begin{equation}
i \mathcal{M}_{i \rightarrow f}=[ \mathcal{E}_1 \mathcal{A}_1 + \mathcal{E}_2 \mathcal{A}_2 ] \omega .
\end{equation} 

Similarly, the amplitude for the process corresponding to its antiparticle counterpart is given as,
\begin{equation}
i \mathcal{M}_{\bar{i} \rightarrow \bar{f}}=[ \mathcal{E}_1^{*} \mathcal{A}_1 + \mathcal{E}_2^{*} \mathcal{A}_2 ] \omega^{\dagger}. 
\end{equation}
The corresponding amplitude squared terms are given as, 
\begin{equation}
\mid   \mathcal{M}_{i \rightarrow f} \mid^2=\left( \mid \mathcal{E}_1 \mid^2 \mid  \mathcal{A}_1 \mid^2 + \mid \mathcal{E}_2 \mid^2 \mid  \mathcal{A}_2 \mid^2 +2{\rm Re}[\mathcal{E}_1^{*} \mathcal{E}_2]{\rm Re}[\mathcal{A}_1^{*}\mathcal{A}_2]+ 2{\rm Im}[\mathcal{E}_1^{*} \mathcal{E}_2] {\rm Im}[\mathcal{A}_1^{*}\mathcal{A}_2  ] \right) \mid \omega \mid^2
\end{equation} 

\begin{equation}
\mid   \mathcal{M}_{\bar{i} \rightarrow \bar{f}} \mid^2=\left( \mid \mathcal{E}_1 \mid^2 \mid  \mathcal{A}_1 \mid^2 + \mid \mathcal{E}_2 \mid^2 \mid  \mathcal{A}_2 \mid^2 +2 {\rm Re}[\mathcal{E}_1 \mathcal{E}_2^{*}] {\rm Re}[\mathcal{A}_1^{*}\mathcal{A}_2]+ 2 {\rm Im}[\mathcal{E}_1 \mathcal{E}_2^{*}] {\rm Im}[\mathcal{A}_1^{*}\mathcal{A}_2]  \right)  \mid \omega \mid^2
\end{equation}

and therefore, the asymmetry is calculated to be, 
\begin{equation}
\delta = \mid \mathcal{M}_{i \rightarrow f} \mid^2 - \mid \mathcal{M}_{\bar{i} \rightarrow \bar{f}} \mid^2=-4 {\rm Im}[\mathcal{E}_1^{*}\mathcal{E}_2] {\rm Im}[\mathcal{A}_{1}^{*}\mathcal{A}_{2}] \mid \omega \mid^2.
\end{equation}

In the following subsections we first calculate the CP asymmetry by using the resummed propagators and then verify the same by tree-loop diagram interference calculation.We have adopted the two spinor notations from \cite{Dreiner:2008tw} throughout the derivations.

\subsection{Calculation of the CP asymmetry using resummed propagator}
\label{appen1b}
\begin{figure}[h]
 \begin{center}
  \includegraphics[scale=0.4]{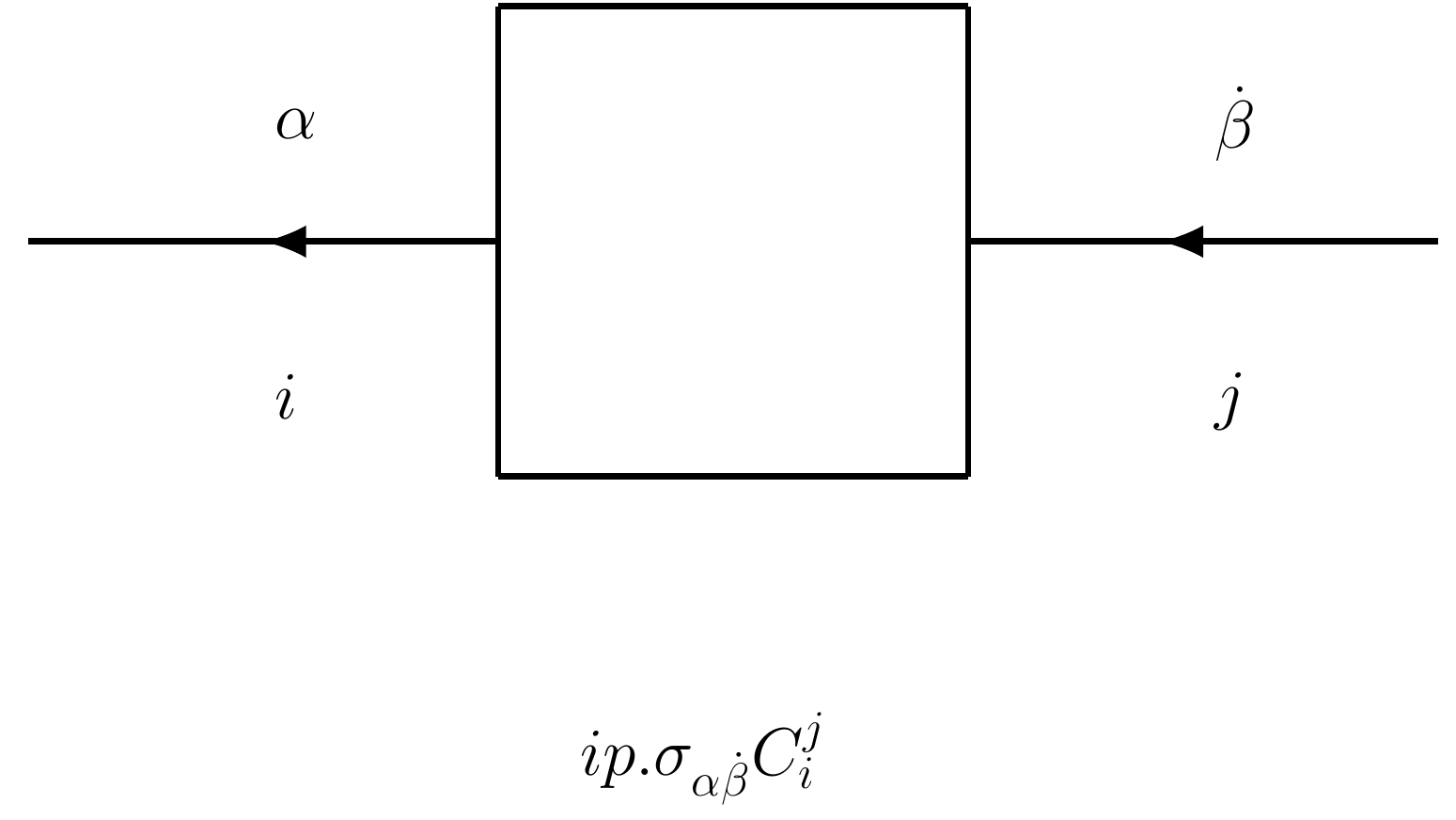}
  \includegraphics[scale=0.4]{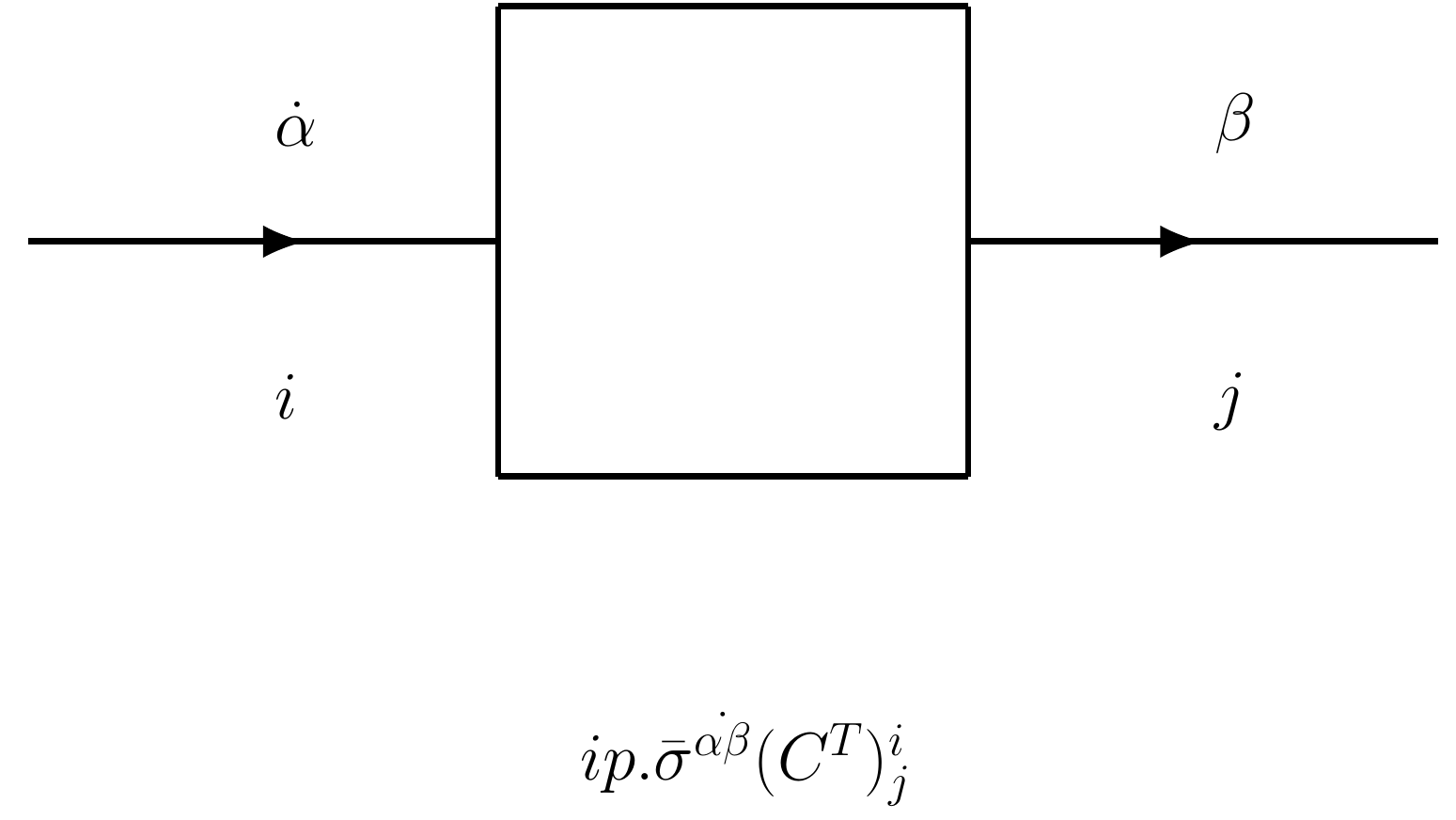}\\
  \vspace{10pt}
  \includegraphics[scale=0.4]{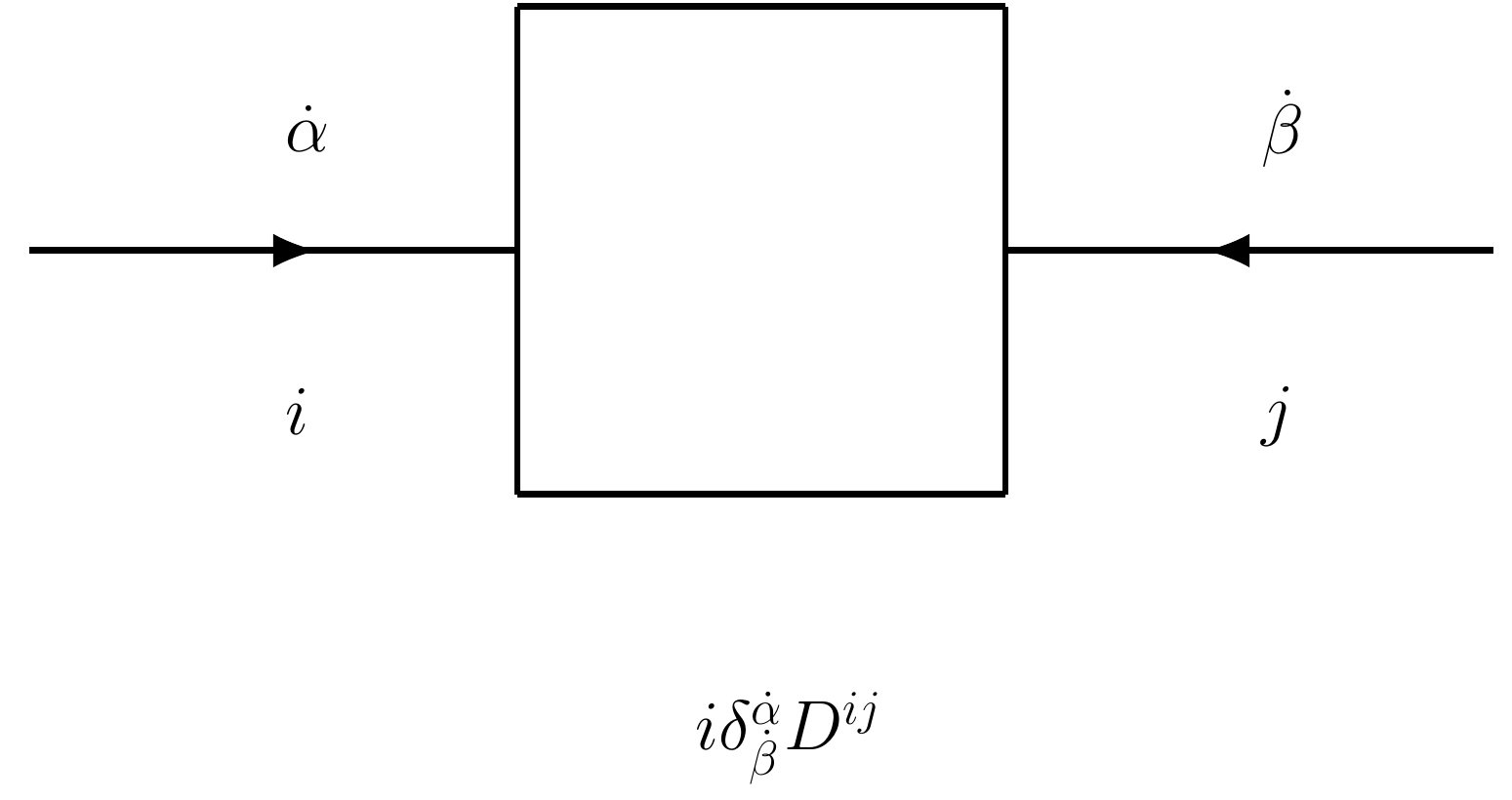}
  \includegraphics[scale=0.4]{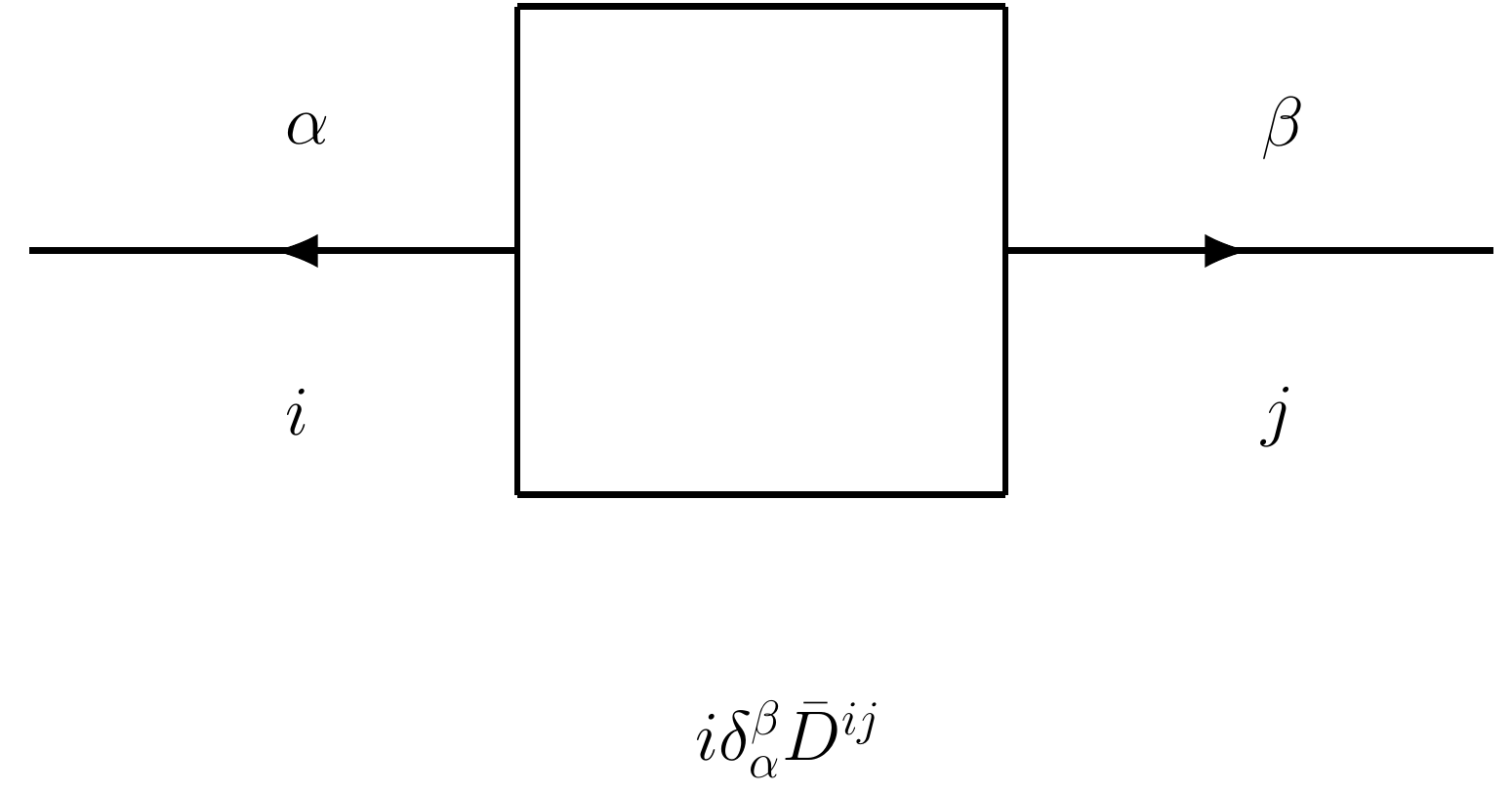}
  \caption{The full loop corrected propagators for two component fermions are associated with functions $C(p^{2})^{j}_{i}$ and its matrix transpose, $D(p^{2})^{ij}$ and $\bar{D}(p^{2})_{ij}$. The square boxes represent all the sum of all connected Feynman diagrams, with external legs included. The four-momentum p flows from right to left.}
  \label{fig:functions}
 \end{center}
 \end{figure}

Before doing the CP asymmetry calculation using resummed propagator let's calculate the resummed propagators first. In figure \ref{fig:functions} we show the diagrammatic representations of the full loop corrected propagators for two component fermions. For a detailed calculation on resummed propagator please see \cite{Dreiner:2008tw}.

 The full propagators can be organised in terms of one particle irreducible (1PI) self-energy functions \cite{Dreiner:2008tw}. These are defined as the sum of Feynman diagrams to all orders in perturbation theory (with the corresponding tree level graph excluded) that contribute to the 1PI two-point Green functions. 
 \begin{align} \label{propagator}
 \begin{pmatrix}
    i\bar{\textbf{D}}  &&& ip.\sigma \textbf{C} \\
    i p.\bar{\sigma} \textbf{C}^{T} &&& i\textbf{D}
 \end{pmatrix}=  
 \begin{pmatrix}
  i (m+\Omega) &&& -ip.\sigma (1-\Xi^{T} ) \\
  -ip.\bar{\sigma}(1-\Xi) &&& i(\bar{m}+\bar{\Omega})
 \end{pmatrix}^{-1}
\end{align}

The right hand side of the above equation can be evaluated by employing the following identity for the inverse of a block diagonal matrix,

\begin{align} \label{inverse}
 \begin{pmatrix}
  P &&& Q \\
  R &&& S
 \end{pmatrix}^{-1}=
 \begin{pmatrix}
  (P-QS^{-1}R)^{-1} &&& (R-SQ^{-1}P)^{-1}\\
  (Q-P R^{-1}S)^{-1} &&& (S-RP^{-1}Q)^{-1}
 \end{pmatrix}
\end{align}
under the assumption that all inverses, appearing in the equation \eqref{inverse} exist. Applying this result to the equation \eqref{propagator} we get,
\begin{eqnarray}
 C^{-1} & = & s(1-\Xi)-(\bar{m}+\bar{\Omega})(1-\Xi^{T})^{-1}(m+\Omega),\\
 D^{-1} & = & s(1-\Xi)(m+\Omega)^{-1}(1-\Xi^{T})-(\bar{m}+\bar{\Omega}), \\
 \bar{D}^{-1} & = & s(1-\Xi^{T})(\bar{m}+\bar{\Omega})^{-1}(1-\Xi)-(m+\Omega),
\end{eqnarray}
where $s=p^{2}$. Taking the inverse and keeping the calculation upto one-loop order, 
\begin{eqnarray}
 C & = & \left[ s(1-\Xi)(1-\Xi^{T})-(\bar{m}m+\bar{m}\Omega+m \bar{\Omega})  \right]^{-1}(1-\Xi^{T}) = C^{T} \nonumber \\
   & = & \dfrac{1-\Xi^{T}}{s(1-(\Xi+\Xi^{T}))-(\bar{m}m+\bar{m}\Omega+m\bar{\Omega})} \nonumber  \\
   & = & \dfrac{1-\Xi^{T}}{\left[ s-m^{2}-(s(\Xi+\Xi^{T})+m\bar{\Xi}+\bar{m}\Omega)  \right]} \nonumber \\
   & = & \dfrac{1-\Xi^{T}}{(s-m^{2})\left[ 1-\dfrac{(s(\Xi+\Xi^{T})+m\bar{\Omega}+\bar{m}\Omega)}{s-m^{2}} \right]} \nonumber \\
   & = & \dfrac{1-\Xi^{T}}{(s-m^{2})}\left[ 1+\dfrac{s(\Xi+\Xi^{T})+m\bar{\Omega}+\bar{m}\Omega}{s-m^{2}} \right] \nonumber \\
   & = & \dfrac{1}{s-m^{2}}+\dfrac{s\Xi+m^{2}\Xi^{T}+m\bar{\Omega}+\bar{m}\Omega}{(s-m^{2})^{2}}
\end{eqnarray}
Similarly, for the mass insertion section it can found out that,
\begin{eqnarray}
 D & = & \dfrac{m}{s-m^{2}}+\dfrac{s(m\Xi+m\Xi^{T}+\Omega)+m^{2}\bar{\Omega}}{(s-m^{2})^{2}} \\
 \bar{D} & = & \dfrac{\bar{m}}{s-m^{2}}+\dfrac{\bar{m}s(\Xi+\Xi^{T})+s\bar{\Omega}+\bar{m}^{2}\Omega}{(s-m^{2})^{2}}= (D)^{*}.
\end{eqnarray}
To calculate the CP asymmetry let us consider the tree level diagrams with the resummed propagators. The amplitudes for the tree level diagrams with the resummed propagators (denoted by subscript $i$ and $j$) can be written as 
\begin{equation}
 \mathcal{M}_{i}=D_{ii}x_{l}^{\dagger}y_{\Psi}^{\dagger}y_{i}^{*}h_{i\alpha}^{*}+C_{ii}x_{l}^{\dagger}\bar{\sigma}px_{\Psi}y_{i}h_{i\alpha}^{*},
\end{equation}
\begin{equation}
 \mathcal{M}_{j}=D_{jj}x_{l}^{\dagger}y_{\Psi}^{\dagger}y_{j}^{*}h_{j\alpha}^{*}+C_{jj}x_{l}^{\dagger}\bar{\sigma}px_{\Psi}y_{j}h_{j\alpha}^{*}.
\end{equation}

With the total amplitude $\mathcal{M}$ being,

\begin{equation}
 \mathcal{M}=\mathcal{M}_{i}+\mathcal{M}_{j}.
\end{equation}

Taking the interference, the asymmetry in the amplitude level can be found to be 
\begin{align}\label{eq:asymmetry_resummed}
  \delta &  \nonumber =  \mid \mathcal{M} \mid^{2}-\mid \bar{\mathcal{M}} 
  \mid^{2} \\
  & \nonumber = {\rm Im}[y_{i}^{*}h_{i\alpha}^{*}y_{j}h_{j\alpha}]{\rm Im}[D_{i}D_{j}^{*}]{\rm Tr}[p_{l}.\sigma p_{\Psi}.\bar{\sigma}] \\  &  \nonumber + {\rm Im}[y_{i}y_{j}^{*}h_{i\alpha}^{*}h_{j\alpha}]{\rm Im}[C_{i}C_{j}^{*}]{\rm Tr}[p_{l}.\sigma \overline{\sigma}.p p_{\Psi}.\sigma \bar{\sigma}.p] \\ & + \left[ {\rm Im}[y_{i}h_{i\alpha}^{*}y_{j}h_{j\alpha}]{\rm Im}[C_{i}D_{j}^{*}]+{\rm Im}[y_{i}^{*}h_{i\alpha}^{*}y_{j}^{*}h_{j\alpha}]{\rm Im}[D_{i}C_{j}^{*}]\right]{\rm Tr}[p_{l}.\sigma \bar{\sigma}.p]m_{\Psi}.
\end{align}
The relevant quantities coming from the interference of the resummed propagators can be found out to be  (keeping the calculation upto one-loop order),
\begin{align}\label{eq:resummed1}
 {\rm Im}[D_{i}D_{j}^{*}] & \nonumber = {\rm Im} \left[ \dfrac{M_{i}}{p^{2}-M_{i}^{2}} \left( \dfrac{p^{2}(M_{j}\Xi_{j}^{*}+M_{j}\Xi_{j})+p^{2}\Omega_{j}^{*}+M_{j}\bar{\Omega_{j}^{*}}}{(p^{2}-M_{j}^{2})^{2}} \right) \right] \\ & +{\rm Im}\left[ \dfrac{M_{j}}{p^{2}-M_{j}^{2}} \left(\dfrac{p^{2}(M_{i}\Xi_{i}+M_{i}\Xi_{i}^{*})+p^{2}\Omega_{j}+M_{i}^{2}\Omega_{i}^{*}}{(p^{2}-M_{i}^{2})^{2}}  \right) \right]
\end{align}

Similarly,  

\begin{align}\label{eq:resummed2}
 {\rm Im}[C_{i}C_{j}^{*}]= & {\rm Im} \left[ \dfrac{p^{2}\Xi_{j}^{*}+M_{j}^{2}\Xi_{j}+M_{j}\Omega_{j}+M_{j}\Omega_{j}^{*}}{(p^{2}-M_{i}^{2})(p^{2}-M_{j}^{2})^{2}}+ \dfrac{p^{2}\Xi_{i}+M_{i}\Xi_{i}^{*}+M_{i}\Omega_{i}^{*}+M_{i}\Omega_{i}}{(p^{2}-M_{i}^{2})^{2}(p^{2}-M_{j}^{2})^{2}} \right],
\end{align}

\begin{align}\label{eq:resummed3}
 {\rm Im}[C_{i}D_{j}^{*}]= & {\rm Im} \left[ \dfrac{p^{2}(M_{i}\Xi_{i}^{*}+M_{i}\Xi_{i}^{*}+\Omega_{i}^{*})+M_{i}\Omega_{i}^{*}}{(p^{2}-M_{j}^{2})(p^{2}-M_{i}^{2})^{2}}+\dfrac{M_{i}(p^{2}\Xi_{j}^{*}+M_{j}^{2}\Xi_{j}+M_{j}\Omega_{j}+M_{j}\Omega_{j}^{*})}{(p^{2}-M_{i}^{2})(p^{2}-M_{j}^{2})^{2}} \right],
\end{align}

\begin{align}\label{eq:resummed4}
 {\rm Im} [D_{i}C_{j}^{*}] = & {\rm Im} \left[ \dfrac{p^{2}(M_{i}\Xi_{i}+M_{i}\Xi_{i}^{*}+\Omega_{i})+M_{i}^{2}\Omega_{j}^{*}}{(p^{2}-M_{j}^{2})(p^{2}-M_{i}^{2})^{2}} + \dfrac{M_{i}(p^{2}\Xi_{j}^{*}+M_{j}^{2}\Xi_{j}+M_{j}\Omega_{j}+M_{j}\Omega_{j}^{*})}{(p^{2}-M_{i}^{2})(p^{2}-M_{j}^{2})^{2}} \right].
\end{align}

\begin{figure}
 \includegraphics[scale=0.4]{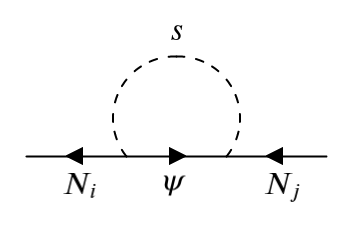}
 \includegraphics[scale=0.4]{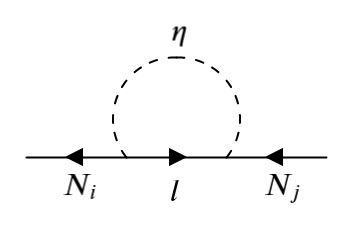}
 \caption{Diagrammatic representation of $(\Xi)_{i}^{j}$.}
 \label{fig:SE1}
\end{figure}

\begin{figure}
 \includegraphics[scale=0.4]{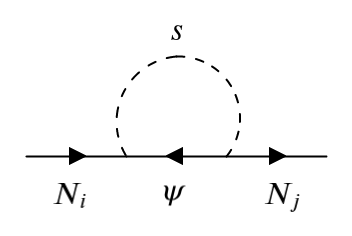}
 \includegraphics[scale=0.4]{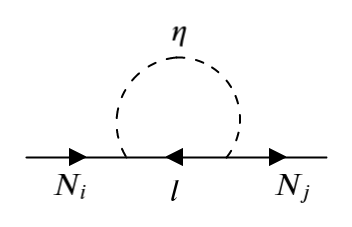}
 \caption{Diagrammatic representation of $(\Xi)^{i}_{j}$.}
 \label{fig:SE2}
\end{figure}

\begin{figure}[h]
 \includegraphics[scale=0.4]{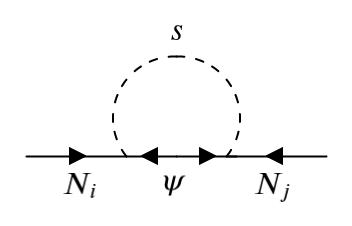}
 \includegraphics[scale=0.4]{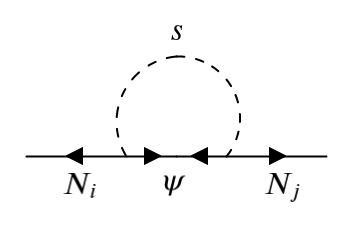}
 \caption{Diagrammatic representation of $(\Omega)^{ij}$ and $(\overline{\Omega})_{ij}$.}
 \label{fig:SE3}.
\end{figure}

For our case the self-energy functions in two component spinor notation are diagrammatically shown in figure \ref{fig:SE1}, \ref{fig:SE2} and \ref{fig:SE3} and they can be found out to be, 

\begin{equation}
 (\Xi)^{i}_{j}=\dfrac{y_{i}^{*}y_{j}}{16\pi^{2}}I_{\rm FS}(s,m_{\Psi}^{2},m_{S}^{2})+ \sum_{l}  \dfrac{h_{il}^{*}h_{jl}}{16\pi^{2}} I_{\rm FS}(s,m_{l}^{2},m_{\eta}^{2})=(\Xi),
\end{equation}

\begin{equation}
 (\Xi^{T})^{i}_{j}= \dfrac{y_{i}y_{j}^{*}}{16\pi^{2}}I_{\rm FS}(s,m_{\Psi}^{2},m_{S}^{2})+\sum_{l} \dfrac{h_{il} h_{jl}^{*}}{16\pi^{2}} I_{\rm FS} (s,m_{l}^{2},m_{\eta}^{2})=(\Xi^{T}),
\end{equation}

\begin{equation}
 (\Omega)^{ij}= \dfrac{y_{i}y_{j}}{16\pi^{2}}m_{\Psi} I_{\overline{\rm FS}}(s,m_{\Psi}^{2},m_{S}^{2})=(\Omega),
\end{equation}

\begin{equation}
 (\bar{\Omega})_{ij}=\dfrac{y_{i}^{*}y_{j}^{*}}{16\pi^{2}}m_{\Psi} I_{\overline{\rm FS}}(s,m_{\Psi}^{2},m_{S}^{2})= (\bar{\Omega}).
\end{equation}

Here,

\begin{align}
I_{\rm FS}(s,x,y)&=\dfrac{1}{2\epsilon}+\left[ \dfrac{(s+x-y)B_{0}(s;x,y)+A_{0}(x)-A_{0}(y)}{2s}  \right]. \\
I_{\overline{\rm FS}}(s,x,y) &= \frac{1}{\epsilon} - B_0(s;x,y)
\end{align} 
where, $A_{0}$ and $B_{0}$ are the Passarino-Veltman functions. For identical right handed neutrinos $N_i$ on external legs, we use the notation $\Xi_i, \Omega_i$, as seen in equation \eqref{eq:resummed1}, \eqref{eq:resummed2}, \eqref{eq:resummed3}, \eqref{eq:resummed4} mentioned above.

Using equations \eqref{eq:resummed1},\eqref{eq:resummed2},\eqref{eq:resummed3} and \eqref{eq:resummed4} in equation \eqref{eq:asymmetry_resummed} we get final expression of the CP asymmetry. To find the CP asymmetry parameter $\epsilon_{\psi}$, defined in equation \eqref{epsilonpsi} we perform the three body phase space integration numerically without any assumption. While we do not write the final CP asymmetry expression in this subsection, it is identical to the one derived using the interference of tree-loop diagrams as we show in the next subsection.

\subsection{CP asymmetry calculation from tree-loop interference}
\label{appen1a}

\begin{figure}[h]
 \begin{center}
 \includegraphics[scale=0.4]{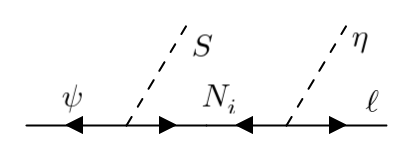}  
 \includegraphics[scale=0.4]{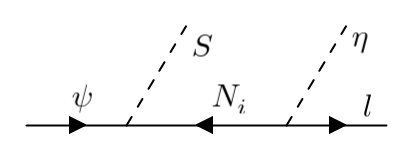}
 \caption{Tree level diagram contributing to the three body decay of $\psi$.}
  \label{fig:treediag}
 \end{center}
\end{figure}

\begin{figure}[h]
 \includegraphics[scale=0.3]{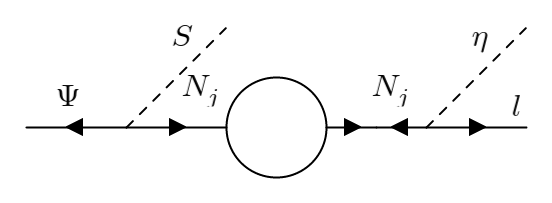}
 \includegraphics[scale=0.3]{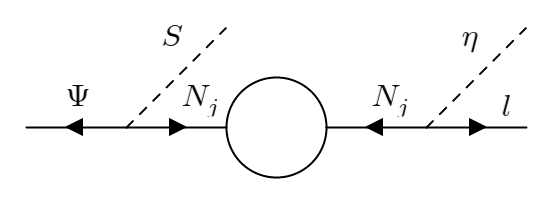}
 \includegraphics[scale=0.3]{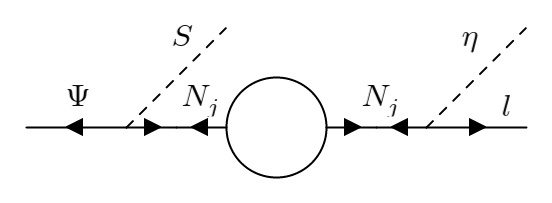}
 \includegraphics[scale=0.3]{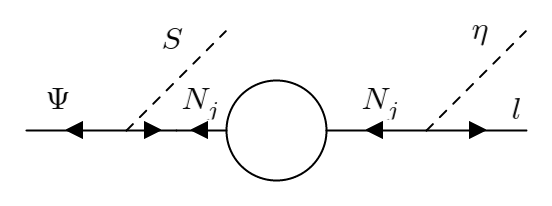}
 \caption{Feynman diagrams contributing to the three body decay of $\Psi$ at one-loop level.}
 \label{fig:loop11}
\end{figure}

\begin{figure}[h]
 \includegraphics[scale=0.3]{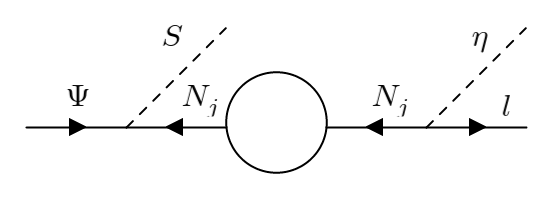}
 \includegraphics[scale=0.3]{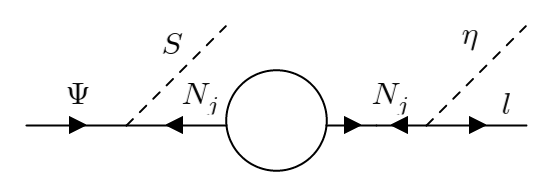}
 \includegraphics[scale=0.3]{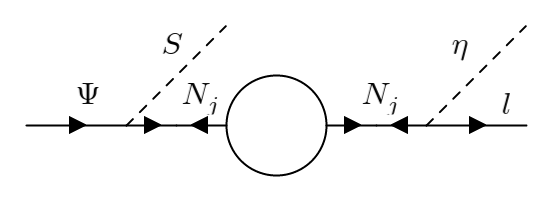}
 \includegraphics[scale=0.3]{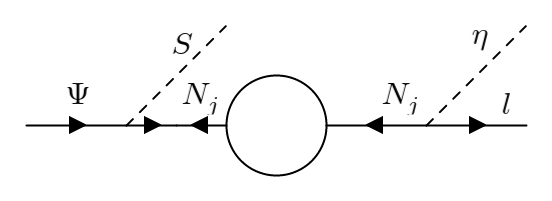}
 \caption{Feynman diagrams contributing to the three body decay of $\Psi$ at one-loop level.}
 \label{fig:loop22}
\end{figure}

 In this appendix we calculate  asymmetry parameter from the interference of tree and one-loop diagrams. The relevant diagrams in the tree level are shown in figure \ref{fig:treediag}. The amplitudes for the tree level diagram can be written as,
\begin{equation}
 i\mathcal{M}_{0}^{i}=y_{i}^{*}h_{i\alpha}^{*}\dfrac{x_{l}^{\dagger}y_{\Psi}^{\dagger}}{(p^{2}-M_{i}^{2})}M_{i}+y_{i}h_{i\alpha}^{*}\dfrac{x_{l}^{\dagger}\overline{\sigma}.p x_{\Psi}}{(p^{2}-M_{i}^{2})}.
\end{equation}
For a Majorana fermion $\psi$ there are two sets of diagrams contributing to the three body decay at one-loop. These diagrams are shown in figure \ref{fig:loop11} and \ref{fig:loop22} respectively. 
For the diagrams in figure \ref{fig:loop11}, the corresponding amplitudes can be written as,
\begin{align}
 i \mathcal{M}_{j}^{1'}& \nonumber =  y_{j}^{*}h_{j\alpha}^{*}\dfrac{x_{l}^{\dagger}\overline{\sigma}.p \sigma.p y_{\Psi}^{\dagger}(\Xi)_{j}^{T}M_{j}}{(p^{2}-M_{j}^{2})^{2}}+ y_{j}^{*}h_{j\alpha}^{*}\dfrac{x_{l}^{\dagger} \overline{\sigma}.p (\Omega)_{j}\sigma.p y_{\Psi}^{\dagger}}{(p^{2}-M_{j}^{2})^{2}} \\ & \nonumber + y_{j}^{*}h_{j\alpha}^{*} \dfrac{x_{l}^{\dagger} \overline{\sigma}.p \sigma.p y_{\Psi}^{\dagger}M_{j}(\Xi)_{j}}{(p^{2}-M_{j}^{2})^{2}}+y_{j}^{*}h_{j\alpha}^{*} \dfrac{x_{l}^{\dagger} M_{j}^{2} (\overline{\Omega})_{j} y_{\Psi}^{\dagger}}{(p^{2}-M_{j}^{2})^{2}} \\  & = y_{j}^{*}h_{j\alpha}^{*}\dfrac{x_{l}^{\dagger}y_{\Psi}^{\dagger}}{(p^{2}-M_{j}^{2})} \left[ p^{2}[M_{j}(\Xi)_{j}^{T}+M_{j}(\Xi)_{j}+(\Omega)_{j}]+M_{j}^{2}(\overline{\Omega})_{j} \right].
\end{align}
Similarly, for the other set of four one-loop diagrams in figure \ref{fig:loop22}, the amplitude can be written as,
\begin{align}
 i \mathcal{M}_{j}^{1''} & \nonumber = y_{j}h_{j\alpha}^{*} \dfrac{x_{l}^{\dagger} \overline{\sigma.p} \sigma.p \overline{\sigma}.p (\Xi)_{j}x_{\Psi}}{(p^{2}-M_{j}^{2})^{2}} + y_{j}h_{j\alpha}^{*}\dfrac{x_{l}^{\dagger} \overline{\sigma}.p M_{j} (\overline{\Omega})_{j}x_{\Psi}}{(p^{2}-M_{j}^{2})^{2}} \\ 
 & +y_{j}h_{j\alpha}^{*} \dfrac{x_{l}^{\dagger} \overline{\sigma}.p x_{\Psi}M_{j}^{2}(\Xi)_{j}^{T}}{(p^{2}-M_{j}^{2})^{2}} + y_{j}h_{j\alpha}^{*} \dfrac{x_{l}^{\dagger} \overline{\sigma}.p x_{\Psi} M_{j}(\Omega)_{j}}{(p^{2}-M_{j}^{2})^{2}} \\ \nonumber & =y_{j}h_{j\alpha}^{*} \dfrac{x_{l}^{\dagger}\overline{\sigma}.p x_{\Psi}}{(p^{2}-M_{j}^{2})} \left[ p^{2}(\Xi)_{j}+M_{j}^{2} (\Xi)_{j}^{T}+M_{j} (\Omega)_{j}+M_{j} (\overline{\Omega})_{j} \right].
\end{align}
Therefore, the total amplitude for the decay at one-loop level can be written as, 
\begin{equation}
 \mathcal{M}_{j}^{1}=\mathcal{M}_{j}^{1'}+\mathcal{M}_{j}^{1''}.
\end{equation}
The asymmetry parameter is given by 
\begin{align}\label{eq:asymmetry_loop}
 \delta  & \nonumber = \mid \mathcal{M} \mid^{2}-\mid \overline{\mathcal{M}} \mid^{2} \\ & \nonumber =4{\rm Im}[\mathcal{M}_{i}^{0}\mathcal{M}_{j}^{1*}+\mathcal{M}_{i}^{1}\mathcal{M}_{j}^{0*}]  \\ & \nonumber = 4{\rm Im}[y_{i}^{*}h_{i\alpha}^{*}y_{j}h_{j\alpha}]  \left[  \dfrac{{\rm Im}[M_{i}(p^{2}[M_{j}(\Xi)_{j}^{T}+M_{j}(\Xi)_{j}^{*}+\Omega_{j}^{*}]+M_{j}^{2}\bar{\Omega}_{j}^{*})]}{(p^{2}-M_{i}^{2})(p^{2}-M_{j}^{2})^{2}} \right] {\rm Tr}[p_{l}.\sigma p_{\Psi}.\overline{\sigma}]  \\ \nonumber & + 4 {\rm Im}[y_{i}^{*}h_{i\alpha}^{*}y_{j}h_{j\alpha}] \left[ \dfrac{ {\rm Im}[M_{j}(p^{2}[M_{i}(\Xi)_{i}^{T}+M_{i}(\Xi)_{i}+\Omega_{i}]+M_{i}^{2}\overline{\Omega}_{i})]}{(p^{2}-M_{j}^{2})(p^{2}-M_{i}^{2})^{2}} \right] {\rm Tr}[p_{l}.\sigma p_{\Psi}.\overline{\sigma}]  \\ \nonumber & + 4 {\rm Im}[y_{i}h_{i\alpha}^{*}y_{j}^{*}h_{j\alpha}] \left[ \dfrac{{\rm Im}[p^{2}(\Xi)_{j}^{*}+M_{j}^{2}(\Xi)_{j}+M_{j}\Omega_{j}^{*}+M_{j}\overline{\Omega_{j}^{*}}]}{(p^{2}-M_{i}^{2})(p^{2}-M_{j}^{2})^{2}} \right] {\rm Tr}[p_{l}.\sigma p.\overline{\sigma} p_{\Psi}.\sigma p.\overline{\sigma}] \\ & \nonumber + 4 {\rm Im}[y_{i}h_{i\alpha}^{*}y_{j}^{*}h_{j\alpha}] \left[ \dfrac{Im[p^{2}(\Xi)_{i}+M_{i}^{2}(\Xi)_{i}^{T}+M_{i}\Omega_{i}+M_{i}\overline{\Omega_{i}}]}{(p^{2}-M_{i}^{2})^{2}(p^{2}-M_{j}^{2})} \right] {\rm Tr}[p_{l}.\sigma p.\overline{\sigma} p_{\Psi}.\sigma p.\overline{\sigma}]  \\ & \nonumber  + 4 {\rm Im}[y_{i}^{*}h_{i\alpha}^{*}y_{j}^{*}h_{j\alpha}] \left[  \dfrac{ {\rm Im}[p^{2}(\Xi)_{j}^{*}+M_{j}^{2}(\Xi^{T})_{j}^{*}+M_{j}\Omega_{j}+M_{j}\overline{\Omega_{j}}]}{(p^{2}-M_{i}^{2})(p^{2}-M_{j}^{2})^{2}} \right]M_{i}m_{\Psi} {\rm Tr}[p_{l}.\sigma \overline{\sigma}.p] \\ & \nonumber  +   4 {\rm Im}[y_{i}^{*}h_{i\alpha}^{*}y_{j}^{*}h_{j\alpha}] \left[  \dfrac{ {\rm Im}[p^{2}(M_{i}(\Xi)_{i}^{T}+M_{i}(\Xi)_{i}+\Omega_{i})+M_{i}^{2}(\overline{\Omega})_{i}]}{(p^{2}-M_{i}^{2})^{2}(p^{2}-M_{j}^{2})} \right]m_{\Psi} {\rm Tr}[p_{l}.\sigma \overline{\sigma}.p] \\ & +\nonumber 4 {\rm Im}[y_{i}h_{i\alpha}^{*}y_{j}h_{j\alpha}] \left[ \dfrac{ {\rm Im}[p^{2}[M_{j}(\Xi)_{j}^{*}+M_{j}(\Xi^{T})^{*}_{j}+\Omega_{j}^{*}]+M_{j}^{2}\bar{\Omega_{j}^{*}}}{(p^{2}-M_{i}^{2})(p^{2}-M_{j}^{2})^{2}]}  \right] {\rm Tr}[p_{l}.\sigma \bar{\sigma}.p] m_{\Psi}  \\   & + 4 {\rm Im}[y_{i}h_{i \alpha}^{*}y_{j}h_{j\alpha}] \left[  \dfrac{ {\rm Im}[p^{2}(\Xi)_{i}+M_{i}^{2}(\Xi)_{i}^{T}+M_{i}\Omega_{i}+M_{i}\overline{\Omega_{i}}]}{(p^{2}-M_{i}^{2})^{2}(p^{2}-M_{j}^{2})} \right] M_{j} m_{\Psi} {\rm Tr}[p_{l}.\sigma \overline{\sigma}.p]
\end{align}

Note that for the full loop corrected propagators $(\Xi^{T})^{*}=\Xi$ and $\overline{\Omega}=\Omega^{*}$. One can easily figure out that the asymmetry we have from tree-loop interference calculation matches with the asymmetry we got in equation \eqref{eq:asymmetry_resummed} after replacing the resummed propagators.



\section{Two body decay of $N_i$}
\label{appen2}
The decay width for the decay $N_{1} \longrightarrow \eta l$ is given by 
\begin{equation}
  \Gamma_{N_{1}\longrightarrow \eta l} =\dfrac{M_{1}}{8 \pi}(h^{\dagger}h)_{11}\left(1-\dfrac{m_{\eta}^2}{M_{1}^2}\right)^2
\end{equation}

The CP asymmetry parameter for $N_1 \rightarrow l_{i} \eta, \bar{l_{i}}\bar{\eta}$ is given by 
\begin{equation}
\epsilon_{(N_1)_i} = \frac{1}{8 \pi (h^{\dagger}h)_{11}} \bigg [ f \left( \frac{M^2_2}{M^2_1}, \frac{m^2_{\eta}}{M^2_1} \right) {\rm Im} [ h^*_{i1} h_{i2} (h^{\dagger} h)_{12}] - \frac{M^2_1}{M^2_2-M^2_1} \left( 1-\frac{m^2_{\eta}}{M^2_1} \right)^2 {\rm Im}[h^*_{i1} h_{i2} H_{12}] \bigg ]
\label{epsilonflav}
\end{equation}
where, the function $f(r_{ji},\eta_{i})$ is coming from the interference of the tree-level and one loop diagrams and has the form
\begin{equation}
f(r_{ji},\eta_{i})=\sqrt{r_{ji}}\left[1+\frac{(1-2\eta_{i}+r_{ji})}{(1-\eta_{i}^{2})^{2}}{\rm ln}(\frac{r_{ji}-\eta_{i}^{2}}{1-2\eta_{i}+r_{ji}})\right]
\end{equation}
with $r_{ji}=M_{j}^{2}/M_{i}^{2}$ and $\eta_{i}=m_{\eta}^{2}/M_{i}^{2}$. The self energy contribution $H_{ij}$ is given by 
\begin{equation}
H_{ij} = (h^{\dagger} h)_{ij} \frac{M_j}{M_i} + (h^{\dagger} h)^*_{ij}
\end{equation}
Now, the CP asymmetry parameter, neglecting the flavour effects (summing over final state flavours $\alpha$) is
\begin{equation}
\epsilon_{N_1}=\frac{1}{8\pi(h^{\dagger}h)_{11}} {\rm Im}[((h^{\dagger}h)_{12})^{2}]\frac{1}{\sqrt{r_{21}}}F(r_{21},\eta_{1})
 \label{eq:14}
\end{equation}
\\
where the function $F(r_{ji},\eta)$ is defined as 
\begin{equation}
F(r_{ji},\eta_{i})=\sqrt{r_{ji}}\left[ f(r_{ji},\eta_{i})-\frac{\sqrt{r_{ji}}}{r_{ji}-1}(1-\eta_{i})^{2} \right].
\end{equation}

\providecommand{\href}[2]{#2}\begingroup\raggedright\endgroup

\end{document}